\documentclass[onefignum,onetabnum]{siamonline190516}

\pdfoutput=1

\usepackage{lipsum}
\usepackage{amsfonts}
\usepackage{amsopn}
\usepackage{bm}                 
\usepackage{array}
\usepackage{booktabs}
\usepackage[section]{placeins}
\usepackage[numbers,sort&compress]{natbib}

\ifpdf
  \DeclareGraphicsExtensions{.eps,.pdf,.png,.jpg}
\else
  \DeclareGraphicsExtensions{.eps}
\fi

\usepackage{enumitem}
\setlist[enumerate]{leftmargin=.5in}
\setlist[itemize]{leftmargin=.5in}

\newsiamremark{remark}{Remark}
\newsiamremark{hypothesis}{Hypothesis}
\crefname{hypothesis}{Hypothesis}{Hypotheses}
\newsiamthm{claim}{Claim}

\headers{Sparse Inpainting with Smoothed Particle Hydrodynamics}{V. Daropoulos, 
M. Augustin, and J. Weickert}

\title{Sparse Inpainting with Smoothed Particle Hydrodynamics\thanks{Submitted 
to the editors DATE%
}}

\author{Viktor Daropoulos\thanks{Interactive Graphics and Simulation Group, 
Department of Computer Science, Universit\"{a}t Innsbruck, 
Technikerstra{\ss}e 21a, 6020 Innsbruck, Austria
(\email{viktor.daropoulos@uibk.ac.at})}
\and Matthias Augustin\thanks{Mathematical Image Analysis Group, 
Faculty of Mathematics and Computer Science,
Campus E1.7, Saarland University, 
66041 Saarbr\"{u}cken, Germany 
(\email{augustin@mia.uni-saarland.de, weickert@mia.uni-saarland.de}).}
\and Joachim Weickert\footnotemark[3]}

\makeatletter
\newcommand*{\addFileDependency}[1]{
  \typeout{(#1)}
  \@addtofilelist{#1}
  \IfFileExists{#1}{}{\typeout{No file #1.}}
}
\makeatother
 
\newcommand*{\myexternaldocument}[1]{%
    \externaldocument{#1}%
    \addFileDependency{#1.tex}%
    \addFileDependency{#1.aux}%
}

\ifpdf
\hypersetup{
  pdftitle={Sparse Inpainting with Smoothed Particle Hydrodynamics},
  pdfauthor={V. Daropoulos, M. Augustin, and J. Weickert}
}
\fi

\myexternaldocument{daropoulos_sph_supplement}

\newfloatcommand{capbtabbox}{table}[][\FBwidth]
\begin{document}

\maketitle

\begin{abstract}
 Digital image inpainting refers to techniques used to reconstruct a damaged or 
 incomplete image by exploiting available image information. The main goal of 
 this work is to perform the image inpainting process from a set of sparsely 
 distributed image samples with the Smoothed Particle Hydrodynamics (SPH) 
 technique. As, in its naive formulation, the SPH technique is not even capable 
 of reproducing constant functions, we modify the approach to obtain 
 an approximation which can reproduce constant and linear functions. 
 Furthermore, we examine the use of Voronoi tessellation for defining the 
 necessary parameters in the SPH method as well as selecting optimally located 
 image samples. In addition to this spatial optimization, optimization of data 
 values is also implemented in order to further improve the results. Apart from 
 a traditional Gaussian smoothing kernel, we assess the performance of other 
 kernels on both random and spatially optimized masks. Since the use of 
 isotropic smoothing kernels is not optimal in the presence of objects with a 
 clear preferred orientation in the image, we also examine anisotropic 
 smoothing kernels. Our final algorithm can compete with well-performing 
 sparse inpainting techniques based on homogeneous or anisotropic diffusion 
 processes as well as with exemplar-based approaches.
\end{abstract}

\begin{keywords}
 Inpainting, Smoothed Particle Hydrodynamics, Data Optimization, Voronoi-based 
 Densification, Mixed Consistency Method, 
 Meshless Interpolation
\end{keywords}

\begin{AMS}
  65D18, 68U10, 76M28, 94A08
\end{AMS}


\section{Introduction}
Image inpainting aims at restoring partially damaged image or missing 
parts of an image in a 
visually appealing manner \cite{BS00}. It has a wide 
number of practical applications such as 
art restoration \cite{KM13,BF08,CA18,RC11},
object removal \cite{CP04}, 
medical imaging \cite{TB19},
inpainting of optical flow fields \cite{RO20},
video inpainting \cite{NA14},
inpainting reflectance/height values in LiDAR images \cite{BA19,CL18},
image compression \cite{GW08,SP14}, and even image 
denoising \cite{AP17}. The term ``inpainting'' itself was introduced for 
digital images by Bertalm{\'{\i}}o et al.~in \cite{BS00}, but similar 
concepts were already explored in earlier work under different names 
such as image restoration, interpolation, disocclusion,
or amodal completion 
\cite{Hu73,OABB85,Fe94,CM98,MM98}. 

Any inpainting model needs to assume some kind of relation between known 
and unknown data. As there is a variety of plausible assumptions for such 
relations, many solutions to an inpainting problem exist. 

Based on the underlying assumptions, the inpainting methods from the 
literature can be grouped into certain main categories \cite{BC14,GL14}. 
One class is based on variational models and partial differential equations 
(PDEs) \cite{Sc15}, 
comprising e.g.~Euler's elastica \cite{NMS93,MM98,CK02,BC10,CP19}, 
transport-like equations \cite{BS00,BM07}, anisotropic diffusion processes
\cite{WW06,GW08,BU13}, harmonic and biharmonic inpainting \cite{CS02,GW08}, 
total variation restoration \cite{CS02}, 
the Mumford-Shah functional \cite{CS02,ES02},
and the Cahn--Hilliard equation \cite{BEG07,BHS09}.
Exemplar-based approaches emerged from texture synthesis and 
exploit the notion of patch similarity \cite{EL99,CP04,FA09,AF11}.
Other techniques 
rely on overcomplete dictionaries and the concept of sparsity 
\cite{ES05,ME08,El10}, and more recently also deep learning 
concepts have been proposed \cite{PKDD16,ISI17,UVL18}. 
Each of these strategies has its advantages and disadvantages depending on the 
type of image it is applied to. For example, exemplar-based techniques perform 
fairly well on highly textured images, PDE-based methods are more 
suited for geometrical structures, and deep learning approaches can capture
high-level semantics from images. This has led to the development of hybrid 
approaches which combine the strengths of different methods 
\cite{BV03,SE05,AL10,PW15}. 

A subclass of inpainting problems deals with the recovery of a whole image from 
a small amount of sparsely distributed data \cite{AA17,BU13,FA09,HM17}. These 
kind of problems are encountered particularly in the context of compression 
\cite{GW08,CR14,SP14,Pe19}. The sparsity of available data makes it 
feasibly to consider scattered data interpolation,
e.g.~by radial basis functions 
\cite{MD05,We05,US06,Fa07,LZ12,CL18} or by Shepard interpolation 
\cite{Sh68,KW93,AA17,Pe19} as inpainting technique.

A key observation for applications in compression is that the data can be 
chosen freely from the image. Thus, a careful selection of the sparse set 
of pixels to store such that it fits a chosen inpainting method is 
essential for a good performance \cite{GW08,MH11,CR14,SP14,HM17,KB18}. 
Astonishingly, simple linear methods such as homogeneous diffusion 
inpainting show remarkable quality if combined with optimally chosen data 
\cite{BB09,GW08,HS13,CR14,PH16,BLPP17} and can even compete with the widely 
used JPEG \cite{PM92} and JPEG2000 \cite{TM02} standards \cite{MB11,HM13,PH16}. 
Anisotropic diffusion approaches perform even better \cite{GW08,SP14,HM17} 
and can outperform JPEG and JPEG2000 for high compression ratios.


\subsection{Goals and Contributions}
The goal of our paper is to show that a hitherto hardly explored class
of scattered data interpolation methods based on Smoothed Particle 
Hydrodynamics (SPH) can provide excellent results on sparse inpainting
problems, if one improves them with a number of refined concepts. 

SPH was originally introduced to solve astrophysical problems \cite{Lu77}, 
but has also been applied to problems that deal with large deformations 
\cite{BS08}, computational fluid mechanics \cite{Mo94}, and soil mechanics 
\cite{MH05}. In SPH, the solution to a given problem is represented by a 
set of particles and functions. Derivatives and integrals are approximated 
using those particles.
 
Di Blasi et al.~\cite{DF11} have introduced the SPH method for sparse
image interpolation problems, and applications to non-sparse inpainting 
are studied in \cite{AP16}. We have not found more work on SPH-based 
image inpainting. One reason for this lack of popularity might lie in the
fact that in the naive formulation, 
not even constant functions are interpolated correctly 
\cite{Fa07}.
However, in our paper we show that 
one can come up with highly competitive approaches by integrating more 
sophisticated concepts.
The modifications that we apply are mostly tailored to the 
particular situation encountered when inpainting is used as a strategy for 
compression. Compared to other, more classical applications of inpainting, the 
key difference when using inpainting for compression is that a ground truth 
image is known, such that the data used for inpainting can be adapted and
optimized with regard to this ground truth. Moreover, compression applications 
are particularly challenging, since they keep only a very sparse subset of the 
original data.

Our key contributions are the following:
\begin{enumerate}
 \item To define a measure for the area of influence of a given 
  particle (mask point), we combine a Voronoi tessellation with the Euclidean 
  distance transform.
 \item We restore particle consistency.
 \item We perform inpainting with a new method that adapts its 
  consistency order to the local approximation error.
 \item We use the Voronoi tessellation to propose a novel 
  strategy for spatial data optimization.
 \item We optimize not only the data locations, but also their
   values (tonal optimization). 
 \item To incorporate anisotropy in the process, we use anisotropic kernels.
 \item We assess the performance of different smoothing kernels and compare to 
  some of the best sparse inpainting methods for optimized data.
\end{enumerate}
In the context of image processing and reconstruction, 
related concepts have been used in combination with kernel regression methods,
e.g., by Takeda et al.\ in \cite{TF07}.


\subsection{Paper Structure}
This paper is organized as follows: In \cref{sec:SPH_basics}, we give a brief 
summary of the ideas behind Smoothed Particle Hydrodynamics. This includes its 
origin from an integral approximation, techniques to restore consistency in a 
discrete setting, and a brief overview on common smoothing kernels used for our 
experiments. \Cref{sec:inpainting} explains how SPH can be used for inpainting. 
Here we introduce Voronoi tessellation to determine parameters of the method.
Further, we compare performance of SPH inpainting with 
diffusion- and exemplar-based methods for examples of sparse inpainting and 
classical inpainting tasks.
For problems in which the ground truth is known, we
show how performance can be enhanced by combining results from methods of 
different consistency order
in \cref{sec:mask_optimization}. Here, we also
explain our data optimization strategies both with respect to data locations 
(spatial optimization) as well as data values (tonal optimization). We proceed 
by comparing results from our method to results from other techniques in
\cref{sec:comparisons} and draw conclusions in \cref{sec:conclusions}.


\section{SPH in a Nutshell}
\label{sec:SPH_basics}


\subsection{Essential Ideas}
\label{sec:formulation}

We are interested in approximating a function $f$ on the domain $\Omega$ in 
$\mathbb{R}^{2}$. The point of departure for SPH is the idea to replace the 
value of $f$ at a point $\bm{q}$ by a weighted average of the function, i.e.
\begin{equation}
 \label{eq:4}
  f\!\left(\bm{q}\right) \approx  
  \left\langle f\!\left(\bm{q}\right) \right\rangle
  \coloneqq
  \int_{\Omega} f\!\left(\bm{p}\right) \, 
   W\!\left(\bm{q}-\bm{p},h\right)\, \mathrm{d}\bm{p}.
\end{equation}
\Cref{eq:4} is also known as the \emph{kernel approximation} of the function
with the smoothing kernel $W(\cdot,h)$ and its smoothing length $h$, which
represents the effective width of $W$. The kernel should be a monotonically 
decreasing positive mollifier, i.e.\ it should have the following properties:
\begin{itemize}
 \item compactness: 
  $W(\bm{q}-\bm{p},h) = 0$ outside 
  a compact domain
  $K \subseteq \Omega$,
 \item unity: 
  $\int_{\Omega} W(\bm{q}-\bm{p},h)\, \mathrm{d}\bm{p} = 1$,
 \item limit behavior: 
  $W(\bm{q}-\bm{p},h) \xrightarrow{h \rightarrow 0} \delta(\bm{q}-\bm{p})$, 
  where $\delta(\bm{q}-\bm{p})$ is Dirac's delta distribution,
 \item positivity: 
  $W(\bm{q}-\bm{p},h) > 0$ over $K \subseteq \Omega$,
 \item monotonicity: 
  $W$ is monotonically decreasing function w.r.t.\
  $\left\lVert \bm{q} - \bm{p}\right\rVert$.
\end{itemize}
Here, $\lVert \cdot \rVert$ denotes the Euclidean distance. Positivity is not 
strictly necessary, but desired in order for the approximated function values to 
have physical meaning. Allowing the kernel to take negative values in parts of 
the domain can lead to unnatural approximated values and corrupt the entire 
computation \cite{LL03}. The same holds for monotonicity, which is connected 
with the usual behavior of physical forces to decrease with increasing distance.

Discretizing the integral of \cref{eq:4} yields the \emph{particle 
approximation} of $f$ given by
\begin{equation}
 \label{eq:particle_approx}
 f\!\left(\bm{q}\right) \approx u\!\left(\bm{q}\right) \coloneqq
 \sum\limits_{j \in \mathcal{N}(\bm{q})} f\!\left(\bm{p}_{j}\right)\, 
  W\!\left(\bm{q}-\bm{p}_{j},h\right)\, V_{j}.
\end{equation}
Here, in order to approximate the function value at point $\bm{q}$ we sum over 
its nearest neighbors $\bm{p}_{j}$, where $\mathcal{N}(\bm{q})$ denotes the 
index set of the nearest neighbors and we assume that $1 \leq j \leq M$ with $M$ 
the total number of particles under consideration. Each of these particles is 
related to a specific area of influence (or weight) $V_{j}$ in this quadrature 
rule. The particle approximation is interpolating at the particles $\bm{p}_{j}$ 
if the kernel $W$ satisfies 
\begin{equation}
  W\!\left(\bm{p}_{k}-\bm{p}_{j},h\right)\, V_{j} = \delta_{k,j} \quad 
  \text{for any} \quad
  j, k = 1,\ldots, M
\end{equation}
with the Kronecker delta $\delta_{k,j}$. This requirement is separate from 
the desired properties in the continuous setting and, in general, not satisfied 
by kernels with said properties. Some modifications which achieve interpolation 
at the particles are discussed in \cref{sec:consistency}.

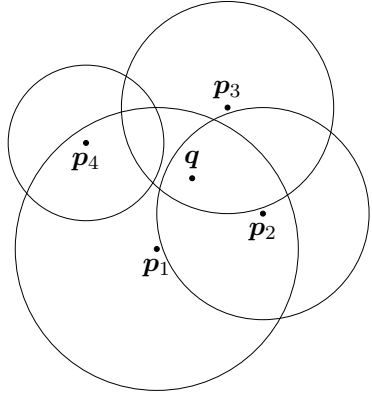
\begin{figure}[htb]
 \centering
 \begin{tikzpicture}
 \draw (2,2) circle (2cm);
 \filldraw (2,2) circle (1pt) node[align=left, below] {$\bm{p}_{1}$};
 \draw (3,4) circle (1.5cm);
 \draw (3.5,2.5) circle (1.5cm);
 \filldraw (3.5,2.5) circle (1pt) node[align=left, below] {$\bm{p}_{2}$};
 \filldraw (3,4) circle (1pt) node[align=left, above] {$\bm{p}_{3}$};
 \draw (1,3.5) circle (1.1cm);
 \filldraw (1,3.5) circle (1pt) node[align=left, below] {$\bm{p}_{4}$};
 \filldraw (2.5,3) circle (1pt) node[align=left, above] {$\bm{q}$};
 \end{tikzpicture}
 \caption[Scatter Approach]{Particles $\bm{p}_{1}$, $\bm{p}_{2}$, and 
  $\bm{p}_{3}$ are  considered neighbors of point $\bm{q}$, while $\bm{p}_{4}$
  is not.}
 \label{fig:Scatter}
\end{figure}

In order to determine the nearest neighbors, we use the so-called scatter 
approach. Here, the neighbors of a point $\bm{q}$, are the particles 
$\bm{p}_{j}$ that include $\bm{q}$ in the support domain of the kernels 
centered at these particles $\bm{p}_{j}$, cf.\ \cref{fig:Scatter}.

An alternative to the scatter approach is the gather approach \cite{Li09} in 
which a disk of a predetermined radius around $\bm{q}$ is considered and 
neighbors are determined by checking which particles $\bm{p}_{j}$ are located
within the disk. The scatter approach is preferable for our purpose as it is 
less dependent on the particle distribution and allows to consider different 
smoothing lengths $h_{j}$ for each particle.


\subsection{Restoring Consistency}
\label{sec:consistency}

Consistency in SPH is defined in the sense of (local) polynomial reproduction
\cite{We05}. While the restriction of kernels to positive mollifiers guarantees 
that linear polynomials are reconstructed in the continuous formulation 
\cref{eq:4}, this is no longer the case in the discrete formulation 
\cref{eq:particle_approx}; a phenomenon known as particle inconsistency 
\cite{Mo96}. Thus, \cref{eq:particle_approx} needs to be modified in order to 
restore consistency in the discrete setting. 

For the particle approximation to satisfy zero order consistency, it needs to 
be able to reproduce constants. This requirement is satisfied by the well known
Shepard interpolation formula \cite{Sh68,CB99,We05}
\begin{equation}
 \label{eq:new_discrete_shepard}
 u\!\left(\bm{q}\right) = 
 \frac{\sum\limits_{j \in \mathcal{N}(\bm{q})}  
  f\!\left(\bm{p}_{j}\right) \, 
  W\!\left(\bm{q}-\bm{p}_{j},h\right) \,V_{j}}{
  \sum\limits_{j \in \mathcal{N}(\bm{q})} W\!\left(\bm{q}-\bm{p}_{j},h\right)\, 
V_{j}}.
\end{equation}
The fact that zero order consistency already requires to modify 
\cref{eq:particle_approx} shows that, in general, using the particle 
approximation directly cannot even reproduce constant functions, i.e.\ 
\cref{eq:particle_approx} has no consistency.

One way to interpret Shepard interpolation is that the kernel $W$ is replaced by 
a modified kernel $\widetilde{W}$ such that
\begin{equation}
 \label{eq:mod_kernel_zero01} 
 \widetilde{W}\!\left(\bm{q}-\bm{p}_{j},h\right) = 
 b_{0}\!\left(\bm{q}\right)\, W\!\left(\bm{q}-\bm{p}_{j},h\right),
\end{equation}
and the original interpolation formula is used with the modified kernel, i.e.,
\begin{equation}
 \label{eq:SPH_mod_kernel}
 u\!\left(\bm{q}\right) =
 \sum\limits_{j \in \mathcal{N}(\bm{q})} f\!\left(\bm{p}_{j}\right)\, 
  \widetilde{W}\!\left(\bm{q}-\bm{p}_{j},h\right)\, V_{j}.
\end{equation}
By comparison with \cref{eq:new_discrete_shepard}, we see that 
$b_{0}\!\left(\bm{q}\right)$ is given by
\begin{equation}
 \label{eq:mod_kernel_zero02}
 b_{0}\!\left(\bm{q}\right) = \frac{1}{
  \sum\limits_{j \in \mathcal{N}(\bm{q})} W\!\left(\bm{q}-\bm{p}_{j},h\right)\, 
  V_{j}}.
\end{equation}
Thus, $b_{0}\!\left(\bm{q}\right)$ is a function of $\bm{q}$, but is constant 
with respect to the particle $\bm{p}_{j}$ for a fixed $\bm{q}$. In other words, 
zero order consistency can be restored by multiplying the kernel by a constant. 
This motivates the attempt to restore first order consistency by multiplying 
the kernel with a function which is linear in the difference 
$\bm{q}-\bm{p}_{j}$, i.e.
\begin{equation}
 \label{eq:mod_kernel_first01}
 \widetilde{W}\!\left(\bm{q}-\bm{p}_{j},h\right) = 
 \left(b_{0}\!\left(\bm{q}\right) 
 + b_{1}\!\left(\bm{q}\right) \left(x_{\bm{q}} - x_{\bm{p}_{j}}\right)
 + b_{2}\!\left(\bm{q}\right) \left(y_{\bm{q}} - y_{\bm{p}_{j}}\right) \right)
 \, W\!\left(\bm{q}-\bm{p}_{j},h\right),
\end{equation}
where 
$\bm{q} = (x_{\bm{q}}, y_{\bm{q}})^{T}$ and 
$\bm{p}_{j} = (x_{\bm{p}_{j}}, y_{\bm{p}_{j}})^{T}$.
For a fixed $\bm{q}$, this 
means that we have to determine three coefficients such that we can reproduce 
linear polynomials in $\mathbb{R}^{2}$. This yields the system of equations
\begin{align}
  \sum\limits_{j \in \mathcal{N}(\bm{q})} 
\widetilde{W}\!\left(\bm{q}-\bm{p}_{j},h\right)\, V_{j}
  =&\ 1,
  \label{eq:first_order01}\\
  \sum\limits_{j \in \mathcal{N}(\bm{q})} x_{\bm{p}_{j}} \, 
   \widetilde{W}\!\left(\bm{q}-\bm{p}_{j},h\right)\, V_{j}
  =&\ x_{\bm{q}},
  \label{eq:first_order02}\\
  \sum\limits_{j \in \mathcal{N}(\bm{q})} y_{\bm{p}_{j}} \, 
   \widetilde{W}\!\left(\bm{q}-\bm{p}_{j},h\right)\, V_{j}
  =&\ y_{\bm{q}}.
  \label{eq:first_order03}
\end{align}
\Cref{eq:first_order01} allows to multiply the right-hand sides of
\cref{eq:first_order02,eq:first_order03} by the left-hand side of 
\cref{eq:first_order01} without changing the equations, such that this linear 
system can be recast as
\begin{equation}
 \label{eq:first_order_comp}
 \bm{D}\!\left(\bm{q}\right)\, \bm{b}\!\left(\bm{q}\right) = \bm{e}
\end{equation}
if we define
\begin{equation}
 \bm{b}\!\left(\bm{q}\right) = 
  \begin{pmatrix}
   b_{0}\!\left(\bm{q}\right) \\
   b_{1}\!\left(\bm{q}\right) \\
   b_{2}\!\left(\bm{q}\right) 
 \end{pmatrix}, 
 \qquad
 \bm{e} = 
  \begin{pmatrix}
   1 \\
   0 \\
   0 
 \end{pmatrix}, 
 \qquad
 \bm{v}_{j}\!\left(\bm{q}\right) = 
  \begin{pmatrix}
   1\\
  x_{\bm{p}_j}-x_{\bm{q}}\\
  y_{\bm{p}_j}-y_{\bm{q}}
 \end{pmatrix}.
\end{equation}
The matrix $\bm{D}\!\left(\bm{q}\right)$ can be expressed as a sum of matrices 
of rank $1$ in the form
\begin{equation}
 \bm{D}\!\left(\bm{q}\right) = 
 \sum\limits_{j \in \mathcal{N}(\bm{q})} W\!\left(\bm{p}_{j}-\bm{q},h\right) \, 
V_{j} 
\, 
 \bm{v}_{j}\!\left(\bm{q}\right) \, \bm{v}_{j}^{T}\!\left(\bm{q}\right).
\end{equation}
It is positive semidefinite as for any $\bm{z} \in \mathbb{R}^{3}$ holds 
\begin{equation}
 \bm{z}^T \bm{D} \bm{z} = 
 \bm{z}^T \left(
 \sum\limits_{j \in \mathcal{N}(\bm{q})} W\!\left(\bm{p}_{j}-\bm{q},h\right) \, 
V_{j} 
\, 
 \bm{v}_{j} \bm{v}_{j}^{T} \right)
 \bm{z} = 
 \sum\limits_{j \in \mathcal{N}(\bm{q})} W\!\left(\bm{p}_{j}-\bm{q},h\right) \, 
V_{j} 
\, 
 \left(\bm{v}_{j}^{T}\, \bm{z} \right)^{2} \geq 0,
\end{equation}
since the smoothing kernel $W(\cdot)$ is nonnegative and $V_{j} > 0$. However, 
$\bm{D}\!\left(\bm{q}\right)$ is singular unless $\bm{q}$ has at least three 
nearest neighbors which are not collinear. 

In order to achieve an SPH interpolation with first order consistency at a pixel 
$\bm{q}$, it is necessary to solve the $3 \times 3$ linear system 
\cref{eq:first_order_comp}. To inpaint a whole image from first order SPH 
interpolation, \cref{eq:first_order_comp} needs to be solved for each unknown 
pixel. With the solution $\bm{b}\!\left(\bm{q}\right)$ of 
\cref{eq:first_order_comp}, a first order consistent SPH approximation can be 
written as
\begin{equation}
 \label{eq:28}
  u\!\left(\bm{q}\right) = 
  \sum\limits_{j \in \mathcal{N}(\bm{q})} f\!\left(\bm{p}_{j}\right) \, 
  \widetilde{W}\!\left(\bm{q}-\bm{p}_{j},h\right) \,V_{j} =
  \sum\limits_{j \in \mathcal{N}(\bm{q})} f\!\left(\bm{p}_{j}\right) \,
  \bm{v}_{j}^{T}\!\left(\bm{q}\right)\, \bm{b}\!\left(\bm{q}\right) \,
  W\!\left(\bm{q}-\bm{p}_{j},h\right) \,V_{j}.
\end{equation}

The particular method used here to restore first order consistency was derived 
in a longer way in \cite{ZB09}, whereas other methods that modify the kernel to 
restore first order consistency can be found in \cite{LJ95,Li09}. The method 
described here has the advantage that it does not involve derivatives of $f$ to 
restore first order consistency.
For image processing, similar techniques were derived from
kernel regression in \cite{TF07}.

The benefit of a higher order consistency does not come without a 
price.
The zero order consistent Shepard interpolation \cref{eq:new_discrete_shepard} 
only modifies the kernel such that it satisfies a discrete partition of unity 
property. With this modified kernel, Shepard interpolation produces the value 
of $u$ at $\bm{q}$ as a convex combination of the values of $f$ at the 
neighboring particles $\bm{p}_{j}$. Thus, it prevents over- and undershoots. If 
we want the first order consistent method \cref{eq:28} to prevent over- and 
undershoots, we have to put a restriction on the positions of particles, since 
we have to satisfy \cref{eq:first_order02,eq:first_order03}. These equations can 
be written in a compact way as
\begin{equation}
 \label{eq:pos_require_first_order}
 \bm{q} = 
 \sum\limits_{j \in \mathcal{N}(\bm{q})} \bm{p}_{j} \, 
   \widetilde{W}\!\left(\bm{q}-\bm{p}_{j},h\right)\, V_{j} .
\end{equation}
In an ideal situation, $u$ at $\bm{q}$ would be a convex combination of the 
values of $f$ at the neighboring particles $\bm{p}_{j}$ with weights given by 
$\widetilde{W}\!\left(\bm{q}-\bm{p}_{j},h\right)\, V_{j}$. However,
\cref{eq:pos_require_first_order} along with \cref{eq:first_order01} implies 
that this can only be the case if the position $\bm{q}$ is also a convex
combination of the particle positions $\bm{p}_{j}$ with the same weights. In 
most cases, this condition on the positions of particles is violated. In order 
to achieve first order consistency regardless of the spatial distribution of 
particles, the modified kernel $\widetilde{W}$ violates some of the properties 
defined in \cref{sec:formulation}. In particular violation of 
the positivity requirement results in visible artifacts as can be seen in the 
bottom left block of images in \cref{fig:random_5}.
This phenomenon is also mentioned as violation of a 
maximum-minimum principle in the context of inpainting in \cite{HH20}.


\subsection{Common Smoothing Kernels}
\label{sec:kernels}

In SPH, most smoothing kernels $W$ incorporate the smoothing length as a scaling 
parameter, such that they can be expressed in the form
\begin{equation}
 \label{eq:eta}
 W\!\left(\bm{q}-\bm{p},h\right) = W\!\left(\frac{\bm{q}-\bm{p}}{h}\right) = 
 W\!\left(\bm{\eta}\right) 
 \qquad \text{with} \qquad
 \bm{\eta} \coloneqq \frac{\bm{q}-\bm{p}}{h}.
\end{equation}
Further, it is common to choose radial kernels such that they can be written in 
the form
\begin{equation}
 \label{eq:kernel_rbf}
 W\!\left(\bm{q}-\bm{p},h\right) = W\!\left(\bm{\eta}\right) =
 \frac{\rho}{h^{2}} \, \Phi\!\left(\left\lVert \bm{\eta} \right\rVert\right).
\end{equation}
Here, $\rho$ is a normalization factor to satisfy the continuous unity property.
Probably the most common kernel is of Gaussian type:
\begin{equation}
 \Phi\!\left(\left\lVert \bm{\eta} \right\rVert\right) = 
 \exp\!\left(-\epsilon\, \left\lVert \bm{\eta} \right\rVert^{2}\right).
\end{equation}
However, as the Gaussian does not have a compact support, it is truncated at 
$\left\lVert \bm{\eta} \right\rVert=1$. Thus, the parameter $\epsilon$ should be 
chosen in a way that values of the resulting kernel $W\!\left(\bm{\eta}\right)$ 
for $\left\lVert \bm{\eta} \right\rVert > 1$ can be safely neglected. For the 
value of the Gaussian at $\left\lVert \bm{\eta} \right\rVert=1$, we obtain
\begin{equation}
 W\!\left(\bm{\eta}\right) = \frac{\rho}{h^{2}}
 \qquad \text{if} \qquad
 \left\lVert \bm{\eta} \right\rVert = 1,
\end{equation}
which inspires the condition
\begin{equation}
 W\!\left(\bm{\eta}\right) 
 \leq
 \frac{0.01}{h^{2}}
 \qquad \text{if} \qquad
 \left\lVert \bm{\eta} \right\rVert = 1.
\end{equation}
Together with the continuous unity property, this allows us to determine both 
parameters $\rho$ and $\epsilon$, such that we use the Gaussian in the form
\begin{equation}
 W\!\left(\bm{\eta}\right) = 
 \frac{\epsilon}{\pi\, h^{2}} \, 
 \exp\!\left(-\epsilon\, \left\lVert \bm{\eta}\right\rVert^{2}\right)
\end{equation}
with $\epsilon = 5.09$. Here, we have expressed $\rho$ as a function of 
$\epsilon$.

An alternative to the Gaussian are Mat\'{e}rn kernels \cite{Fa07}. Contrary to 
the Gaussian which is arbitrarily often continuously differentiable, Mat\'{e}rn 
kernels differ in smoothness. The $C^{0}$-Mat\'{e}rn kernel, which is not 
differentiable but just continuous at $\bm{\eta} = \bm{0}$, is given by
\begin{equation}
 \label{eq:matern_1}
 W\!\left(\bm{\eta}\right) = \frac{\epsilon^{2}}{2 \pi\, h^{2}}\, 
 \exp\!\left(-\epsilon \left\lVert \bm{\eta} \right\rVert\right),
\end{equation} 
for which we chose $\epsilon = 6.52$. A higher regularity at 
$\bm{\eta} = \bm{0}$ can be achieved with the $C^{2}$-Mat\'{e}rn kernel
\begin{equation}
 \label{eq:matern_2}
 W\!\left(\bm{\eta}\right) = 
 \frac{\epsilon^{2}}{6 \pi\, h^{2}}\, 
 \left(1 + \epsilon \left\lVert \bm{\eta} \right\rVert\right)
 \exp\!\left(-\epsilon \left\lVert \bm{\eta} \right\rVert\right),
\end{equation}
which we use in our experiments with $\epsilon = 8.04$.

Although the truncated Gaussian is a common choice, the original SPH paper 
\cite{Lu77} already introduced a kernel with a compact support, 
namely
\begin{equation}
 W\!\left(\bm{\eta}\right) = \frac{5}{\pi\, h^{2}}
 \begin{cases} 
  \left(1 + 3 \left\lVert \bm{\eta} \right\rVert\right)
   \left(1 - \left\lVert \bm{\eta} \right\rVert\right)^{3},
  &\ \left\lVert \bm{\eta} \right\rVert \leq 1, \\
  0, &\ \left\lVert \bm{\eta} \right\rVert > 1, 
 \end{cases}
\end{equation}
which we will call Lucy kernel. Other commonly used kernels with compact support 
are the cubic spline \cite{LL03}
\begin{equation}
 W\!\left(\bm{\eta}\right) = \frac{120}{14 \pi\, h^{2}}
 \begin{cases}
  \frac{2}{3} - 4 \left\lVert \bm{\eta} \right\rVert^{2} + 
   4 \left\lVert \bm{\eta} \right\rVert^{3}, 
   &\ \left\lVert \bm{\eta} \right\rVert \leq \frac{1}{2}, \\
  \frac{1}{6} \left(2 - 2 \left\lVert \bm{\eta} \right\rVert\right)^{3}, 
  &\ \frac{1}{2} <  \left\lVert \bm{\eta} \right\rVert \leq 1, \\
  0, &\ \left\lVert \bm{\eta} \right\rVert > 1,
 \end{cases}
\end{equation}
and the Wendland $C^{4}$ kernel \cite{We05}
\begin{equation}
 W\!\left(\bm{\eta}\right) = \frac{3}{\pi\, h^{2}}
 \begin{cases}
 \left(35 \left\lVert \bm{\eta} \right\rVert^{2} + 
       18 \left\lVert \bm{\eta} \right\rVert + 3\right)
 \left(1 - \left\lVert \bm{\eta} \right\rVert\right)^{6}, 
 &\ \left\lVert \bm{\eta} \right\rVert \leq 1, \\
 0, &\ \left\lVert \bm{\eta} \right\rVert > 1.
 \end{cases}
\end{equation}


\section{SPH Inpainting}
\label{sec:inpainting}

For the majority of inpainting problems discussed in this 
paper,
we consider the reconstruction of an image $f$ from a sparse set of values at
scattered pixel locations. These locations, called mask points, take the role 
of particles $\bm{p}_{j}$ for our SPH-inspired inpainting procedure. The set of 
all mask points is the inpainting mask $\bm{c}$. 
Exceptions from this setting are the examples of scratch and 
text removal in \cref{sec:scratch}, which we include to investigate
how SPH inpainting performs for some classical inpainting problems.


\subsection{Choosing Influence Areas and Smoothing Lengths}

In order to use \cref{eq:SPH_mod_kernel} for inpainting, whether with the 
original particle approximation, Shepard interpolation, or the first order 
consistent method, we still have to determine an area of influence $V_{j}$ for 
each given mask point $\bm{p}_{j}$. Further, we want to enhance the adaptivity 
of the method by allowing for different smoothing lengths $h_{j}$ of the kernels 
centered at the individual particles. This adaptivity is motivated by the 
results in \cite{DF11}.

A reasonable idea is to assume that the area of influence of a given mask point 
$\bm{p}_{j}$ is the set of all points which are closer to $\bm{p}_{j}$ than to 
any other mask point in $\bm{c}$. This idea leads to a Voronoi tessellation of 
the domain $\Omega$ with seeds given by the mask points. Voronoi cells have 
been used before in the context of SPH 
\cite{GX16,SA17} with promising 
results. 

As we are working in a discrete setting where the smallest unit of area is a 
pixel, a method which determines approximate Voronoi diagrams based on the 
squared Euclidean distance transform is our tool of choice for this task. This 
is a rather natural approach since the Voronoi cell $\Omega_{j}$ associated to 
the mask point $\bm{p}_{j}$ is defined as
\begin{equation}
 \Omega_{j} = \left\{ \bm{q} \in \Omega\, \vert \, 
  \left\lVert \bm{q}-\bm{p}_{j} \right\rVert \leq 
  \left\lVert \bm{q}-\bm{p}_{k} \right\rVert \text{ for all } 
  1 \leq k \leq M,\, k \neq j \right\}.
\end{equation}

\begin{figure}[htb]
 \captionsetup[subfigure]{justification=centering}
 \centering
 \begin{subfigure}[t]{0.29\textwidth}
  \centering
  \includegraphics[width=\textwidth]{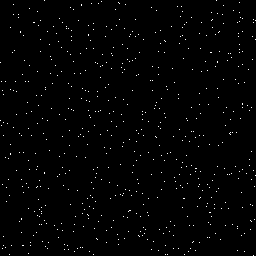}
  \caption[Voronoi Seeds]{Image with seeds marked in white.}
  \label{fig:mask}
 \end{subfigure}
 \quad
 \begin{subfigure}[t]{0.29\textwidth}
  \centering
  \includegraphics[width=\textwidth]{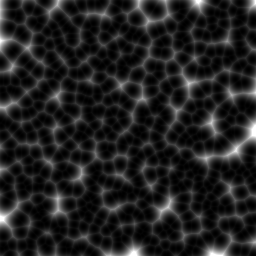}
  \caption[Distance transform in $2D$]{Distance transform.}
  \label{fig:dt_new}
 \end{subfigure}
 \quad
 \begin{subfigure}[t]{0.29\textwidth}
  \centering
  \includegraphics[width=\textwidth]{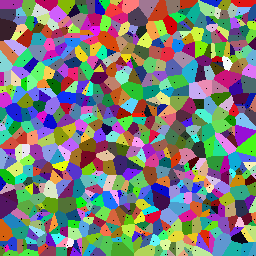}
  \caption[Voronoi Diagram]{Resulting Voronoi diagram with seeds marked in 
   black.}
  \label{fig:color}
 \end{subfigure}
 \caption[Voronoi Tessellation]{Voronoi tessellation using distance transform 
  for a given set of mask points $\bm{p}_{j}$ depicted as white seeds in 
  \cref{fig:mask}. The resulting distance transform is depicted in 
  \cref{fig:dt_new}. Corresponding Voronoi cells can be seen in 
  \cref{fig:color}, each depicted in a different color with seeds in black 
  now.} 
 \label{fig:tessellation}
\end{figure}

Given a binary image $g$ which only takes the values $0$ and $\infty$ 
throughout a domain $\Omega$, the distance 
transform
assigns to each pixel 
$\bm{q}$ in $\Omega$ its squared distance to the nearest pixel $\bm{p}$ with 
$g(\bm{p})=0$. In our case, $g$ takes the value $0$ at the mask points and 
$\infty$ everywhere else. For practical applications, $\infty$ can be replaced 
by a sufficiently large number. In a two-dimensional domain, the squared 
distance transform is given by
\begin{equation}
 \label{eq:dist}
 \begin{split}
  \mathcal{D}(x,y) =&\
   \min_{x',y'}\left\{(x-x')^2 + (y-y')^2 + g(x',y')\right\} \\
   =&\ 
   \min_{x'}\left\{(x-x')^2 + \min_{y'}\left\{(y-y')^2+g(x',y')\right\}\right\}
 \end{split}
\end{equation}
such that it can be computed by two consecutive squared distance transforms in 
one dimension. We used the algorithm from \cite{Bo92,FH12} to compute the 
distance transform, which is shown to have a complexity of $\mathcal{O}(n_{x}\, 
n_{y})$, where $n_{x}$ and $n_{y}$ are the number of pixels in the image domain 
$\Omega$ in $x$- and $y$-direction, respectively. For visualization purposes 
each and every Voronoi cell is depicted with a different color per Voronoi cell 
in \cref{fig:color}.

After the Voronoi tessellation, each mask point $\bm{p}_{j}$ is assigned as 
area of influence the area of its corresponding Voronoi cell. In our discrete 
setting, this area is defined as the sum of pixels that belong to that cell, 
i.e., pixels are assumed to have area equal to $1$.

It seems natural to determine the smoothing length $h_{j}$ in relation to the 
volume $V_{j}$, e.g.\ half the diameter of the Voronoi cell associated to 
$\bm{p}_{j}$. However, this particular choice is prone to result in pixels for 
which the requirements of the minimal necessary number of nearest neighbors 
are not satisfiable. If a pixel does not lie within the support of any kernel, 
it cannot be inpainted.  

A straightforward remedy would be to multiply the diameter of each Voronoi cell 
with a constant factor chosen such that each pixel has at least the desired 
minimum number of nearest neighbors. Unfortunately, this would result in 
oversmoothing and blurring as the resulting kernel supports would be rather 
large. Instead, we follow the adaptive, iterative approach of \cite{DF11} for 
the choice of smoothing lengths, which also enforces that any pixel is inpainted 
with at least a specified minimum number of neighbors. The scheme starts by 
assigning each mask point $\bm{p}_{j}$ an initial smoothing length 
$h_{j,\textrm{init}}$. Using the corresponding kernels, unknown pixels are 
inpainted, but only if they lie within the support of a least a fixed number of 
kernels. All pixels which do not satisfy this requirement are not inpainted. 
Afterwards we check whether there are still pixels with no assigned value left. 
If so, we increase all smoothing lengths according to a certain rule and try 
again to inpaint those pixels which are not yet assigned a value. This procedure 
is repeated iteratively until each pixel is inpainted. As growing strategy for 
the smoothing lengths, we increase them linearly with the number of iterations. 

The original method in \cite{DF11} assigned initial smoothing lengths which are 
connected to the choice of $V_{j}$ as made there. However, our experiments 
showed that it is beneficial if each kernel starts with a minimal smoothing 
length of $h_{j,\textrm{init}} = 1$. Thus, mask points can initially only be 
recognized as neighbors within a $3\times 3$-patch around them such that the 
process starts with the smallest sensible isotropic support for each kernel. The 
smoothing length $h_{j}$ is in each step equal to the number of iterations.

The last parameter that we have to set is the required minimal number of nearest 
neighbors. For Shepard interpolation, we need at least one neighbor to perform 
an inpainting, whereas for the method with first order consistency, any pixel 
must be contained in the support of at least three kernels. For all methods, it 
is reasonable to choose a slightly larger necessary minimal number of nearest 
neighbors as this improves results. In particular for the method of first order 
consistency that reduces the chance to encounter cases in which all mask points 
closest to an unknown pixel are collinear. For our experiments we require a 
minimal number of five nearest neighbors. 


\subsection{Sparse Inpainting on Regular and Random Masks}
\label{sec:improvement}

We follow here a didactic approach and consider the test image ``trui'' of 
size $256 \times 256$ pixels as an example. Results and comparisons for further
images can be found in \cref{sec:comparisons} and in the supplementary material.
All experiments in this paper were performed on an Intel Core 
i7-9700K CPU @ 3.6GHz.

\begin{figure}[htb]
 \captionsetup[subfigure]{justification=centering}
 \centering
 \begin{subfigure}[t]{0.29\textwidth}
  \centering
  \includegraphics[width=\textwidth]{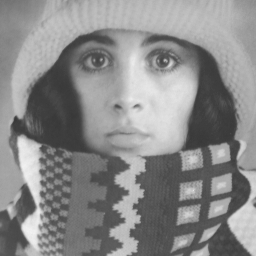}
  \caption{Original image}
  \label{fig:trui}
 \end{subfigure}
 \quad
 \begin{subfigure}[t]{0.29\textwidth}
  \centering
  \includegraphics[width=\textwidth]{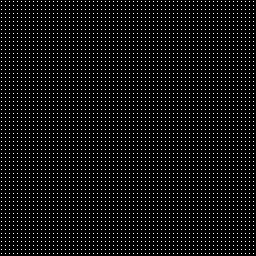}
  \caption{6.25 \% regular mask}
 \end{subfigure}
 \quad
 \begin{subfigure}[t]{0.29\textwidth}
  \centering
  \includegraphics[width=\textwidth]{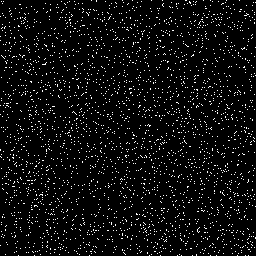}
  \caption{5 \% random mask}
 \end{subfigure}
 \caption[Test image, regular mask ($6.25 \%$ density) and random mask 
  ($5 \%$ density)]{The $256 \times 256$ test image ``trui'', 
  a regular grid of mask points, and a random selection of 
  $5 \%$ of image pixels.}
 \label{fig:trui_mask}
\end{figure}

To get a first impression, we equip the test image with 
two different types of sparse masks
$\bm{c}$. 
In one case, we choose for $\bm{c}$ a regular mask with a 
density of $6.25 \%$, i.e., pixels on a square grid with a grid width of $4$
pixels are taken as mask points. In the other case, we randomly selected $5 \%$ 
of all image pixels as mask points. Test image ``trui'' and both masks are 
shown in \cref{fig:trui_mask}. 

In this setting, we perform SPH inpainting with a required minimal number of 
five neighbors, starting with the kernels given in \cref{sec:kernels} and 
modifying them either according to Shepard interpolation 
\cref{eq:new_discrete_shepard} or the first order consistent method given by 
\cref{eq:first_order_comp,eq:28}. 

\begin{figure}[p]
 \captionsetup[subfigure]{justification=centering}
 \centering
 \begin{tabular}{m{0.7\textwidth}m{0.2\textwidth}}
 \begin{subfigure}[t]{0.20\textwidth}
  \centering
  \includegraphics[width=\textwidth]{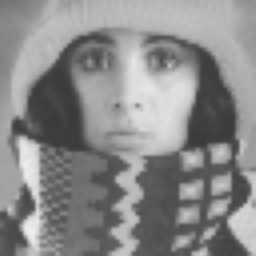}
  \caption{Gaussian\\
   $\textrm{MSE} = 83.28$}
  \label{fig:zero_reco_6_25_gaussian}
 \end{subfigure}
 \quad
 \begin{subfigure}[t]{0.20\textwidth}
  \centering
  \includegraphics[width=\textwidth]{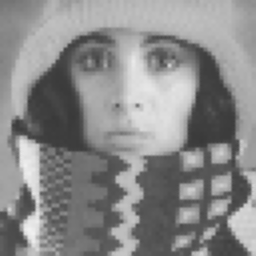}
  \caption{$C^{0}$-Mat\'{e}rn\\
   $\textrm{MSE} = 83.22$}
 \end{subfigure}
 \quad
 \begin{subfigure}[t]{0.20\textwidth}
  \centering
  \includegraphics[width=\textwidth]{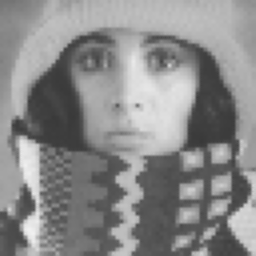}
  \caption{$C^{2}$-Mat\'{e}rn\\
   $\textrm{MSE} = 82.03$}
  \label{fig:zero_reco_6_25_matern_2}
 \end{subfigure}

 \begin{subfigure}[t]{0.20\textwidth}
  \centering
  \includegraphics[width=\textwidth]{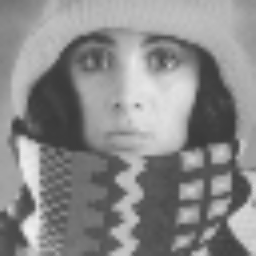}
  \caption{Lucy\\
   $\textrm{MSE} = 81.15$}
  \label{fig:zero_reco_6_25_lucy}
 \end{subfigure}
 \quad
 \begin{subfigure}[t]{0.20\textwidth}
  \centering
  \includegraphics[width=\textwidth]{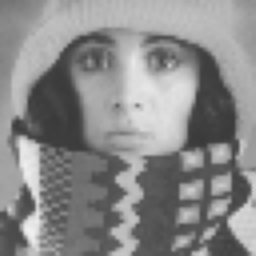}
  \caption{cubic spline\\
   $\textrm{MSE} = 78.64$}
  \label{fig:zero_reco_6_25_cubic}
 \end{subfigure}
 \quad
 \begin{subfigure}[t]{0.20\textwidth}
  \centering
  \includegraphics[width=\textwidth]{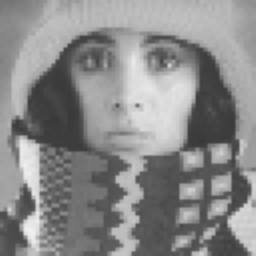}
  \caption{$C^{4}$-Wendland\\
   $\textrm{MSE} = 81.94$}
  \label{fig:zero_reco_6_25_wendland}
 \end{subfigure}
 \vspace{5ex}
 
 \begin{subfigure}[t]{0.20\textwidth}
  \centering
  \includegraphics[width=\textwidth]{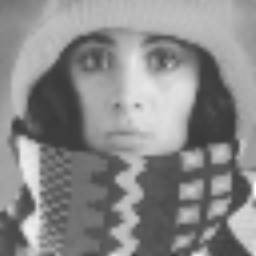}
  \caption{Gaussian\\
   $\textrm{MSE} = 85.01$}
  \label{fig:first_reco_6_25_gaussian}
 \end{subfigure}
 \quad
 \begin{subfigure}[t]{0.20\textwidth}
  \centering
  \includegraphics[width=\textwidth]{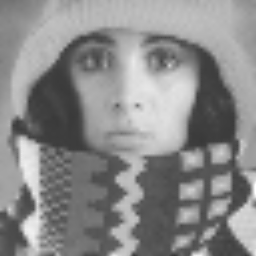}
  \caption{$C^{0}$-Mat\'{e}rn\\
   $\textrm{MSE} = 80.57$}
 \end{subfigure}
 \quad
 \begin{subfigure}[t]{0.20\textwidth}
  \centering
  \includegraphics[width=\textwidth]{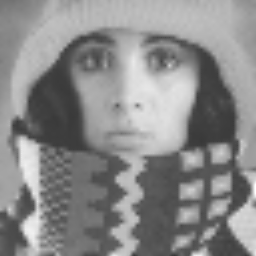}
  \caption{$C^{2}$-Mat\'{e}rn\\
   $\textrm{MSE} = 80.99$}
  \label{fig:first_reco_6_25_matern_2}
 \end{subfigure}
 
 \begin{subfigure}[t]{0.20\textwidth}
  \centering
  \includegraphics[width=\textwidth]{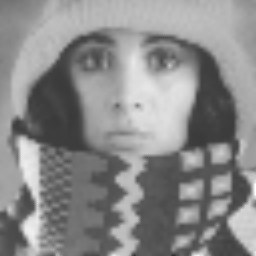}
  \caption{Lucy\\
   $\textrm{MSE} = 82.71$}
  \label{fig:first_reco_6_25_lucy}
 \end{subfigure}
 \quad
 \begin{subfigure}[t]{0.20\textwidth}
  \centering
  \includegraphics[width=\textwidth]{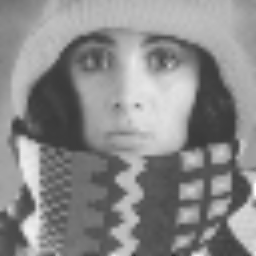}
  \caption{cubic spline\\
   $\textrm{MSE} = 80.94$}
  \label{fig:first_reco_6_25_cubic}
 \end{subfigure}
 \quad
 \begin{subfigure}[t]{0.20\textwidth}
  \centering
  \includegraphics[width=\textwidth]{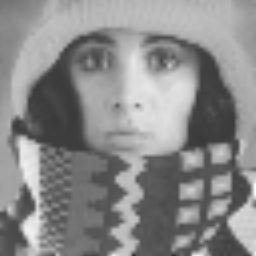}
  \caption{$C^{4}$-Wendland\\
   $\textrm{MSE} = 79.53$}
  \label{fig:first_reco_6_25_wendland}
 \end{subfigure}
& 
 \begin{subfigure}[t]{0.20\textwidth}
  \centering
  \includegraphics[width=\textwidth]{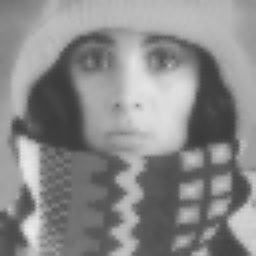}
  \caption{Harmonic\\
   $\textrm{MSE} = 121.96$}
 \end{subfigure}
 
 \begin{subfigure}[t]{0.20\textwidth}
  \centering
  \includegraphics[width=\textwidth]{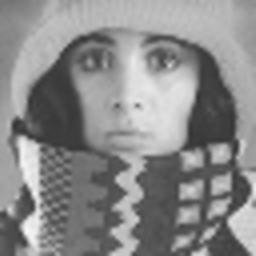}
  \caption{Biharmonic\\
   $\textrm{MSE} = 67.95$}
 \end{subfigure}
 
 \begin{subfigure}[t]{0.20\textwidth}
  \centering
  \includegraphics[
   width=\textwidth]{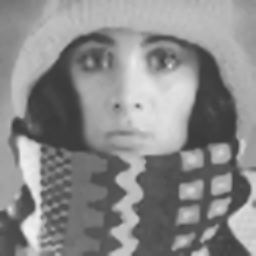}
  \caption{EED\\
   $\textrm{MSE} = 60.68$}
 \end{subfigure}
 \end{tabular}
 \caption[Inpainting of ``trui'' with a 6.25\% regular mask ]{Inpainting of 
  ``trui'' with the 6.25\% regular mask from \cref{fig:trui_mask} with a zero 
  order consistency method (top left, (\textbf{a})-(\textbf{f})), first order 
  consistency method (bottom left, (\textbf{g})-(\textbf{l})), and 
  diffusion-based inpainting (right, (\textbf{m})-(\textbf{o})).
  Inpainting runtime for zero order consistency method was 16.35 s; 
  for first order consistency method 18.26 s (each averaged across all kernels).
  Parameters for EED are $\lambda=0.2$ and $\sigma=0.8$.}
 \label{fig:regular_6_25}
\end{figure}

\begin{figure}[p]
 \captionsetup[subfigure]{justification=centering}
 \centering
 \begin{tabular}{m{0.7\textwidth}m{0.2\textwidth}}
 \begin{subfigure}[t]{0.20\textwidth}
  \centering
  \includegraphics[width=\textwidth]{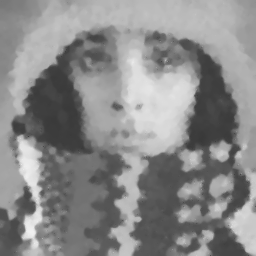}
  \caption{Gaussian\\
   $\textrm{MSE} = 208.48$}
  \label{fig:zero_reco_5_gaussian}
 \end{subfigure}
 \quad
 \begin{subfigure}[t]{0.20\textwidth}
  \centering
  \includegraphics[width=\textwidth]{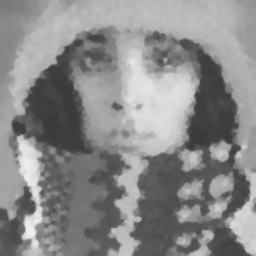}
  \caption{$C^{0}$-Mat\'{e}rn\\
   $\textrm{MSE} = 206.82$}
 \end{subfigure}
 \quad
 \begin{subfigure}[t]{0.20\textwidth}
  \centering
  \includegraphics[width=\textwidth]{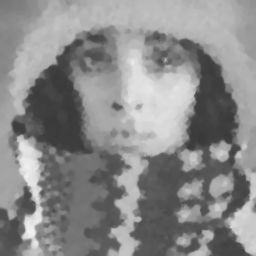}
  \caption{$C^{2}$-Mat\'{e}rn\\
   $\textrm{MSE} = 207.85$}
  \label{fig:zero_reco_5_matern_2}
 \end{subfigure}

 \begin{subfigure}[t]{0.20\textwidth}
  \centering
  \includegraphics[width=\textwidth]{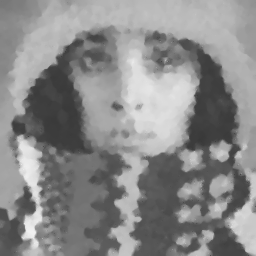}
  \caption{Lucy\\
   $\textrm{MSE} = 219.29$}
  \label{fig:zero_reco_5_lucy}
 \end{subfigure}
 \quad 
 \begin{subfigure}[t]{0.20\textwidth}
  \centering
  \includegraphics[width=\textwidth]{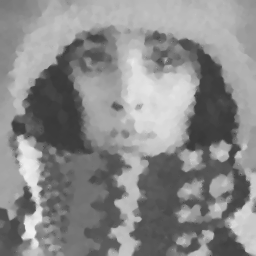}
  \caption{cubic spline\\
   $\textrm{MSE} = 223.02$}
  \label{fig:zero_reco_5_cubic}
 \end{subfigure}
 \quad
 \begin{subfigure}[t]{0.20\textwidth}
  \centering
  \includegraphics[width=\textwidth]{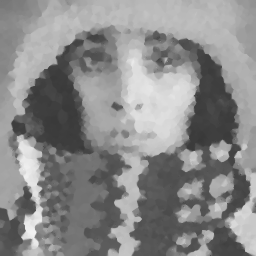}
  \caption{$C^{4}$-Wendland\\
   $\textrm{MSE} = 244.35$}
  \label{fig:zero_reco_5_wendland}
 \end{subfigure}
 \vspace{5ex}
 
 \begin{subfigure}[t]{0.20\textwidth}
  \centering
  \includegraphics[width=\textwidth]{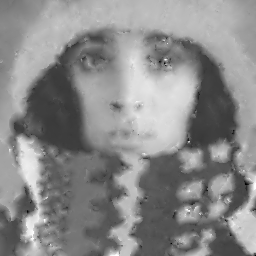}
  \caption{Gaussian\\
   $\textrm{MSE} = 197.58$}
  \label{fig:first_reco_5_gaussian}
 \end{subfigure}
 \quad
 \begin{subfigure}[t]{0.20\textwidth}
  \centering
  \includegraphics[width=\textwidth]{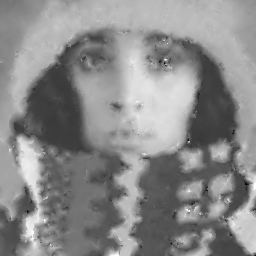}
  \caption{$C^{0}$-Mat\'{e}rn\\
   $\textrm{MSE} = 194.80$}
 \end{subfigure}
 \quad
 \begin{subfigure}[t]{0.20\textwidth}
  \centering
  \includegraphics[width=\textwidth]{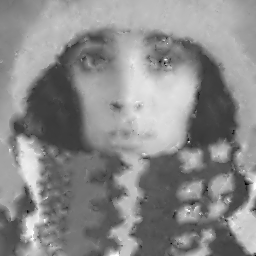}
  \caption{$C^{2}$-Mat\'{e}rn\\
   $\textrm{MSE} = 195.21$}
  \label{fig:first_reco_5_matern_2}
 \end{subfigure}
 
 \begin{subfigure}[t]{0.20\textwidth}
  \centering
  \includegraphics[width=\textwidth]{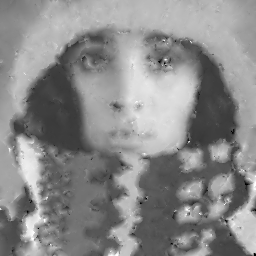}
  \caption{Lucy\\
   $\textrm{MSE} = 220.55$}
  \label{fig:first_reco_5_lucy}
 \end{subfigure}
 \quad
 \begin{subfigure}[t]{0.20\textwidth}
  \centering
  \includegraphics[width=\textwidth]{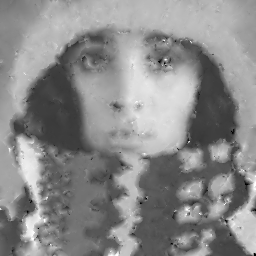}
  \caption{cubic spline\\
   $\textrm{MSE} = 223.26$}
  \label{fig:first_reco_5_cubic}
 \end{subfigure}
 \quad
 \begin{subfigure}[t]{0.20\textwidth}
  \centering
  \includegraphics[width=\textwidth]{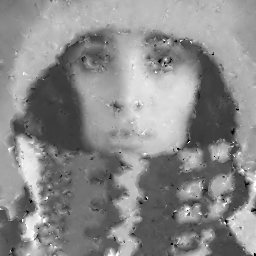}
  \caption{$C^{4}$-Wendland\\
   $\textrm{MSE} = 287.41$}
  \label{fig:first_reco_5_wendland}
 \end{subfigure}
 & 
 \begin{subfigure}[t]{0.20\textwidth}
  \centering
  \includegraphics[width=\textwidth]{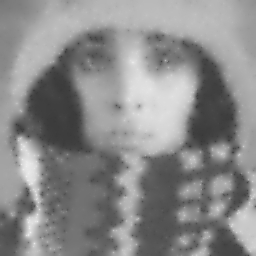}
  \caption{Harmonic\\
   $\textrm{MSE} = 226.06$}
 \end{subfigure}

 \begin{subfigure}[t]{0.20\textwidth}
  \centering
  \includegraphics[width=\textwidth]{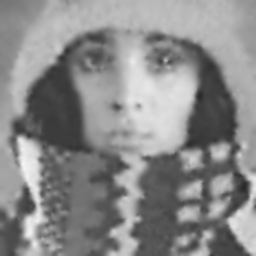}
  \caption{Biharmonic\\
   $\textrm{MSE} = 146.46$}
 \end{subfigure}

 \begin{subfigure}[t]{0.20\textwidth}
  \centering
  \includegraphics[
   width=\textwidth]{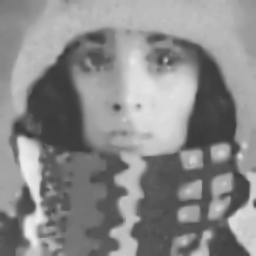}
  \caption{EED\\
   $\textrm{MSE} = 134.92$}
 \end{subfigure}
 \end{tabular}
 \caption[Inpainting of ``trui'' with a 5 \% random mask]
  {Inpainting of ``trui'' 
  with the 5 \% random mask from \cref{fig:trui_mask} with a zero order 
  consistency method (top left, (\textbf{a})-(\textbf{f})), first order 
  consistency method (bottom left, (\textbf{g})-(\textbf{l})), and 
  diffusion-based inpainting (right, (\textbf{m})-(\textbf{o})).
  Inpainting runtime for zero order consistency method was 3.95 s; for first 
  order consistency method 5.15 s (each averaged across all kernels). 
  Parameters for EED are $\lambda=0.2$ and $\sigma=0.8$.}
 \label{fig:random_5}
\end{figure}

Corresponding results for the case of having a regular mask are 
depicted in \cref{fig:regular_6_25}, whereas the results based
on the randomly chosen mask points are given in \cref{fig:random_5}. 
In order to compare results, all figures give the corresponding mean square 
errors (MSEs) between the inpainting result and the original image.
Further, we have included the runtimes of each set of 
experiments, averaged across the six different kernels under consideration.

First of all, comparing the results on the regular mask, we note
that the differences are not significant. Further, the higher order of 
consistency in the bottom left block of images in \cref{fig:regular_6_25} 
is not always beneficial. Indeed, 
the Gaussian, Lucy kernel, and cubic spline achieve lower MSEs if used in the 
zero order consistent Shepard interpolation method. Hence, improving consistency
in the sense of polynomial reproduction does not automatically yield overall 
better results. If we only regard the MSE, this stays true for the setting of a 
random inpainting mask. Here, 
the performance gains with first order consistency compared to zero order 
consistency are also not significant whereas for kernels with compact support 
(Lucy, cubic spline, and $C^{4}$-Wendland) the results are even worse with the 
first order consistency method. 

Comparing the results in 
\cref{fig:regular_6_25,fig:random_5}, we observe that further
problems arise in the case of a random mask. While for a regular mask, 
\cref{eq:pos_require_first_order} is satisfied in a way that gives a  convex 
combination on the right-hand side for every unknown pixel $\bm{q}$, this is 
no longer the case for our random mask.
In other words, the modified kernels for a higher order 
consistency, at pixels whose position cannot be expressed as a convex 
combination of mask point positions
do no longer obey the positivity requirements, resulting in over- and 
undershoots which can become quite severe. Moreover, the $3 \times 3$-system 
that needs to be solved at each pixel for each modified kernel may become 
almost singular, leading to further instabilities. 
For the zero order consistency method, i.e., Shepard 
interpolation, the $C^{0}$-Mat\'{e}rn kernel yields the best result. This is 
in line with recent findings by Dell'Accio et~al.\ in \cite{DD20}. 


\subsection{Classical Inpainting Applications: Scratch and 
Text Removal}
\label{sec:scratch}

Among the classical applications for inpainting \cite{BS00} are 
the removal of scratches or text from an image. Alves Mazzini and Petronetto do 
Carmo already used SPH inpainting for such tasks in \cite{AP16}. Thus, we also
briefly address such problems here.

\begin{figure}[p]
 \captionsetup[subfigure]{justification=centering}
 \begin{subfigure}[t]{0.29\textwidth}
  \centering
  \includegraphics[width=\textwidth]{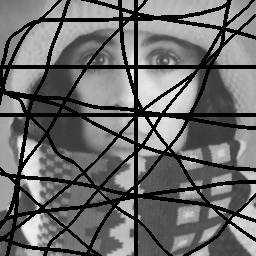}
  \caption{``trui'' with scratches}
 \end{subfigure}
 \quad
 \begin{subfigure}[t]{0.29\textwidth}
  \centering
  \includegraphics[width=\textwidth]{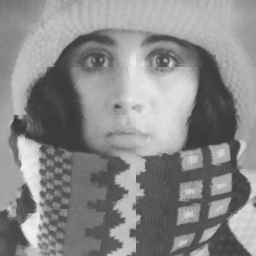}
  \caption{Zero order\\
   $\textrm{MSE} = 28.93$}
 \end{subfigure}
  \quad 
 \begin{subfigure}[t]{0.29\textwidth}
  \centering
  \includegraphics[width=\textwidth]{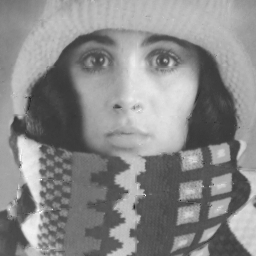}
  \caption{First order\\
   $\textrm{MSE} = 23.97$}
 \end{subfigure}
 
 \begin{subfigure}[t]{0.29\textwidth}
  \centering
  \includegraphics[width=\textwidth]{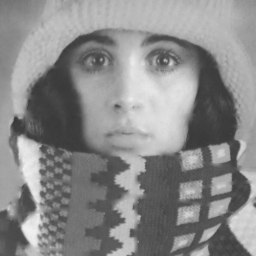}
  \caption{Harmonic\\
   $\textrm{MSE} = 23.32$}
 \end{subfigure}
 \quad
 \begin{subfigure}[t]{0.29\textwidth}
  \centering
  \includegraphics[width=\textwidth]{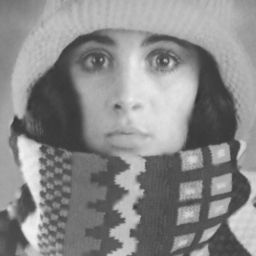}
  \caption{Biharmonic\\
   $\textrm{MSE} = 12.86$}
 \end{subfigure}
 \quad 
 \begin{subfigure}[t]{0.29\textwidth}
  \centering
  \includegraphics[
   width=\textwidth]{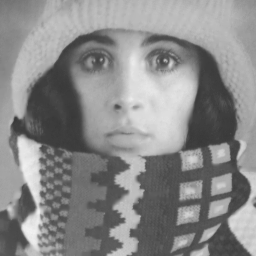}
  \caption{EED\\
   $\textrm{MSE} = 12.28$}
 \end{subfigure}
  
 \begin{subfigure}[t]{0.29\textwidth}
  \centering
  \includegraphics[width=\textwidth]{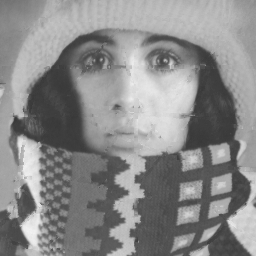}
  \caption{Exemplar-based\\
   $\textrm{MSE} = 49.58$}
 \end{subfigure}
 \caption[Inpainting of ``trui'' for scratch removal ]{Image ``trui'' damaged 
  by scratches (\textbf{a}) and corresponding SPH inpaintings of zero order
  consistency (\textbf{b}) as well as first order consistency (\textbf{c})
  using a Gaussian kernel. Diffusion- and exemplar-based inpainting results 
  are shown in (\textbf{d}-\textbf{g}). 
  Parameters for EED are $\lambda=0.2$ and $\sigma=0.6$.}
 \label{fig:trui_scratch}
\end{figure}

As a first example, we consider a damaged version of the image 
``trui'' with scratches, see \cref{fig:trui_scratch}. To repair those scratches,
we consider SPH inpainting of zero and first order consistency for the 
commonly used Gaussian kernel.

Regarding MSEs, both methods yield similar results. The 
differences become clearer when we look at particular areas of the image. For
the horizontal scratch below the eyes, we observe visible artifacts in the
zero order method while the first order method produces a more pleasing, though
not perfect visual impression. On the other hand, for scratches crossing the 
scarf, the zero order inpainting shows fewer artifacts than the first order
inpainting.

\begin{figure}[p]
 \captionsetup[subfigure]{justification=centering}
 \begin{subfigure}[t]{0.29\textwidth}
  \centering
  \includegraphics[width=\textwidth]{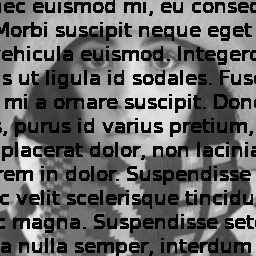}
  \caption{``trui'' with text}
 \end{subfigure}
 \quad
 \begin{subfigure}[t]{0.29\textwidth}
  \centering
  \includegraphics[width=\textwidth]{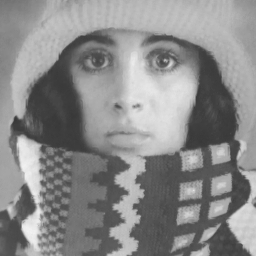}
  \caption{Zero order\\
   $\textrm{MSE} = 18.58$}
 \end{subfigure}
  \quad 
 \begin{subfigure}[t]{0.29\textwidth}
  \centering
  \includegraphics[width=\textwidth]{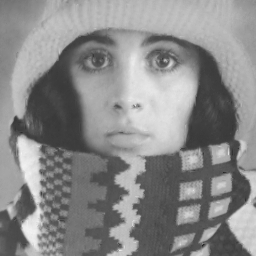}
  \caption{First order\\
   $\textrm{MSE} = 14.27$}
 \end{subfigure}
 
 \begin{subfigure}[t]{0.29\textwidth}
  \centering
  \includegraphics[width=\textwidth]{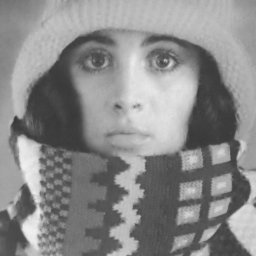}
  \caption{Harmonic\\
   $\textrm{MSE} = 16.29$}
 \end{subfigure}
 \quad
 \begin{subfigure}[t]{0.29\textwidth}
  \centering
  \includegraphics[width=\textwidth]{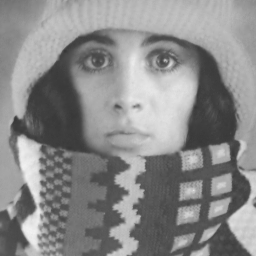}
  \caption{Biharmonic\\
   $\textrm{MSE} = 8.72$}
 \end{subfigure}
 \quad 
 \begin{subfigure}[t]{0.29\textwidth}
  \centering
  \includegraphics[
   width=\textwidth]{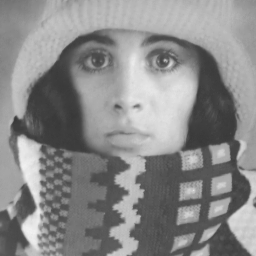}
  \caption{EED\\
   $\textrm{MSE} = 7.80$}
 \end{subfigure}
 
 \begin{subfigure}[t]{0.29\textwidth}
  \centering
  \includegraphics[width=\textwidth]{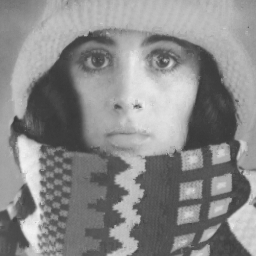}
  \caption{Exemplar-based\\
   $\textrm{MSE} = 27.76$}
 \end{subfigure}
 \caption[Inpainting of ``trui'' with overlaid text]{Image ``trui'' with 
  overlaid text (\textbf{a}) and corresponding SPH inpaintings with zero order 
  consistency (\textbf{b}) as well as first order consistency (\textbf{c})
  using a Gaussian kernel. Diffusion- and exemplar-based inpainting results are 
  shown in (\textbf{d}-\textbf{g}). 
  Parameters for EED are $\lambda=0.5$ and $\sigma=0.6$.}
 \label{fig:trui_text}
\end{figure}

As a second example, we consider ``trui'' overlaid with some 
text which we attempt to remove in \cref{fig:trui_text}. As before, we use a
Gaussian and compare SPH inpainting of zero and first order consistency.
For this example, both zero and first order consistency method
yield good results, though the first order method is overall slightly better, 
for example at the boundary between hair and hat on the left-hand side. 
Overall, both results looks visually pleasant.
 
\subsection{Comparisons with Diffusion-Based and Non-Local 
Inpainting Methods}

To put the results that we have seen so far in perspective, we 
compare them with the performance of other inpainting methods. As simple 
representatives of diffusion-based methods, we consider harmonic and biharmonic 
inpainting \cite{CS02,GW08}. A more sophisticated method is 
edge-enhancing diffusion (EED). Although introduced as a denoising technique 
\cite{We98}, it turned out to be also a powerful inpainting method 
\cite{GW08,PH16}. For all results shown here and in the 
supplement, we have used a discretisation of EED which corresponds to the one 
given in \cite{WW13} for $\alpha=0$ and $\beta=0$. Contrast parameter $\lambda$ 
and noise scale $\sigma$ were adapted to the image at hand.

Let us first consider how these methods perform in the case of sparse 
inpainting masks. The results for the different diffusion based 
inpainting methods when using the regular inpainting mask from 
\cref{fig:trui_mask} are included in \cref{fig:regular_6_25} on the right. 
Comparing all results in \cref{fig:regular_6_25}, we see that, with regard
to MSEs, SPH inpainting performs better than harmonic, but worse than 
biharmonic inpainting. As can be expected, EED shows the best results, both 
visually and in terms of MSE.

When it comes to inpainting on random masks, the situation is 
slightly different as \cref{fig:random_5} illustrates. Again, the results of 
the three diffusion-based inpainting methods for the random inpainting mask 
from \cref{fig:trui_mask} are included on the right. Comparing all results in 
\cref{fig:random_5} shows that the performance of SPH inpainting is similar to 
harmonic inpainting in terms of MSE, but closer to biharmonic inpainting in 
terms of visual impression. Again, EED achieves the best MSE and visually 
smoothest inpainting.

We also compare the results achieved by diffsion-based methods 
in case of the image damaged by scratches in \cref{fig:trui_scratch}. 
Furthermore, we considered the exemplar-based inpainting approach by Criminisi 
et al.~\cite{CP04} as an example of a non-local inpainting method. For this 
method, we have considered disc-shaped patches and adapted the patch radius to 
the image. As is evident from \cref{fig:trui_scratch}, the first order 
consistency SPH inpainting can achieve an MSE similar to harmonic inpainting, 
but shows more artifacts. The exemplar-based method on the other hand is 
worse with regard to both MSE and creation of artifacts.

Results of the diffusion- and exemplar-based methods in case of
the text removal task are included in \cref{fig:trui_text}. The results of SPH 
inpainting are, once more, similar to the results obtained by harmonic 
inpainting. The exemplar-based method again produces artifacts, in particular 
around the eyes.

\begin{figure}[t]
 \captionsetup[subfigure]{justification=centering}
 \centering
 \begin{subfigure}[t]{0.29\textwidth}
  \centering
  \includegraphics[width=\textwidth]{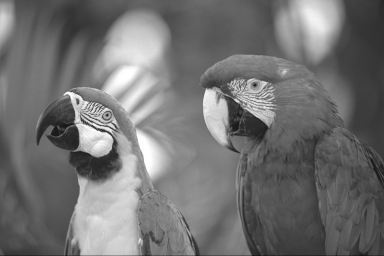}
  \caption{Ground truth}
 \end{subfigure}
 \quad
 \begin{subfigure}[t]{0.29\textwidth}
  \centering
  \includegraphics[width=\textwidth]{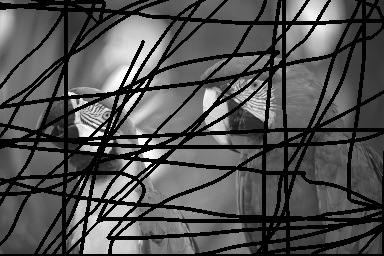}
  \caption{Damaged by scratches}
 \end{subfigure}
 \quad
 \begin{subfigure}[t]{0.29\textwidth}
  \centering
  \includegraphics[width=\textwidth]{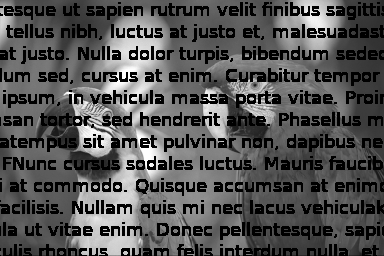}
  \caption{Overlaid with text}
 \end{subfigure}

 \begin{subfigure}[t]{0.29\textwidth}
  \centering
  \includegraphics[width=\textwidth]{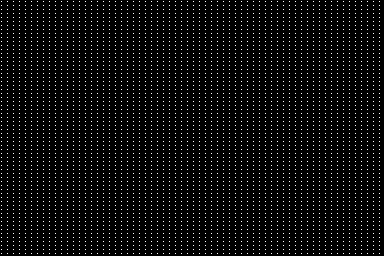}
  \caption{4.16 \% regular mask}
 \end{subfigure}
 \quad
 \begin{subfigure}[t]{0.29\textwidth}
  \centering
  \includegraphics[width=\textwidth]{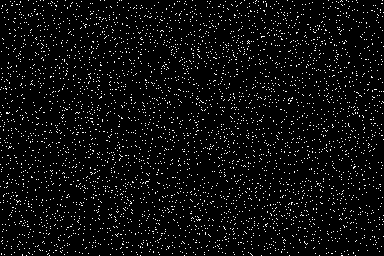}
  \caption{5 \% random mask}
 \end{subfigure}
 \caption[Image ``parrots'' with various damages and sparse inpainting masks]{
  Image ``parrots'' with various inpainting tasks.}
 \label{fig:parrots}
\end{figure}

\begin{figure}[htb]
 \captionsetup[subfigure]{justification=centering}
 \centering
 \begin{subfigure}[t]{0.29\textwidth}
  \centering
  \includegraphics[width=\textwidth]{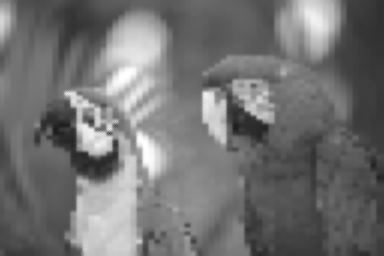}
  \caption{Zero order consistency\\
   $\textrm{MSE} = 132.79$}
 \end{subfigure}
 \quad
 \begin{subfigure}[t]{0.29\textwidth}
  \centering
  \includegraphics[width=\textwidth]{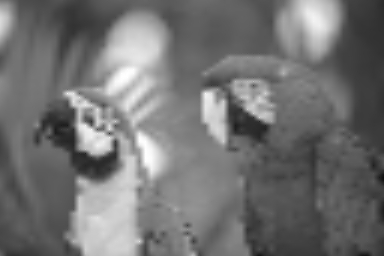}
  \caption{First order consistency\\
   $\textrm{MSE} = 124.81$}
 \end{subfigure}

 \begin{subfigure}[t]{0.29\textwidth}
  \centering
  \includegraphics[width=\textwidth]{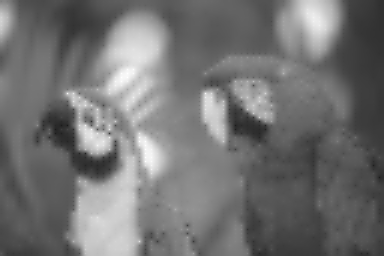}
  \caption{Harmonic\\
   $\textrm{MSE} = 139.10$}
 \end{subfigure}
 \quad
 \begin{subfigure}[t]{0.29\textwidth}
  \centering
  \includegraphics[width=\textwidth]{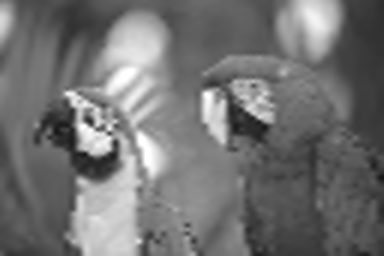}
  \caption{Biharmonic\\
   $\textrm{MSE} = 128.18$}
 \end{subfigure}
 \quad
 \begin{subfigure}[t]{0.29\textwidth}
  \centering
  \includegraphics[
   width=\textwidth]{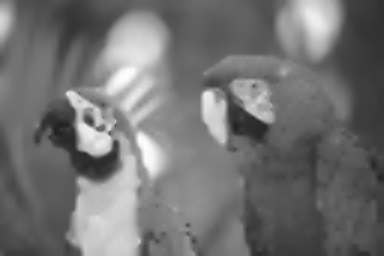}
  \caption{EED\\
   $\textrm{MSE} = 118.79$}
 \end{subfigure}
 \caption[Inpainting of ``parrots'' on regular mask]{Inpainting
  of ``parrots'' for the regular inpainting mask given in \cref{fig:parrots} for 
  various inpainting methods. The SPH inpainting uses a Gaussian kernel. 
  Parameters for EED are $\lambda=1.5$ and $\sigma=2.0$.}
 \label{fig:parrots_reg}
\end{figure}

\begin{figure}[htb]
 \captionsetup[subfigure]{justification=centering}
 \centering
 \begin{subfigure}[t]{0.29\textwidth}
  \centering
  \includegraphics[width=\textwidth]{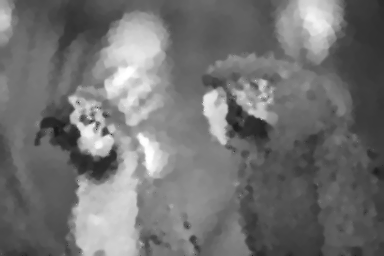}
  \caption{Zero order consistency\\
   $\textrm{MSE} = 169.62$}
 \end{subfigure}
 \quad
 \begin{subfigure}[t]{0.29\textwidth}
  \centering
  \includegraphics[width=\textwidth]{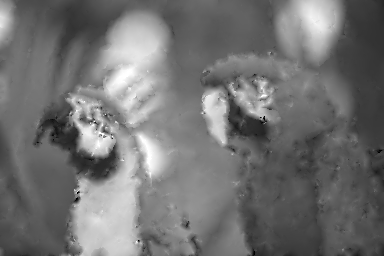}
  \caption{First order consistency\\
   $\textrm{MSE} = 173.71$}
 \end{subfigure}

 \begin{subfigure}[t]{0.29\textwidth}
  \centering
  \includegraphics[width=\textwidth]{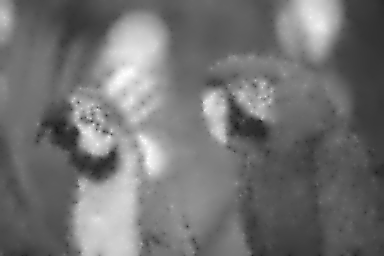}
  \caption{Harmonic\\
   $\textrm{MSE} = 162.53$}
 \end{subfigure}
 \quad
 \begin{subfigure}[t]{0.29\textwidth}
  \centering
  \includegraphics[width=\textwidth]{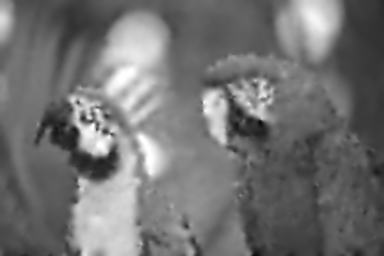}
  \caption{Biharmonic\\
   $\textrm{MSE} = 147.75$}
 \end{subfigure}
 \quad
 \begin{subfigure}[t]{0.29\textwidth}
  \centering
  \includegraphics[
   width=\textwidth]{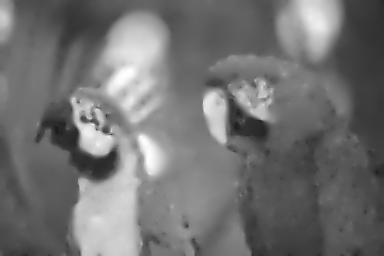}
  \caption{EED\\
   $\textrm{MSE} = 137.04$}
 \end{subfigure}
 \caption[Inpainting of ``parrots'' on regular mask]{Inpainting
  of ``parrots'' for the random inpainting mask given in \cref{fig:parrots} for 
  various inpainting methods. The SPH inpainting uses a Gaussian kernel.
  Parameters for EED are $\lambda=1.2$ and $\sigma=2.0$.}
 \label{fig:parrots_ran}
\end{figure}

As a second example, we consider the ``parrots'' image from the
Kodak database, downscaled to size $384 \times 256$. As inpainting tasks, we 
consider the removal of scratches or overlaid texts as well as inpainting 
based on a sparse regular and a sparse random mask, respectively (see 
\cref{fig:parrots}).
\Cref{fig:parrots_reg} shows the results for the inpainting of
``parrots'' with the regular inpainting masks for SPH inpainting with an 
isotropic Gaussian and diffusion-based inpainting, whereas results for the 
random inpainting mask are given in \cref{fig:parrots_ran}.
For the regular mask, the best MSE is achieved by EED, 
followed by the first order consistency SPH inpainting. For the random
mask, diffusion-based methods show a better performance with EED inpainting
taking the lead both with respect to MSE and visual impression. The first order 
consistency SPH inpainting suffers from the aforementioned artifacts. 

\begin{figure}[htb]
 \captionsetup[subfigure]{justification=centering}
 \centering
 \begin{subfigure}[t]{0.29\textwidth}
  \centering
  \includegraphics[width=\textwidth]{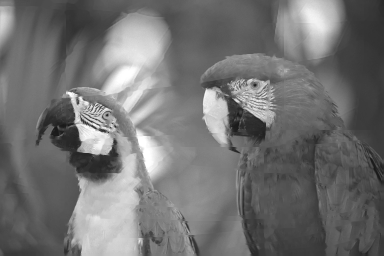}
  \caption{Zero order consistency\\
   $\textrm{MSE} = 37.55$}
 \end{subfigure}
  \quad 
 \begin{subfigure}[t]{0.29\textwidth}
  \centering
  \includegraphics[width=\textwidth]{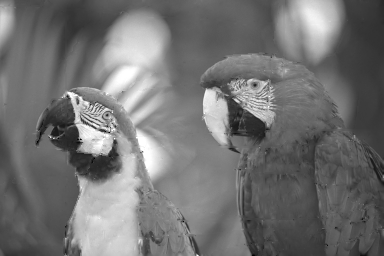}
  \caption{First order consistency\\
   $\textrm{MSE} = 44.92$}
 \end{subfigure}
   \quad 
 \begin{subfigure}[t]{0.29\textwidth}
  \centering
  \includegraphics[width=\textwidth]{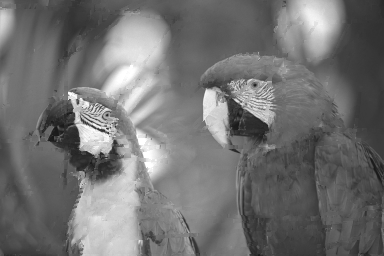}
  \caption{Exemplar-based\\
   $\textrm{MSE} = 76.95$}
 \end{subfigure}
 
 \begin{subfigure}[t]{0.29\textwidth}
  \centering
  \includegraphics[width=\textwidth]{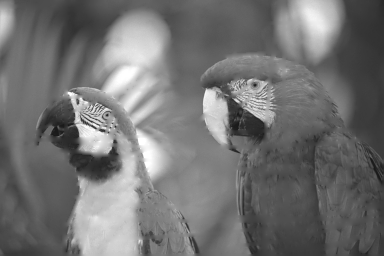}
  \caption{Harmonic\\
   $\textrm{MSE} = 32.94$}
 \end{subfigure}
 \quad 
 \begin{subfigure}[t]{0.29\textwidth}
  \centering
  \includegraphics[width=\textwidth]{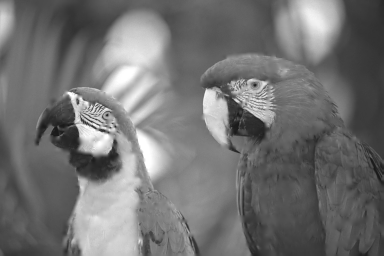}
  \caption{Biharmonic\\
   $\textrm{MSE} = 28.42$}
 \end{subfigure}
   \quad 
 \begin{subfigure}[t]{0.29\textwidth}
  \centering
  \includegraphics[
   width=\textwidth]{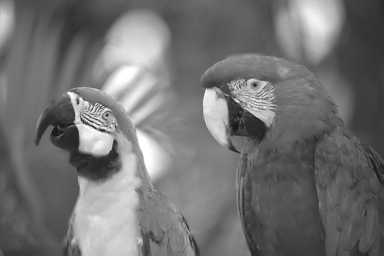}
  \caption{EED\\
   $\textrm{MSE} = 25.70$}
 \end{subfigure}
 \caption[Inpainting of ``parrots'' for scratch removal]{
  Inpainting of ``parrots'' damaged by scratches 
  (cf.~\cref{fig:parrots}). The SPH inpainting uses a Gaussian kernel.
  Parameters for EED are $\lambda=0.8$ and $\sigma=1.8$.}
 \label{fig:parrots_scratches}
\end{figure}

\begin{figure}[htb]
 \captionsetup[subfigure]{justification=centering}
 \centering
 \begin{subfigure}[t]{0.29\textwidth}
  \centering
  \includegraphics[width=\textwidth]{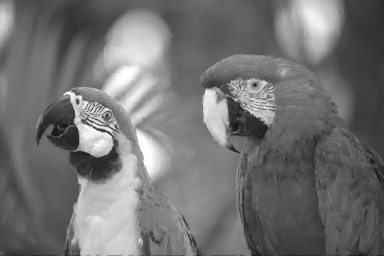}
  \caption{Zero order consistency\\
   $\textrm{MSE} = 24.00$}
 \end{subfigure}
  \quad 
 \begin{subfigure}[t]{0.29\textwidth}
  \centering
  \includegraphics[width=\textwidth]{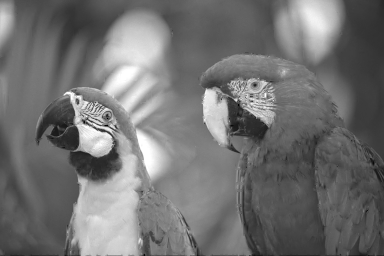}
  \caption{First order consistency\\
   $\textrm{MSE} = 26.41$}
 \end{subfigure}
   \quad 
 \begin{subfigure}[t]{0.29\textwidth}
  \centering
  \includegraphics[width=\textwidth]{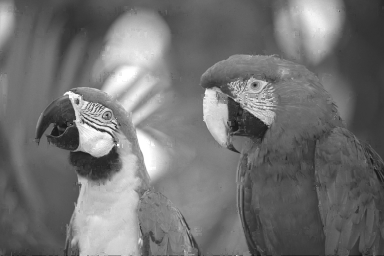}
  \caption{Exemplar-based\\
   $\textrm{MSE} = 35.90$}
 \end{subfigure}
 
 \begin{subfigure}[t]{0.29\textwidth}
  \centering
  \includegraphics[width=\textwidth]{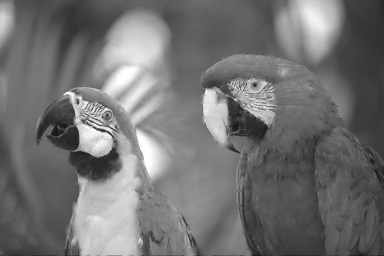}
  \caption{Harmonic\\
   $\textrm{MSE} = 21.20$}
 \end{subfigure}
 \quad 
 \begin{subfigure}[t]{0.29\textwidth}
  \centering
  \includegraphics[width=\textwidth]{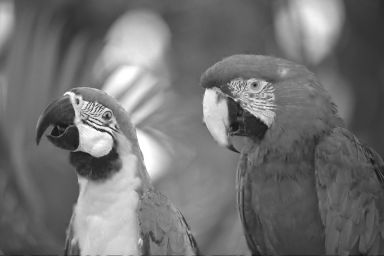}
  \caption{Biharmonic\\
   $\textrm{MSE} = 20.26$}
 \end{subfigure}
   \quad 
 \begin{subfigure}[t]{0.29\textwidth}
  \centering
  \includegraphics[
   width=\textwidth]{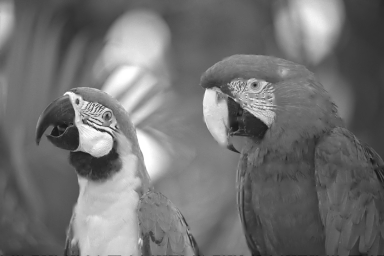}
  \caption{EED\\
   $\textrm{MSE} = 17.76$}
 \end{subfigure}
 \caption[Inpainting of ``parrots'' for text removal]{
  Inpainting of ``parrots'' overlaid by text 
  (cf.~\cref{fig:parrots}). The SPH inpainting uses a Gaussian kernel.
  Parameters for EED are $\lambda=1.2$ and $\sigma=2.0$.}
 \label{fig:parrots_text}
\end{figure}

Results for the inpainting of image ``parrots'' damaged by
scratches can be found in \cref{fig:parrots_scratches} whereas 
\cref{fig:parrots_text} shows the results for text removal.
In both tasks, zero order consistency SPH inpainting performs
better than the first order consistency method. Both are preferable to the 
exemplar-based inpainting, but cannot quite achieve the same quality as the 
diffusion-based approaches. Further results and comparisons are included in the 
supplementary material.

Overall, we see that SPH inpainting is better suited for 
inpainting problems with sparse masks, which are closer in nature to the 
original applications of SPH. Some further remarks on how SPH inpainting may 
be adapted to non-sparse inpainting tasks, which are out of the scope of this 
paper, can be found in \cref{sec:conclusions}. Instead, we focus on how the 
performance of SPH inpainting can be enhanced in settings which allow data
optimization, as they are encountered, e.g., in compression.


\section{Optimized Inpainting for Known Ground Truths}
\label{sec:mask_optimization}


\subsection{A Mixed Order Consistency Method}

The goal of mixed order consistency is to combine zero order and first order 
consistency to get the best possible result. For this purpose, two inpaintings 
are done, one with zero order consistency and one with first order consistency. 
Both results are compared, and the one with the better reconstruction error is 
kept. The new method is described in \cref{alg:new_inpainting}.
\begin{center}
 \begin{algorithm}[htb]
  \DontPrintSemicolon
  \SetKwInOut{Input}{Input}
  \SetKwInOut{Output}{Output}
  \SetKwInOut{Initialize}{Initialize}
  \Input{Original image $\bm{f}$, mask $\bm{c}$}
  \Output{Reconstruction $\bm{u}$}
  \Initialize{Perform Voronoi tessellation. Assign areas of influence $V_{j}$ 
   to mask points. Assign initial smoothing lengths 
   $h_{j} = h_{j,\textrm{init}} = 1$ to mask points. Set $k = 1$.}
  \While{\normalfont{not all pixels $\bm{q}$ have been inpainted}}{
   \For{\normalfont{each pixel} $\bm{q}$} {
    Detect neighboring mask points $\bm{p}_{j}$, $j\in \mathcal{N}(\bm{q})$ of 
    $\bm{q}$.\;
    \eIf{\normalfont{number of neighbors is larger than or equal to required
     minimum and neighbors are not collinear}}{
     Inpaint $\bm{q}$ with zero order consistency
     according to
      \cref{eq:SPH_mod_kernel,eq:mod_kernel_zero01,eq:mod_kernel_zero02}.\;
     Inpaint $\bm{q}$ with first order consistency
     according to 
     \cref{eq:mod_kernel_first01,eq:first_order_comp,eq:28}.\;
     \eIf{\normalfont{error of zero order consistency} is less than 
     \normalfont{error of first order consistency}}
     {
      keep inpainting of $\bm{q}$ with zero order consistency,\;
     }
     {
      keep inpainting of $\bm{q}$ with first order consistency.\;
     }
    }{continue\;}
   }
   $k = k+1$\;
   $h_{j} = k \cdot h_{j,\textrm{init}}$\;
  }
  \caption[Mixed Consistency Algorithm]{Mixed Consistency Algorithm}
  \label{alg:new_inpainting}
 \end{algorithm}
\end{center}

\begin{figure}[htb]
 \captionsetup[subfigure]{justification=centering}
 \centering
 \begin{subfigure}[t]{0.29\textwidth}
  \centering
  \includegraphics[width=\textwidth]{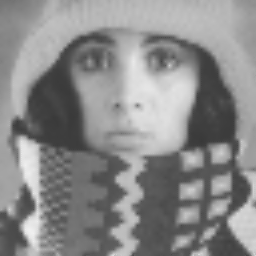}
  \caption{Gaussian\\
   $\textrm{MSE} = 78.37$}
  \label{fig:mixed_reco_6_25_gaussian}
 \end{subfigure}
 \quad
 \begin{subfigure}[t]{0.29\textwidth}
  \centering
  \includegraphics[width=\textwidth]{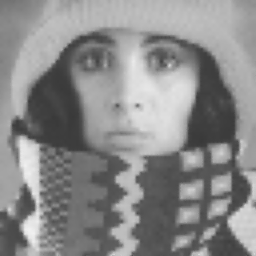}
  \caption{$C^{0}$-Mat\'{e}rn\\
   $\textrm{MSE} = 67.03$}
 \end{subfigure}
 \quad
 \begin{subfigure}[t]{0.29\textwidth}
  \centering
  \includegraphics[width=\textwidth]{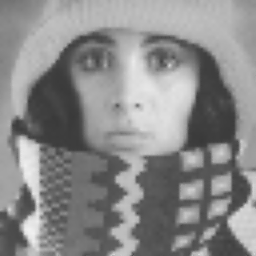}
  \caption{$C^{2}$-Mat\'{e}rn\\
   $\textrm{MSE} = 68.12$}
  \label{fig:mixed_reco_6_25_matern_2}
 \end{subfigure}

 \begin{subfigure}[t]{0.29\textwidth}
  \centering
  \includegraphics[width=\textwidth]{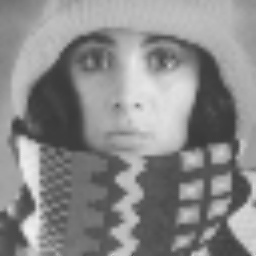}
  \caption{Lucy\\
   $\textrm{MSE} = 78.86$}
  \label{fig:mixed_reco_6_25_lucy}
 \end{subfigure}
 \quad
 \begin{subfigure}[t]{0.29\textwidth}
  \centering
  \includegraphics[width=\textwidth]{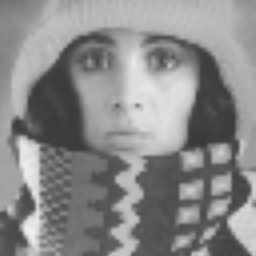}
  \caption{cubic spline\\
   $\textrm{MSE} = 72.78$}
  \label{fig:mixed_reco_6_25_cubic}
 \end{subfigure}
 \quad
 \begin{subfigure}[t]{0.29\textwidth}
  \centering
  \includegraphics[width=\textwidth]{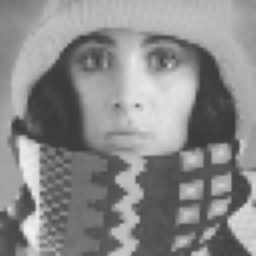}
  \caption{$C^{4}$-Wendland\\
   $\textrm{MSE} = 62.08$}
  \label{fig:mixed_reco_6_25_wendland}
 \end{subfigure}
 \caption[Inpainting of ``trui'' with a 6.25 \% regular mask with a mixed order 
  consistency method]{Inpainting of ``trui'' with the 6.25 \%
  regular mask from \cref{fig:trui_mask} with a mixed order consistency method. 
  Inpainting runtime 19.18 s (averaged across  all kernels).} 
 \label{fig:mixed_regular_6_25}
\end{figure}

\begin{figure}[htb]
 \captionsetup[subfigure]{justification=centering}
 \centering
 \begin{subfigure}[t]{0.29\textwidth}
  \centering
  \includegraphics[width=\textwidth]{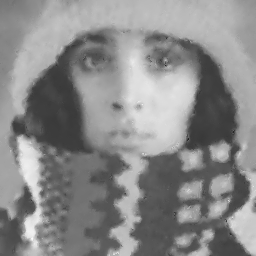}
  \caption{Gaussian\\
   $\textrm{MSE} = 128.37$}
  \label{fig:zero_reco_mixed_5_gaussian}
 \end{subfigure}
 \quad
 \begin{subfigure}[t]{0.29\textwidth}
  \centering
  \includegraphics[width=\textwidth]{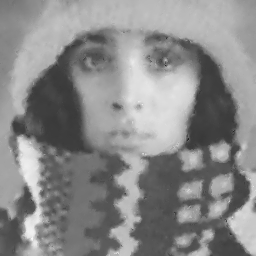}
  \caption{$C^{0}$-Mat\'{e}rn\\
   $\textrm{MSE} = 126.43$}
 \end{subfigure}
 \quad
 \begin{subfigure}[t]{0.29\textwidth}
  \centering
  \includegraphics[width=\textwidth]{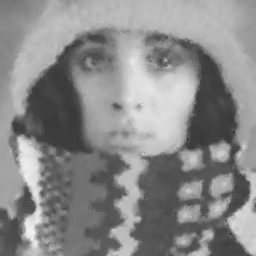}
  \caption{$C^{2}$-Mat\'{e}rn\\
   $\textrm{MSE} = 126.36$}
 \end{subfigure}

 \begin{subfigure}[t]{0.29\textwidth}
  \centering
  \includegraphics[width=\textwidth]{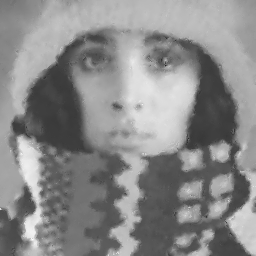}
  \caption{Lucy\\
   $\textrm{MSE} = 127.07$}
  \label{fig:zero_reco_mixed_5_lucy}
 \end{subfigure}
 \quad
 \begin{subfigure}[t]{0.29\textwidth}
  \centering
  \includegraphics[width=\textwidth]{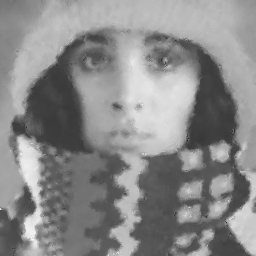}
  \caption{cubic spline\\
   $\textrm{MSE} = 125.51$}
  \label{fig:zero_reco_mixed_5_cubic}
 \end{subfigure}
 \quad
 \begin{subfigure}[t]{0.29\textwidth}
  \centering
  \includegraphics[width=\textwidth]{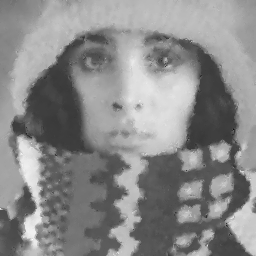}
  \caption{$C^{4}$-Wendland\\
   $\textrm{MSE} = 125.02$}
  \label{fig:zero_reco_mixed_5_wendland}
 \end{subfigure}
 \caption[Inpainting of ``trui'' with a 5 \% random mask with a mixed 
  consistency method]{Inpainting of ``trui'' with 5 \% random mask from 
  \cref{fig:trui_mask} with a mixed consistency method.
  Inpainting runtime 5.13 s (averaged across all kernels).} 
 \label{fig:mixed_random_5}
\end{figure}

Based on the 6.25 \% regular mask from \cref{fig:trui_mask}, 
we get the results depicted in \cref{fig:mixed_regular_6_25}, whereas the 
results for the 5 \% random mask are depicted in \cref{fig:mixed_random_5}. For
both masks, we have used a minimum of five neighbors. The obtained results
are clearly superior to the results achievable with either the zero or first 
order consistent method, especially for the sparser random mask, showing better
MSEs and no visible over- or undershoots. 


\subsection{Spatial Optimization}

Our spatial optimization relies on a novel densification strategy. Instead of a 
probabilistic approach as in \cite{HM13}, we base our method on Voronoi 
tessellation. The algorithm starts with an empty mask and, as an initial step, 
inserts the minimum number of neighbors required at random positions. After this 
step, an initial inpainting takes place and a Voronoi tessellation is performed 
with the initial mask points as ``seeds''. Once this is done, we detect the 
Voronoi cell with the highest error and insert a new mask point at the pixel 
with the highest error within the cell. The error of the reconstruction at a 
pixel $\bm{q}_{j,k}$ in the Voronoi cell $\Omega_{j}$ is defined as 
\begin{equation}
 E_{\bm{q}_{j,k}} = 
 \left\lvert f\!\left(\bm{q}_{j,k}\right) - u\!\left(\bm{q}_{j,k}\right)
  \right\rvert^{2},
\end{equation}
with $f$ being the original image and $u$ the reconstruction, whereas the error 
for the Voronoi cell $\Omega_{j}$ is given by the sum of the reconstruction 
errors at all pixels in the cell, 
i.e., 
\begin{equation}
 E_{\Omega_{j}} = \sum\limits_{\bm{q}_{j,k} \in \Omega_{j}} E_{\bm{q}_{j,k}} =
 \sum\limits_{\bm{q}_{j,k} \in \Omega_{j}} 
  \left\lvert f\!\left(\bm{q}_{j,k}\right) - u\!\left(\bm{q}_{j,k}\right)
  \right\rvert^{2}.
\end{equation}
A new inpainting as well as a new Voronoi tessellation are then computed with 
the new mask and the process continues in the same manner until the required 
mask density is achieved. The densification algorithm is described in 
\cref{alg:densification}. We remark that it is possible to insert more than one 
new mask point in each step to speed up the procedure. However, inserting too 
many mask points at once deteriorates the quality of the final mask. For the 
sake of completeness, we also mention that a densification approach using the 
$\mathrm{L}^{1}$-error within Voronoi cells has been used in \cite{SB00} for 
nearest-neighbor and piecewise constant interpolation. In our experiments, 
using the $\mathrm{L}^{1}$-error always yielded inferior results.
\begin{center}
 \begin{algorithm}
  \DontPrintSemicolon
  \SetKwInOut{Input}{Input}
  \SetKwInOut{Output}{Output}
  \SetKwInOut{Initialize}{Initialize}
  \Input{Original image $\bm{f}$, minimum number of neighbors, number of mask 
   points to add per iteration, required density}
  \Output{Mask $\bm{c}$, reconstruction $\bm{u}$}
  \Initialize{Insert minimum number of neighbors at random positions. Perform 
   Voronoi tessellation and initial inpainting.}
  \While{\normalfont{mask density $<$ required mask density}}{
   Find Voronoi cell(s) $\Omega_{j}$ with highest error $E_{\Omega_{j}}$.\;
   Find pixel(s) $\bm{q}_{j,k}$ in cell(s) $\Omega_{j}$ with highest error(s) 
   $E_{\bm{q}_{j,k}}$.\;
   Add mask point(s) at position(s) $\bm{q}_{j,k}$.\;
   Perform Voronoi tessellation.\;
   Perform inpainting.\;
  }
  \caption{Densification Algorithm}
  \label{alg:densification}
 \end{algorithm}
\end{center}


\subsection{Tonal Optimization}
\label{sec:tonal_opt}

Apart from spatial optimization, we also incorporate a gray value optimization 
of the mask points for a fixed mask $\bm{c}$. The goal of this process is to 
find the optimal gray values $\bm{g}$ such that the mean square error of the 
reconstructed image is minimal. Indeed, for a fixed mask $\bm{c}$, the 
inpainting is given by \cref{eq:SPH_mod_kernel}. However, due to the adaptive 
nature of the smoothing length $h_{j}$ which is not only determined in 
dependence of the mask point $\bm{p}_{j}$, but also in dependence of the pixel 
$\bm{q}$ currently under consideration for inpainting, we have to perform one 
inpainting for our final mask first to determine all necessary smoothing 
lengths. Once this is done, \cref{eq:SPH_mod_kernel} can be written as a matrix 
vector multiplication of the form $\bm{u} = \bm{A} \widetilde{\bm{f}}$ where 
$\bm{u}$ is a vector containing the values of the reconstruction at every 
pixel, $\bm{A}$ is a matrix containing the values of all modified kernels at 
all pixels multiplied with their area of influence $V_{j}$, and 
$\widetilde{\bm{f}}$ is a vector containing the values of the original image at 
all mask points $\bm{p}_{j}$. Thus, as $\bm{A}$ is fixed now, $\bm{u}$ can be 
interpreted as the solution to an interpolation problem at the mask points. 
With this interpretation, it is straightforward to consider the corresponding 
least-squares problem. With the above form for $\bm{u}$, it can be written as
\begin{equation}
 \min\limits_{\bm{g}} \left\lVert \bm{A}\bm{g} - \bm{f} \right\rVert^{2},
\end{equation}
where $\bm{f}$ denotes the vector with the original image values at all pixels 
and $\bm{g}$ is a vector with gray values at the mask points, which can be 
determined by solving the normal equations
\begin{equation}
 \bm{A}^{T} \bm{A} \bm{g} = \bm{A}^{T} \bm{f}.
\end{equation}
Although the least squares problem can be solved directly, we prefer an 
iterative solver instead, specifically the conjugate gradient on the normal 
residual (CGNR) method \cite{Sa03}. This variant of conjugate gradients avoids
the explicit computation of $\bm{A}^{T} \bm{A}$ to reduce runtime and 
circumvent the larger condition number of $\bm{A}^{T} \bm{A}$ compared to 
$\bm{A}$. We 
always initialize by choosing for $\bm{g}_{0}$ the zero vector. As stopping 
criterion we use a threshold on the relative residual defined such that
\begin{equation}
 \frac{\left\lVert \bm{A}^{T} \bm{f} - \bm{A}^{T} \bm{A} \bm{g}_{k} 
  \right\rVert}{\left\lVert \bm{A}^{T} \bm{f}\right\rVert}  \leq 10^{-8}.
\end{equation}
The above procedure is clear for the zero and first order consistency method as 
they use the same kind of modified kernel in every pixel. It stays valid for 
the mixed order consistency method as in this approach, the kernel that is used 
at each pixel $\bm{q}$ is of the same type for all mask points $\bm{p}_{j}$ 
contributing to the inpainting at $\bm{q}$. Thus, for the mixed consistency 
method, the type of the modified kernel changes with the rows in $\bm{A}$, but 
stays the same within each row over all columns. We can still write the whole 
inpainting process as a matrix-vector-multiplication and solve the associated 
least-square problem to perform tonal optimization.


\subsection{Inpainting on Spatially and Tonally Optimized Data with Isotropic 
Kernels}
\label{sec:inp_isotropic}

\begin{figure}[htb]
 \captionsetup[subfigure]{justification=centering}
 \centering
 \begin{subfigure}{0.29\textwidth}
  \centering
  \includegraphics[
   width=\textwidth]{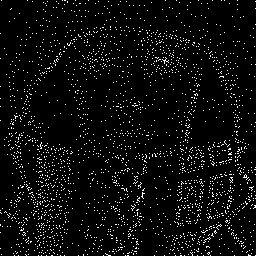}
  \caption{Optimized 5 \% mask}
  \label{fig:5_mask}
 \end{subfigure}
 \quad 
 \begin{subfigure}{0.29\textwidth}
  \centering
  \includegraphics[
   width=\textwidth]{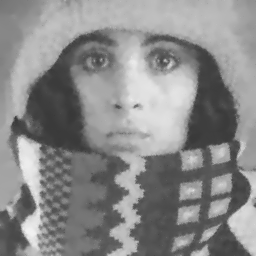}
  \caption{$\textrm{MSE} = 30.65$}
  \label{fig:zero_reco_5}
 \end{subfigure}
 \quad
 \begin{subfigure}{0.29\textwidth}
  \centering
  \includegraphics[
   width=\textwidth]{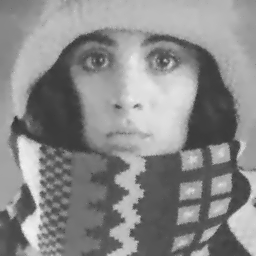}
  \caption{$\textrm{MSE} = 19.62$}
 \end{subfigure}
 \caption{Inpainting of ``trui'' with spatially and tonally 
  optimized mask with a zero order consistency method with an isotropic 
  Gaussian kernel. 
  We show the 5 \% optimized zero order consistency mask (\textbf{a}), zero 
  order consistency inpainting result with this mask without tonal optimization 
  (\textbf{b}), and zero order consistency inpainting result on this mask with 
  tonal optimization (\textbf{c}). 
  Runtimes were 89.09 min for densification and 2.59 min for 
  tonal optimization.}
 \label{fig:zero_dense_5}
\end{figure}

As an example for inpainting on an optimized mask with zero order consistency, 
we present the results produced with a Gaussian kernel in 
\cref{fig:zero_dense_5}. Even without tonal optimization, the MSE improves by 
roughly a factor $6.5$. With tonal optimization, the MSE improves by a factor 
$10$ with respect to the random mask and by almost $35 \%$ with respect to the 
result on the spatially optimized mask without tonal optimization. As far as 
spatial optimization is concerned, the densification process prefers to capture 
the geometry of the image, by adding more mask points near edges compared to 
rather homogeneous regions of the image.

\begin{figure}[htb]
 \captionsetup[subfigure]{justification=centering}
 \centering
 \begin{subfigure}[t]{0.29\textwidth}
  \centering
  \includegraphics[
   width=\textwidth]{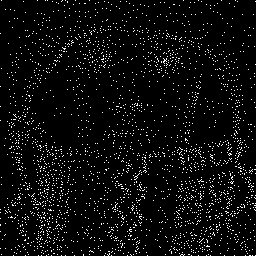}
  \caption{Optimized 5 \% mask}
  \label{fig:5_mixed_isotropic_mask}
 \end{subfigure}
 \quad
 \begin{subfigure}[t]{0.29\textwidth}
  \centering
  \includegraphics[
   width=\textwidth]{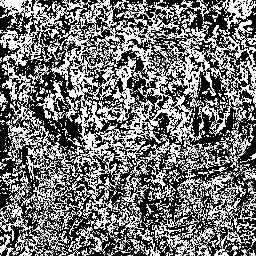}
  \caption{Mixed order consistency map}
  \label{fig:consistency_map}
 \end{subfigure}

 \begin{subfigure}[t]{0.29\textwidth}
  \centering
  \includegraphics[
   width=\textwidth]{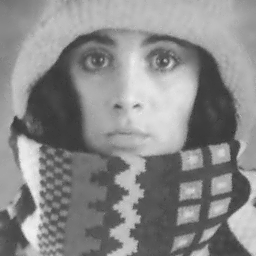}
  \caption{$\textrm{MSE} = 16.30$}
  \label{fig:mixed_isotropic_reco_5}
 \end{subfigure}
 \quad 
 \begin{subfigure}[t]{0.29\textwidth}
  \centering
  \includegraphics[
   width=\textwidth]{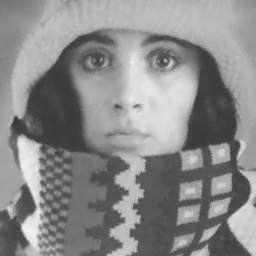}
  \caption{$\textrm{MSE} = 11.68$}
 \end{subfigure}
 \caption{Inpainting of ``trui'' with spatially and tonally 
  optimized mask with a mixed order consistency method with an isotropic 
  Gaussian kernel. (\textbf{a}) shows an optimized 5 \% mask for mixed order 
  consistency SPH inpainting. (\textbf{b}) shows a mixed order consistency map 
  with white areas denoting first order consistency reconstruction and black 
  areas denoting zero order consistency reconstruction. (\textbf{c}) shows a 
  mixed order consistency inpainting result on the mask from (\textbf{a}) 
  without tonal optimization. (\textbf{d}) shows a mixed order consistency 
  inpainting on the same mask with tonal optimization. 
  Runtimes were 111.28 min for densification and 2.66 min for tonal 
  optimization.} 
 \label{fig:mixed_dense_5}
\end{figure}

Changing the SPH inpainting method from zero order consistency to the mixed 
order consistency method improves the result even further as can be seen in
\cref{fig:mixed_dense_5}. Using the same isotropic Gaussian kernel as before, 
spatial optimization improves the MSE by almost a factor $8$ compared to the the 
random mask result in \cref{fig:zero_reco_mixed_5_gaussian}. A comparison with 
respect to the inpainting method instead of with respect to the mask shows 
improvement of almost a factor $2$ with respect to the MSE compared to the 
results in \cref{fig:zero_dense_5}. Unfortunately, there seems to be no 
structure in the distribution of pixels for which a first order consistency and 
for which a zero order consistency method performs better, respectively, as can 
be seen in \cref{fig:consistency_map}.

\Cref{table:2} summarizes MSEs of inpainting results for 
``trui'' with the other isotropic kernels used in \cref{fig:random_5} if 
these kernels are equipped with optimized masks containing $5 \%$ of all pixels 
and tonal optimization is performed. 
\begin{table}[htb]
 \begin{tabular}{lcc}
  Kernel & Zero order consistency & mixed order consistency \\ 
  \midrule
  Gaussian & 19.62 & 11.68 \\ 
  $C^{0}$-Mat\'{e}rn & 18.57 & \textbf{9}.\textbf{95} \\
  $C^{2}$-Mat\'{e}rn & \textbf{17}.\textbf{93} & 9.99 \\ 
  Lucy & 21.83 & 12.15 \\
  cubic spline & 21.19 & 11.71 \\
  $C^{4}$- Wendland & 23.33 & 11.83
 \end{tabular}
 \caption[MSE comparison between zero and mixed order consistency optimized 
  data inpainting with isotropic kernels of ``trui'']{MSE 
  comparison between zero and mixed order consistency optimized inpainting 
  with isotropic kernels on  ``trui'' for 5 \% masks.}
 \label{table:2}
\end{table}
Once again, the benefits of mixed order consistency are quite substantial 
since a significant decrease of the MSE has been achieved in all cases compared 
to zero order consistency. Even the best performing kernel for the zero order 
consistency method, the $C^{2}$-Mat\'{e}rn kernel, has an MSE which is 
approximately 50 \% larger than the MSE of the worst performing kernel in the 
mixed consistency setting and almost double of the MSE of the best performing 
kernel in the mixed consistency setting 
which is the $C^{0}$-Mat\'{e}rn kernel. 
Further, it appears that compactly supported kernels perform worse in 
this setting for the zero order consistency SPH inpainting than truncated 
kernels, but are competitive in the mixed order consistency method. 


\subsection{Optimized Inpainting with Anisotropic Kernels}
\label{sec:opt_inp_aniso_kernel}

The observation that optimized mask points tend to cluster around edges and the 
fact that edges are clearly oriented structures suggest to adapt the support of 
kernels to account for this by incorporating anisotropy. This is further 
supported by results which show that incorporating anisotropy in other 
inpainting strategies can improve reconstruction quality compared to the 
related isotropic method \cite{SP14}. 

For SPH, anisotropic kernels have been used in the so-called Adaptive Smoothed 
Particle Hydrodynamics (ASPH) formulation \cite{SM96} to better account for the 
actual distribution of particles. Here, we replace the smoothing length $h$ by 
a symmetric positive definite tensor $\bm{G} \in \mathbb{R}^{2 \times 2}$ for a 
two-dimensional problem and redefine $\bm{\eta}$ as
\begin{equation}
 \label{eq:after_new}
 \bm{\eta} = \bm{G} \left(\bm{q}-\bm{p}\right).
\end{equation}
$\bm{G}$ has units of inverse length and in the isotropic case it is given by a 
diagonal matrix with each diagonal element equal to $\frac{1}{h}$. This 
observation makes it clear that we also have to adapt the normalization of our 
kernels from a factor $\frac{\rho}{h^{2}}$ in \cref{eq:kernel_rbf} to a factor
$\rho\, \det(\bm{G})$. 

For SPH inpainting, we determine the anisotropy from the distribution of mask 
points. For this purpose, we follow the approach of \cite{YT13} by constructing 
a weighted local covariance matrix $\bm{C}$ within a fixed predetermined window 
around each mask point. For a known mask point $\bm{p}_{j}$, the covariance 
matrix is given by
\begin{equation}
 \bm{C}_{j} = 
 \frac{\sum\limits_{\ell} 
  w_{j,\ell} \left(\bm{p}_{\ell}-\widetilde{\bm{p}}_{j}\right) 
  \left(\bm{p}_{\ell} - \widetilde{\bm{p}}_{j} \right)^{T}}{
  \sum\limits_{\ell} w_{j,\ell}},
  \qquad \text{ with } \qquad
  \widetilde{\bm{p}}_{j} = 
  \frac{\sum\limits_{\ell} w_{j,\ell}\, \bm{p}_{\ell}}{\sum\limits_{\ell} 
   w_{j,\ell}}.
\end{equation}
Here, $\ell$ numbers the mask points within a neighborhood of $\bm{p}_{j}$. It 
is necessary to restrict the set of mask points under consideration to such a 
neighborhood to catch the locally prevalent direction of structures in the 
image. Next, we perform a singular value decomposition (SVD). As $\bm{C}_{j}$ is 
symmetric and positive semidefinite by construction, this is the same as the 
eigenvalue decomposition
\begin{equation}
 \bm{C}_{j} = \bm{Q} \bm{D} \bm{Q}^{T},
\end{equation}
with a rotation matrix $\bm{Q}$ and a matrix $\bm{D}$ with nonnegative 
eigenvalues along the diagonal in decreasing order. As $\bm{C}_{j}$ is 
constructed from the positions of mask points, its eigenvalues can be assigned a 
unit of length. The eigenvectors in $\bm{Q}$ correspond to the directions of 
major and minor axis of an ellipse whose orientation is in line with the locally 
prevalent orientation in the distribution of mask points. Hence, the tensor 
$\bm{G}$ is given by
\begin{equation}
 \bm{G} = \bm{Q} \bm{D}^{-1} \bm{Q}^{T},
\end{equation}
such that it has units of inverse length as desired.

In the context of kernel regression, the matrix $\bm{C}$ that
we have introduced above is related to the so-called ``steering matrix'' of an
anisotropic regression kernel \cite{TF07}.

\begin{figure}[htb]
 \captionsetup[subfigure]{justification=centering}
 \centering
 \begin{subfigure}[t]{0.29\textwidth}
  \centering
  \includegraphics[
   width=\textwidth]{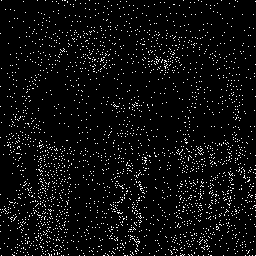}
  \caption{Optimized 5 \% mask}
  \label{fig:5_mixed_anisotropic_mask}
 \end{subfigure}
 \quad
 \begin{subfigure}[t]{0.29\textwidth}
  \centering
  \includegraphics[
 width=\textwidth]{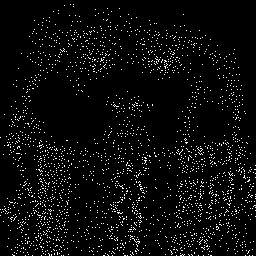}
  \caption{Anisotropic mask points of the 5 \% mask}
 \end{subfigure}
 \quad
 \begin{subfigure}[t]{0.29\textwidth}
  \centering
  \includegraphics[
   width=\textwidth]{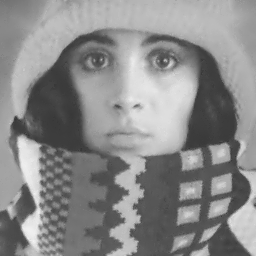}
  \caption{$\textrm{MSE} = 10.49$}
 \end{subfigure}
 \caption{Inpainting of ``trui'' with spatially and tonally 
  optimized mask with a mixed order consistency method with an anisotropic 
  Gaussian kernel.
  From left to right: Optimized 5 \% mask for mixed order consistency method 
  (\textbf{a}), mask points which incorporate anisotropic kernels in white 
  (\textbf{b}),  and mixed order consistency inpainting result with given mask 
  with tonal optimization (\textbf{c}). 
  Runtimes were 129.44 min for densification and 2.51 min for  
  tonal optimization.}
 \label{fig:mixed_anisotropic_dense_5}
\end{figure}

In our experiments, we incorporate anisotropy after spatially optimizing mask 
points for isotropic kernels. We fix the window size for construction of 
covariance matrices to $25 \times 25$ pixels and demand a minimum number of 
15 mask points within that window. If this minimum number of mask points is 
not satisfied, the corresponding kernel stays isotropic. This behavior is 
desirable since the densification process results in masks where the majority 
of mask points are placed near discontinuities rather than in homogeneous areas 
of the image. Thus, a low local density of mask points implies homogeneous 
areas of the image. The results achieved with mixed order consistency and an 
anisotropic Gaussian kernel are depicted in 
\cref{fig:mixed_anisotropic_dense_5}.

Compared to the result in \cref{fig:mixed_dense_5}, 
we observe an improvement in MSE of only 1.19, which is 
roughly 11 \%, whereas a large amount of mask 
points is now equipped with anisotropic kernels. This relatively moderate 
improvement may be explained by the fact that our method of determining 
anisotropy relies on the local spatial distribution of mask points whereas many 
important structures in the given image live on a mesoscale. This behavior 
cannot be captured by increasing the size of the search window as the covariance 
matrix becomes prone to incorporating the orientations of neighboring 
structures, resulting in a more isotropic behavior instead of a better 
orientation along the mesoscale structures. \Cref{table:anisotropy} summarizes 
MSEs of inpainting results for ``trui'' with the other anisotropic kernels used 
if these kernels are equipped with optimized masks containing $5 \%$ of all 
pixels and tonal optimization is performed. 
\begin{table}[htb]
 \begin{tabular}{lr}
  Kernel & MSE \\ 
  \midrule
  Gaussian & 10.49 \\ 
  $C^{0}$-Mat\'{e}rn & \textbf{9}.\textbf{51}  \\
  $C^{2}$-Mat\'{e}rn & 9.81  \\ 
  Lucy & 11.57  \\
  cubic spline & 11.26 \\
  $C^{4}$- Wendland & 11.95
 \end{tabular}
 \caption[MSE with mixed order consistency optimized data inpainting with 
  anisotropic kernels of ``trui'']{MSE with mixed order 
  consistency optimized inpainting with anisotropic kernels on  ``trui'' 
  for 5 \% masks.}
 \label{table:anisotropy}
\end{table}
%


\subsection{Performance Compared to Diffusion-based and Exemplar-based 
Inpainting Methods}
\label{sec:comparisons}

\begin{figure}[htb]
 \captionsetup[subfigure]{justification=centering}
 \centering
 \begin{subfigure}[t]{0.29\textwidth}
  \centering
  \includegraphics[
   width=\textwidth]{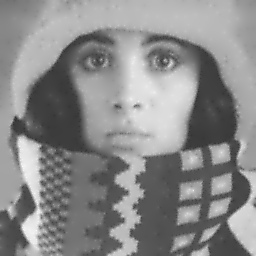}
  \caption{Harmonic\\
  $\textrm{MSE} = 20.18$}
 \end{subfigure}
 \quad
 \begin{subfigure}[t]{0.29\textwidth}
  \centering
  \includegraphics[
   width=\textwidth]{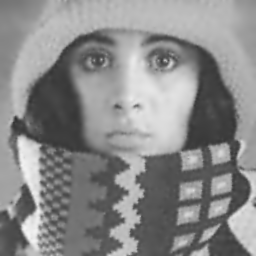}
  \caption{Biharmonic\\
  $\textrm{MSE} = 15.00$}
 \end{subfigure}
 \caption{Inpainting of ``trui'' with spatially and tonally 
  optimized 5 \% mask for harmonic and biharmonic inpainting.}
 \label{fig:trui_diff_opt}
\end{figure}

To assess the performance of SPH inpainting, we combine our
implementations of harmonic and biharmonic inpainting with our Voronoi-based
densification strategy and a tonal optimization approach similar in spirit to 
\cref{sec:tonal_opt}. The results achieved by these two inpainting methods 
are depicted in \cref{fig:trui_diff_opt} together with the corresponding
MSEs.
As we can see from \cref{table:2}, SPH inpainting with mixed
order consistency performs better than these two diffusion-based methods even
if we consider only isotropic kernels. \Cref{table:3} shows that we can obtain
an MSE which is less than half that of harmonic inpainting if we incorporate
anisotropy.

\begin{figure}[htb]
 \captionsetup[subfigure]{justification=centering}
 \centering
 \begin{subfigure}[t]{0.29\textwidth}
  \centering
  \includegraphics[
   width=\textwidth]{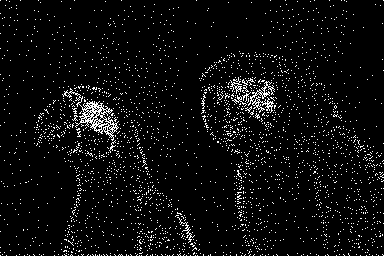}
  \caption{Isotropic mask}
 \end{subfigure}
 \quad 
 \begin{subfigure}[t]{0.29\textwidth}
  \centering
  \includegraphics[
   width=\textwidth]{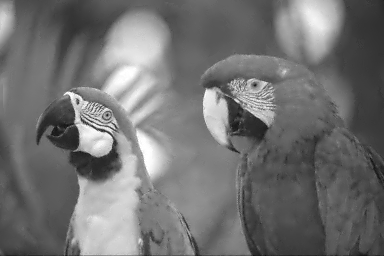}
  \caption{Isotropic, mixed order\\
  $\textrm{MSE} = 10.07$}
 \end{subfigure}

 \begin{subfigure}[t]{0.29\textwidth}
  \centering
  \includegraphics[
   width=\textwidth]{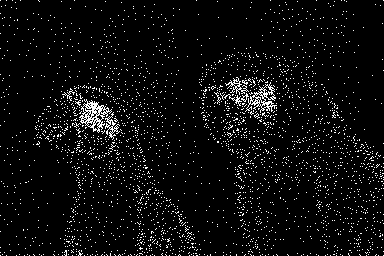}
  \caption{Anisotropic mask}
 \end{subfigure}
 \quad 
 \begin{subfigure}[t]{0.29\textwidth}
  \centering
  \includegraphics[
   width=\textwidth]{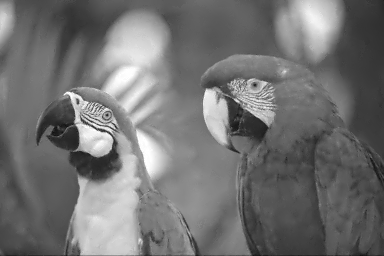}
  \caption{Anisotropic, mixed order\\
  $\textrm{MSE} = 8.51$}
 \end{subfigure}
 \caption[SPH inpainting of ``parrots'' with spatially and tonally optimized 
  mask]{Inpainting of ``parrots'' with a 5 \% spatially and 
  tonally optimized mask with a mixed order consistency method and Gaussian
  kernels. Left column shows the masks. Right column shows the inpaintings. 
  Top row is the isotropic case. Bottom row is the anisotropic case.} 
 \label{fig:parrots_SPH_opt} 
\end{figure}

\begin{figure}[htb]
 \captionsetup[subfigure]{justification=centering}
 \centering
 \begin{subfigure}[t]{0.29\textwidth}
  \centering
  \includegraphics[
   width=\textwidth]{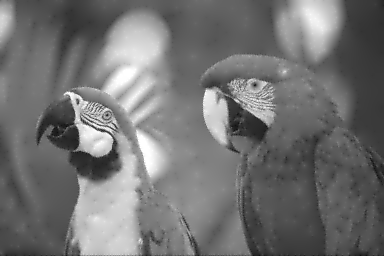}
  \caption{Harmonic\\
   $\textrm{MSE} = 16.44$}
 \end{subfigure}
 \quad
 \begin{subfigure}[t]{0.29\textwidth}
  \centering
  \includegraphics[
   width=\textwidth]{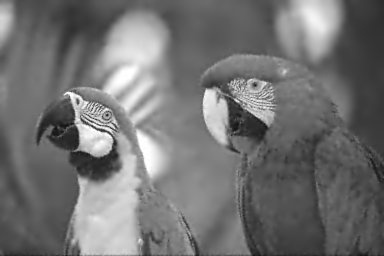}
  \caption{Biharmonic\\
   $\textrm{MSE} = 15.32$}
 \end{subfigure}
 \caption[Inpainting of ``parrots'' with spatially and tonally optimized 
  mask]{Inpainting of ``parrots'' with a 5 \% spatially and 
  tonally optimized mask with harmonic (\textbf{a}) and biharmonic 
  (\textbf{b}) inpainting.}
 \label{fig:parrots_diff_opt} 
\end{figure}

As another example, consider the ``parrots'' image from 
\cref{fig:parrots}. \Cref{fig:parrots_SPH_opt} shows the results obtained by 
Voronoi densification for a 5 \% mask for mixed order consistency SPH 
inpainting 
with isotropic and anisotropic Gaussian kernels, including tonal optimization.
The corresponding results achieved by harmonic and biharmonic
inpainting are shown in \cref{fig:parrots_diff_opt}.
Already the isotropic variant of SPH inpainting reduces the MSE of the 
diffusion-based methods by at least 34 \%, whereas the anisotropic version
achieves a reduction by 44 \%. Further comparisons between SPH inpainting and
diffusion-based strategies when equipped with Voronoi-based densification are
included in the supplement.

\begin{figure}[htb]
 \captionsetup[subfigure]{justification=centering}
 \centering
 \begin{subfigure}[t]{0.29\textwidth}
  \centering
  \includegraphics[
   width=\textwidth]{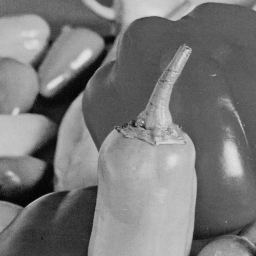}
  \caption{Original image ``peppers''}
 \end{subfigure}
 \quad
 \begin{subfigure}[t]{0.29\textwidth}
  \centering
  \includegraphics[
   width=\textwidth]{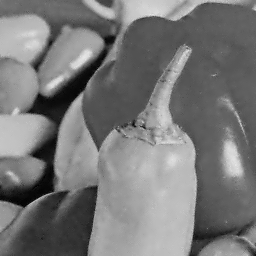}
  \caption{Inpainted ``peppers''}
 \end{subfigure}

 \begin{subfigure}[t]{0.29\textwidth}
  \centering
  \includegraphics[
   width=\textwidth]{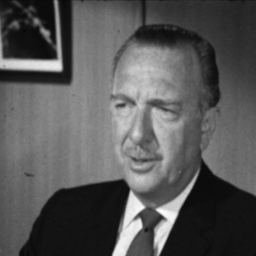}
  \caption{Original image ``walter''}
 \end{subfigure}
 \quad 
 \begin{subfigure}[t]{0.29\textwidth}
  \centering
  \includegraphics[
   width=\textwidth]{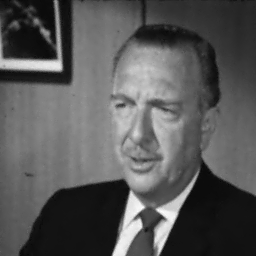}
  \caption{Inpainted ``walter''}
 \end{subfigure}
 \caption{Images ``peppers'' and ``walter'' (left column) and
  mixed order consistency SPH inpaintings with isotropic Gaussian kernels
  on spatially and tonally optimized 5 \% masks (right column).} 
 \label{fig:peppers_walter_opt_mixed_iso_gauss}
\end{figure}

In order to assess the performance of our inpainting method 
further,
we compare it with other existing methods in the literature. As a first example, 
we consider results for harmonic inpainting on ``trui'' for an optimized 5 \% 
mask from \cite{MH11}. There, spatial optimization was done with a probabilistic 
sparsification and further improved with a Nonlocal Pixel Exchange (NLPE). With 
optimally chosen mask points and gray values, harmonic inpainting 
shows an impressive quality in reconstructing the original image. 
We compare these results with the ones we got for a mixed order 
consistency SPH inpainting with $C^{0}$-Mat\'{e}rn kernels. 
To be fair, we only consider isotropic kernels since harmonic inpainting has 
no way to incorporate anisotropy. The results are summarized in \cref{table:4}.
\begin{table}[htb]
 \begin{tabular}{lcc}
  Method & Spatially Optimized & Spatially \& Tonally Optimized \\ \midrule
  Harmonic Inpainting & 23.21 (with NLPE) & 17.17 (with NLPE) \\ 
  Mixed SPH (Isotropic) & \textbf{13.88} & \textbf{9.95}
 \end{tabular}
 \caption[MSE of 5 \% ``trui'']{MSE for inpaintings of ``trui'' on optimized 
  5 \% masks. Compared are results achieved with harmonic inpainting
  in \cite{MH11} and results from our method with an 
  isotropic $C^{0}$-Mat\'{e}rn kernel with mixed order consistency.}
 \label{table:4}
\end{table}
Evidently, we can outperform harmonic inpainting, both without and with tonal
optimization. 
Further results in \cite{MH11} report MSEs for spatially and 
tonally optimized harmonic inpainting on the images ``peppers'' and ``walter''.
We include these images together with results obtained with mixed order 
consistency SPH inpainting with isotropic Gaussians in 
\cref{fig:peppers_walter_opt_mixed_iso_gauss}. The MSEs are reported together
with those from \cite{MH11} in \cref{table:MSE_peppers_walter_mainb}. Again, 
the results obtained with our method are 16 \% and 37 \% better, respectively.
\begin{table}[htb]
 \begin{tabular}{lcc}
  Image & Harmonic Inpainting & Mixed SPH (Isotropic) \\ \midrule
  ``peppers'' & 19.38 & \textbf{16.29} \\ 
  ``walter'' & 8.14 & \textbf{5.15}
 \end{tabular}
 \caption[MSE of 5 \% ``peppers'' and ``walter'']{MSE for
  inpaintings on optimized 5 \% masks including tonal optimization for images
  ``peppers'' and ``walter''. Compared are results achieved with harmonic 
  inpainting including NLPE from \cite{MH11} and results from our method with 
  an isotropic Gaussian kernel with mixed order consistency.}
 \label{table:MSE_peppers_walter_mainb}
\end{table}

To evaluate SPH inpainting with anisotropic kernels, we consider the results 
from \cite{HM17} achieved with edge-enhancing diffusion (EED) inpainting for 
``trui'' with a mask of density 4 \% that is constructed by probabilistic 
sparsification. The authors report the MSE of inpainting on this mask without 
tonal optimization and improve the location of mask points further with NLPE 
before considering tonal optimization. As competitor, we used a mixed order 
consistency SPH inpainting with anisotropic $C^{0}$-Mat\'{e}rn kernels on an 
optimized 4 \% mask. We also tried to improve our inpaintings with NLPE. However, 
the reduction on MSE was negligible. Results for both inpainting methods are 
summarized in \cref{table:3}.
\begin{table}[htb]
 \begin{tabular}{lcc}
  Method & Spatially Optimized & Spatially + Tonally Optimized \\ \midrule 
  Edge-Enhancing Diffusion (EED) & 24.20 & \textbf{10.79} (with NLPE) \\
  Mixed SPH (Anisotropic) & \textbf{17.10} & 12.28 
 \end{tabular}
 \caption[MSE of 4 \% ``trui'']{MSE for inpaintings of ``trui'' on optimized 
  4 \% masks. Compared are results from EED inpainting \cite{HM17} 
  and our method with an anisotropic $C^{0}$-Mat\'{e}rn kernel with mixed 
  order consistency.}
 \label{table:3}
\end{table}
As can be seen, we outperform EED if we only incorporate spatial optimization, 
but no tonal optimization. By construction, EED should perform better in 
preserving edges \cite{We98}. Thus, we conjecture that probabilistic 
sparsification, which only relies on pointwise errors, is inferior to our 
Voronoi-based densification method as long as the former is not improved by a 
consecutive NLPE.

\begin{figure}[htb]
 \captionsetup[subfigure]{justification=centering}
 \centering
 \begin{subfigure}[t]{0.29\textwidth}
  \centering
  \includegraphics[
  width=\textwidth]{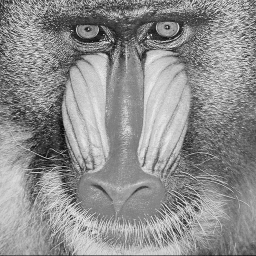}
  \caption{Image ``baboon''}
 \end{subfigure}
 \quad
 \begin{subfigure}[t]{0.29\textwidth}
  \centering
   \includegraphics[
   width=\textwidth]{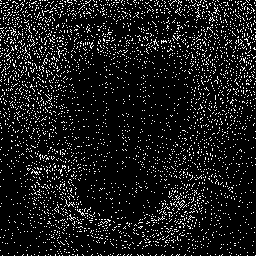}
  \caption{Optimized 10 \% mask}
 \end{subfigure}
 
 \begin{subfigure}[t]{0.29\textwidth}
  \centering
  \includegraphics[
   width=\textwidth]{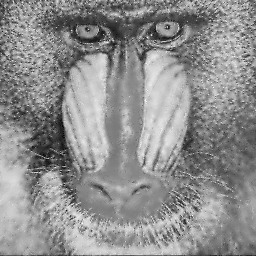}
  \caption{$\textrm{MSE} = 290.64$}
 \end{subfigure}
 \quad
 \begin{subfigure}[t]{0.29\textwidth}
  \centering
  \includegraphics[
   width=\textwidth]{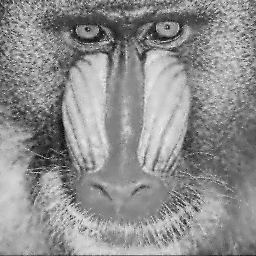}
  \caption{$\textrm{MSE} = 223.37$}
 \end{subfigure}
 \caption{Inpainting of ``baboon'' (rescaled to $256 \times 256$ pixels as in 
  \cite{KB18}) with a spatially optimized mask for a mixed order consistency 
  method with an isotropic Gaussian kernel.
  We show the original image (\textbf{a}), an optimized 
  10 \% mask for mixed order consistency method (\textbf{b}), a mixed order 
  consistency inpainting result with given mask without tonal optimization 
  (\textbf{c}), and a mixed order consistency inpainting result with given mask 
  with tonal optimization.
  Runtimes were 244.18 min for densification and 5.40 min for tonal 
  optimization.}  
 \label{fig:baboon}
\end{figure}

To evaluate the performance of our method for images rich in texture, we 
consider the exemplar-based inpainting technique from \cite{KB18} and the 
results given there for an inpainting of a gray value version of the ``baboon'' 
image. In this setting, the authors report an MSE of 518.52 on a mask 
constructed with ``densification by dithering'' and a consecutive NLPE. As tonal 
optimization is not considered in \cite{KB18}, we compare to the result our 
method could achieve for mixed order consistency inpainting with isotropic 
Gaussians on an optimized 10 \% mask in \cref{fig:baboon}. 
Already without tonal optimization, the MSE is 290.64, which 
means we outperform the exemplar-based inpainting method by almost 44 \%. 
Including tonal optimization improves the result further to an MSE of 223.37,
less than half of the MSE the exemplar-based method could achieve.

For the sake of completeness, we also consider the results on 
``trui'' reported for a 10 \% masked in \cite{KB18}. Here, the exemplar-based
approach could achieve an MSE of 12.99 with spatial optimization including NLPE.
Our mixed consistency SPH inpainting equipped with isotropic Gaussian kernels
and a spatially optimized 10 \% mask can inpaint ``trui'' with an MSE of 6.65
which translates to an improvement of 49 \%. Tonal optimization decreases the 
MSE further to 4.68 or an improvement of 64 \% compared to the exemplar-based 
method.


\section{Conclusions and Outlook}
\label{sec:conclusions}

We have shown that smoothed particle hydrodynamics is a highly 
competitive method for the challenging problem of sparse data inpainting. It 
can produce results on par or even better than other, better explored PDE- or 
exemplar-based inpainting strategies.

The success of SPH for sparse inpainting relies on several novel modifications. 
With regard to the interpolation procedure, we presented a way to combine the 
strength of first and zero order methods into a mixed order method. Moreover, 
we presented a better approach to choose method parameters based on Voronoi 
tesselations.

The main ingredient to reveal the potential of SPH is the use of optimally 
chosen data. We proposed a new densification process based on Voronoi 
tesselations, which lead naturally to a strategy based on a regional error 
instead of a completely local pointwise error. Thus, a larger amount of data is 
considered in the optimization procedure, yielding better suitable inpainting 
masks. Furthermore, we introduced a so far unexplored formulation that allows 
to use SPH for least-square approximation, i.e., the optimization of data 
not only in the spatial, but also in the tonal domain. 
What remains an ongoing topic of research is the question how 
to choose anisotropies. Taking into account the superior performance of EED, it
seems natural to determine anisotropies based on gradient data or the structure 
tensor. While this is straightforward for regular masks, it becomes more of 
a challenge for randomly distributed or optimized mask points. In the context
of data optimization, one could think about computing anisotropies from the 
original image. However, for compression purposes, storing this additional data 
would reduce the rate of compression or necessitate to consider sparser mask,
such that there is overall less data to store. On the other hand, when it comes
to non-sparse inpainting tasks as briefly touched on in \cref{sec:scratch},
anisotropies could be determined from the known parts of the image, e.g. by 
considering gradients similar to \cite{TF07}. An alternative used for object 
removal in \cite{HH20} is to detect edges and their orientation to determine
anisotropies from these structures. Both strategies look promising to us when 
it comes to improving the performance of SPH regarding classical inpainting
problems in future research.

We hope that our work will help to give SPH-based inpainting the attention
that it deserves. Moreover, we believe that some of our novel concepts, 
e.g.~the Voronoi-based densification for data optimization, will also 
be useful in applications beyond SPH-based inpainting.


\section*{Acknowledgments}
While working on this article, Matthias Augustin and Joachim Wei\-ckert
have received funding from the European Research Council (ERC)
under the European Union's Horizon 2020 research and innovation programme
(grant agreement no. 741215, ERC Advanced Grant INCOVID).

We thank Vassillen Chizhov for his support regarding programs 
with spatial and tonal optimization for harmonic and biharmonic inpainting.


\bibliographystyle{siamplain}
\bibliography{daropoulos_sph_references}

\begin{thebibliography}{10}

\bibitem{AA17}
{\sc R.~Achanta, N.~Arvanitopoulos, and S.~S{\"{u}}sstrunk}, {\em Extreme image
  completion}, in Proc.~42nd IEEE International Conference on Acoustics, Speech
  and Signal Processing, New Orleans, LA, Mar. 2017, IEEE, pp.~1333--1337.

\bibitem{AP17}
{\sc R.~D. Adam, P.~Peter, and J.~Weickert}, {\em Denoising by inpainting}, in
  Scale Space and Variational Methods in Computer Vision, F.~Lauze, Y.~Dong,
  and A.~B. Dahl, eds., vol.~10302 of Lecture Notes in Computer Science,
  Springer, Cham, Switzerland, 2017, pp.~121--132.

\bibitem{AP16}
{\sc F.~{Alves Mazzini} and F.~{Petronetto do Carmo}}, {\em Digital inpainting
  with {SPH} method}, in Workshop Works in Progress in {SIBGRAPI} 2016 --
  Conference on Graphics, Patterns and Images, J.~P. Gois and F.~Ricardo, eds.,
  S{\~{a}}o Jos{\'{e}}, Brazil, Oct. 2016.

\bibitem{AF11}
{\sc P.~Arias, G.~Facciolo, V.~Caselles, and G.~Sapiro}, {\em A variational
  framework for exemplar-based image inpainting}, International Journal of
  Computer Vision, 93 (2011), pp.~319--347.

\bibitem{AL10}
{\sc J.-F. Aujol, S.~Ladjal, and S.~Masnou}, {\em Exemplar-based inpainting
  from a variational point of view}, SIAM Journal on Mathematical Analysis, 42
  (2010), pp.~1246--1285.

\bibitem{BF08}
{\sc W.~Baatz, M.~Fornasier, P.~A. Markowich, and C.-B. Sch{\"{o}}nlieb}, {\em
  Inpainting of ancient austrian frescoes}, in Bridges Leeuwarden: Mathematical
  Connections in Art, Music, and Science, R.~Sarhangi and C.~H. Sequin, eds.,
  Tarquin Publications, London, July 2008, pp.~163--170.

\bibitem{BB09}
{\sc Z.~Belhachmi, D.~Bucur, B.~Burgeth, and J.~Weickert}, {\em {H}ow to choose
  interpolation data in images}, SIAM Journal on Applied Mathematics, 70
  (2009), pp.~333--352.

\bibitem{BC14}
{\sc M.~Bertalm{\'{\i}}o, V.~Caselles, S.~Masnou, and G.~Sapiro}, {\em
  Inpainting}, in Computer Vision: {A} Reference Guide, K.~Ikeuchi, ed.,
  Springer, Boston, MA, 2014, pp.~401--416.

\bibitem{BS00}
{\sc M.~Bertalm{\'{\i}}o, G.~Sapiro, V.~Caselles, and C.~Ballester}, {\em Image
  inpainting}, in Proc.~27th Annual Conference on Computer Graphics and
  Interactive Techniques (SIGGRAPH), New Orleans, LA, July 2000, ACM
  Press/Addison-Wesley, pp.~417--424.

\bibitem{BV03}
{\sc M.~Bertalm{\'{\i}}o, L.~Vese, G.~Sapiro, and S.~Osher}, {\em Simultaneous
  structure and texture image inpainting}, IEEE Transactions on Image
  Processing, 12 (2003), pp.~882--889.

\bibitem{BEG07}
{\sc A.~L. Bertozzi, S.~Esedoglu, and A.~Gillette}, {\em Inpainting of binary
  images using the {C}ahn–-{H}illiard equation}, IEEE Transactions on Image
  Processing, 16 (2007), pp.~285--291.

\bibitem{BA19}
{\sc P.~Biasutti, J.-F. Aujol, M.~Br{\'{e}}dif, and A.~Bugeau}, {\em Diffusion
  and inpainting of reflectance and height {LiDAR} orthoimages}, Computer
  Vision and Image Understanding, 179 (2019), pp.~31--40.

\bibitem{BLPP17}
{\sc S.~Bonettini, I.~Loris, F.~Porta, M.~Prato, and S.~Rebegoldi}, {\em On the
  convergence of a linesearch based proximal-gradient method for nonconvex
  optimization}, Inverse Problems, 33 (2017).
\newblock Article 055005.

\bibitem{BM07}
{\sc F.~Bornemann and T.~M{\"{a}}rz}, {\em Fast image inpainting based on
  coherence transport}, Journal of Mathematical Imaging and Vision, 28 (2007),
  pp.~259--278.

\bibitem{BU13}
{\sc A.~Bourquard and M.~Unser}, {\em Anisotropic interpolation of sparse
  generalized image samples}, IEEE Transactions on Image Processing, 22 (2013),
  pp.~459--472.

\bibitem{BC10}
{\sc C.~{B}rito {L}oeza and K.~Chen}, {\em Fast numerical algorithms for
  {E}uler’s elastica inpainting model}, International Journal of Modern
  Mathematics, 5 (2010), pp.~157--182.

\bibitem{BS08}
{\sc H.~H. Bui, K.~Sako, R.~Fukagawa, and J.~C. Wells}, {\em {SPH}-based
  numerical simulations for large deformation of geomaterial considering
  soil-structure interaction}, in Proc.~12th International Conference of
  International Association for Computer Methods and Advances in Geomechanics,
  vol.~1, Goa, India, Oct. 2008, pp.~570--578.

\bibitem{BHS09}
{\sc M.~Burger, L.~He, and C.-B. Sch{\"{o}}nlieb}, {\em {C}ahn--{H}illiard
  inpainting and a generalization for grayvalue images}, SIAM Journal on
  Imaging Sciences, 2 (2009), pp.~1129--1167.

\bibitem{CA18}
{\sc L.~Calatroni, M.~d'Autume, R.~Hocking, S.~Panayotova, S.~Parisotto,
  P.~Ricciardi, and C.-B. Sch{\"{o}}nlieb}, {\em Unveiling the invisible:
  {M}athematical models for restoring and interpreting illuminated
  manuscripts}, Heritage Science, 6 (2018), pp.~56:1--21.

\bibitem{CM98}
{\sc V.~Caselles, J.-M. Morel, and C.~Sbert}, {\em An axiomatic approach to
  image interpolation}, IEEE Transactions on Image Processing, 7 (1998),
  pp.~376--386.

\bibitem{CP19}
{\sc A.~Chambolle and T.~Pock}, {\em Total roto-translational variation},
  Numerische Mathematik, 142 (2019), pp.~611--–666.

\bibitem{CK02}
{\sc T.~F. Chan, S.~H. Kang, and J.~Shen}, {\em {E}uler's elastica and
  curvature-based inpainting}, SIAM Journal on Applied Mathematics, 63 (2002),
  pp.~564--592.

\bibitem{CS02}
{\sc T.~F. Chan and J.~Shen}, {\em Mathematical models for local non-texture
  inpaintings}, SIAM Journal on Applied Mathematics, 62 (2002), pp.~1019--1043.

\bibitem{CB99}
{\sc J.~K. Chen, J.~E. Beraun, and T.~C. Carney}, {\em A corrective smoothed
  particle method for boundary value problems in heat conduction},
  International Journal for Numerical Methods in Engineering, 46 (1999),
  pp.~231--252.

\bibitem{CR14}
{\sc Y.~Chen, R.~Ranftl, and T.~Pock}, {\em A bi-level view of inpainting-based
  image compression}, in Proc.~19th Computer Vision Winter Workshop,
  Z.~K{\'{u}}kelov{\'{a}} and J.~Heller, eds., K{\v{r}}tiny, Czech Republic,
  Feb. 2014, pp.~19--26.

\bibitem{CL18}
{\sc C.~Cheng, Y.~Li, N.~Zhao, B.~Guo, and N.~Mou}, {\em Least squares
  compactly supported radial basis function for digital terrain model
  interpolation from airborne {L}idar point clouds}, Remote Sensing, 10 (2019),
  pp.~587:1--24.

\bibitem{CP04}
{\sc A.~Criminisi, P.~P{\'e}rez, and K.~Toyama}, {\em Region filling and object
  removal by exemplar-based image inpainting}, IEEE Transactions on Image
  Processing, 13 (2004), pp.~1200--1212.

\bibitem{DD20}
{\sc F.~Dell'Accio, F.~Di~Tommaso, and D.~Gonnelli}, {\em Comparison of
  {S}hepard's like methods with different basis functions}, in Numerical
  Computations: Theory and Algorithms, Y.~D. Sergeyev and D.~E. Kvasov, eds.,
  vol.~11973 of Lecture Notes in Computer Science, Springer, Cham, Switzerland,
  2020, pp.~47--55.

\bibitem{DF11}
{\sc G.~Di~Blasi, E.~Francomano, A.~Tortorici, and E.~Toscano}, {\em A smoothed
  particle image reconstruction method}, Calcolo, 48 (2011), pp.~61--74.

\bibitem{EL99}
{\sc A.~A. Efros and T.~K. Leung}, {\em Texture synthesis by non-parametric
  sampling}, in Proceedings of the Seventh IEEE International Conference on
  Computer Vision, vol.~2, Kerkyra, Greece, Sept. 1999, IEEE, pp.~1033--1038.

\bibitem{El10}
{\sc M.~Elad}, ed., {\em Sparse and Redundant Representations: From Theory to
  Applications in Signal and Image Processing}, Springer, New York, NY, 2010.

\bibitem{ES05}
{\sc M.~Elad, J.-L. Starck, P.~Querre, and D.~L. Donoho}, {\em Simultaneous
  cartoon and texture image inpainting using morphological component analysis
  ({MCA})}, Applied and Computational Harmonic Analysis, 19 (2005),
  pp.~340--358.

\bibitem{ES02}
{\sc S.~Esedoglu and J.~Shen}, {\em Digital inpainting based on the
  {M}umford-{S}hah-{E}uler image model}, European Journal of Applied
  Mathematics, 13 (2002), pp.~353--370.

\bibitem{FA09}
{\sc G.~Facciolo, P.~Arias, V.~Caselles, and G.~Sapiro}, {\em Exemplar-based
  interpolation of sparsely sampled images}, in Energy Minimization Methods in
  Computer Vision and Pattern Recognition, D.~Cremers, Y.~Boykov, A.~Blake, and
  F.~R. Schmidt, eds., vol.~5681 of Lecture Notes in Computer Science,
  Springer, Berlin, Germany, Aug. 2009, pp.~331--344.

\bibitem{Fa07}
{\sc G.~E. Fasshauer}, {\em Meshfree Approximation Methods with {MATLAB}},
  vol.~6 of Interdisciplinary Mathematical Sciences, World Scientific, River
  Edge, NJ, 2007.

\bibitem{Fe94}
{\sc H.~G. Feichtinger and T.~Strohmer}, {\em Recovery of missing segments and
  lines in images}, Optical Engineering, 33 (1994), pp.~3283--3289.

\bibitem{FH12}
{\sc P.~F. Felzenszwalb and D.~P. Huttenlocher}, {\em Distance transforms of
  sampled functions}, Theory of Computing, 8 (2012), pp.~415--428.

\bibitem{GW08}
{\sc I.~Gali{\'{c}}, J.~Weickert, M.~Welk, A.~Bruhn, A.~Belyaev, and H.-P.
  Seidel}, {\em Image compression with anisotropic diffusion}, Journal of
  Mathematical Imaging and Vision, 31 (2008), pp.~255--269.

\bibitem{GX16}
{\sc M.~A. Ghaffari and S.~Xiao}, {\em Smoothed particle hydrodynamics with
  stress points and centroid {V}oronoi tessellation ({CVT}) topology
  optimization}, International Journal of Computational Methods, 13 (2016),
  pp.~1650031:1--23.

\bibitem{GL14}
{\sc C.~Guillemot and O.~Le~Meur}, {\em Image inpainting: {O}verview and recent
  advances}, IEEE Signal Processing Magazine, 31 (2014), pp.~127--144.

\bibitem{HH20}
{\sc L.~R. Hocking, T.~Holding, and C.-B. Sch{\"{o}}nlieb}, {\em Analysis of
  artifacts in shell-based image inpainting: {W}hy they occur and how to
  eliminate them}, Foundations of Computational Mathematics, 20 (2020),
  pp.~1549--1651.

\bibitem{HM17}
{\sc L.~Hoeltgen, M.~Mainberger, S.~Hoffmann, J.~Weickert, C.~H. Tang,
  S.~Setzer, D.~Johannsen, F.~Neumann, and B.~Doerr}, {\em Optimising spatial
  and tonal data for {PDE}-based inpainting}, in Variational Methods in Imaging
  and Geometric Control, M.~Bergounioux, G.~Peyr{\'{e}}, C.~Schn{\"{o}}rr,
  J.-B. Caillau, and T.~Haberkorn, eds., vol.~18 of Radon Series on
  Computational and Applied Mathematics, De Gruyter, Berlin, 2017, pp.~35--83.

\bibitem{HS13}
{\sc L.~Hoeltgen, S.~Setzer, and J.~Weickert}, {\em An optimal control approach
  to find sparse data for {L}aplace interpolation}, in Energy Minimization
  Methods in Computer Vision and Pattern Recognition, A.~Heyden, F.~Kahl,
  C.~Olsson, M.~Oskarsson, and X.-C. Tai, eds., vol.~8081 of Lecture Notes in
  Computer Science, Springer, Berlin, Germany, Aug. 2013, pp.~151--164.

\bibitem{HM13}
{\sc S.~Hoffmann, M.~Mainberger, J.~Weickert, and M.~Puhl}, {\em Compression of
  depth maps with segment-based homogeneous diffusion}, in Scale Space and
  Variational Methods in Computer Vision, A.~Kuijper, K.~Bredies, T.~Pock, and
  H.~Bischof, eds., vol.~7893 of Lecture Notes in Computer Science, Springer,
  Berlin, Germany, June 2013, pp.~319--330.

\bibitem{Hu73}
{\sc B.~R. Hunt}, {\em The application of constrained least squares estimation
  to image restoration by digital computer}, IEEE Transactions on Computers,
  C-22 (1973), pp.~2856--2869.

\bibitem{ISI17}
{\sc S.~Iizuka, E.~Simo-Serra, and H.~Ishikawa}, {\em Globally and locally
  consistent image completion}, {ACM} Transactions on Graphics, 36 (2017).
\newblock Article No.~107.

\bibitem{KM13}
{\sc N.~Karianakis and P.~Maragos}, {\em An integrated system for digital
  restoration of prehistoric {T}heran wall paintings}, in Proc.~18th
  International Conference on Digital Signal Processing, Fira, Greece, July
  2013, IEEE, pp.~1--6.

\bibitem{KB18}
{\sc L.~Karos, P.~Bheed, P.~Peter, and J.~Weickert}, {\em Optimising data for
  exemplar-based inpainting}, in Advanced Concepts for Intelligent Vision
  Systems, J.~Blanc-Talon, D.~Helbert, W.~Philips, D.~Popescu, and
  P.~Scheunders, eds., vol.~11182 of Lecture Notes in Computer Science,
  Springer, Cham, Switzerland, Sept. 2018, pp.~547--558.

\bibitem{KW93}
{\sc H.~Knutsson and C.~Westin}, {\em Normalized and differential convolution},
  in Proc.~1993 IEEE Computer Society Conference on Computer Vision and Pattern
  Recognition, New York City, NY, June 1993, IEEE Computer Society Press,
  pp.~515--523.

\bibitem{LZ12}
{\sc B.~Lipu{\v{s}} and B.~{\v{Z}}alik}, {\em Efficient reconstruction of
  images with deliberately corrupted pixels}, Informatica, 23 (2012),
  pp.~47--63.

\bibitem{Li09}
{\sc G.-R. Liu}, {\em Meshfree Methods: Moving Beyond the Finite Element
  Method}, CRC Press, Boca Raton, FL, 2009.

\bibitem{LL03}
{\sc G.-R. Liu and M.~B. Liu}, {\em Smoothed Particle Hydrodynamics: A Meshfree
  Particle Method}, World Scientific, Singapore, 2003.

\bibitem{LJ95}
{\sc W.~K. Liu, S.~Jun, and Y.~F. Zhang}, {\em Reproducing kernel particle
  methods}, International Journal for Numerical Methods in Fluids, 20 (1995),
  pp.~1081--1106.

\bibitem{Lu77}
{\sc L.~B. Lucy}, {\em A numerical approach to the testing of the fission
  hypothesis}, The Astronomical Journal, 82 (1977), pp.~1013--1024.

\bibitem{MD05}
{\sc F.~Magoul{\`{e}}s, L.~A. Diago, and I.~Hagiwara}, {\em A two-level
  iterative method for image reconstruction with radial basis functions}, JSME
  International Journal Series C, Mechanical Systems, Machine Elements and
  Manufacturing, 48 (2005), pp.~149--158.

\bibitem{MB11}
{\sc M.~Mainberger, A.~Bruhn, J.~Weickert, and S.~Forchhammer}, {\em Edge-based
  image compression of cartoon-like images with homogeneous diffusion}, Pattern
  Recognition, 44 (2011), pp.~1859--1873.

\bibitem{MH11}
{\sc M.~Mainberger, S.~Hoffmann, J.~Weickert, C.~H. Tang, D.~Johannsen,
  F.~Neumann, and B.~Doerr}, {\em Optimising spatial and tonal data for
  homogeneous diffusion inpainting}, in Scale Space and Variational Methods in
  Computer Vision, A.~M. Bruckstein, B.~M. ter Haar~Romeny, A.~M. Bronstein,
  and M.~M. Bronstein, eds., vol.~6667 of Lecture Notes in Computer Science,
  Springer, Berlin, Germany, June 2011, pp.~26--37.

\bibitem{ME08}
{\sc J.~Mairal, M.~Elad, and G.~Sapiro}, {\em Sparse representation for color
  image restoration}, IEEE Transactions on Image Processing, 17 (2008),
  pp.~53--69.

\bibitem{MM98}
{\sc S.~Masnou and J.-M. Morel}, {\em Level lines based disocclusion}, in
  Proc.~IEEE International Conference on Image Processing, vol.~3, Chicago, IL,
  Oct. 1998, pp.~259--263.

\bibitem{MH05}
{\sc S.~McDougall and O.~Hungr}, {\em Dynamic modelling of entrainment in rapid
  landslides}, Canadian Geotechnical Journal, 42 (2005), pp.~1437--1448.

\bibitem{Mo94}
{\sc J.~J. Monaghan}, {\em Simulating free surface flows with {SPH}}, Journal
  of Computational Physics, 110 (1994), pp.~399--406.

\bibitem{Mo96}
{\sc J.~P. Morris}, {\em Analysis of {S}moothed {P}article {H}ydrodynamics with
  {A}pplications}, PhD thesis, Department of Mathematics, Monash University,
  Melbourne, Australia, 1996.

\bibitem{NA14}
{\sc A.~Newson, A.~Almansa, M.~Fradet, Y.~Gousseau, and P.~P{\'{e}}rez}, {\em
  Video inpainting of complex scenes}, SIAM Journal on Imaging Sciences, 7
  (2014), pp.~1993--2019.

\bibitem{NMS93}
{\sc M.~Nitzberg, D.~Mumford, and T.~Shiota}, {\em Filtering, Segmentation and
  Depth}, vol.~662 of Lecture Notes in Computer Science, Springer, Berlin,
  1993.

\bibitem{OABB85}
{\sc J.~Ogden, E.~Adelson, J.~Bergen, and P.~Burt}, {\em Pyramid-based computer
  graphics}, RCA Engineer, 30 (1985), pp.~4--15.

\bibitem{PKDD16}
{\sc D.~Pathak, P.~Kr{\"{a}}henb{\"{u}}hl, J.~Donahue, T.~Darrell, and A.~A.
  Efros}, {\em Context encorder: Feature learning by inpainting}, in Proc.~2016
  IEEE Computer Society Conference on Computer Vision and Pattern Recognition,
  Las Vegas, NV, June 2016, IEEE Computer Society Press, pp.~2536--2544.

\bibitem{PM92}
{\sc W.~B. Pennebaker and J.~L. Mitchell}, {\em {JPEG}: Still Image Data
  Compression Standard}, Springer, New York, NY, 1992.

\bibitem{Pe19}
{\sc P.~Peter}, {\em Fast inpainting-based compression: {C}ombining {S}hepard
  interpolation with joint inpainting and prediction}, in Proc.~26th IEEE
  International Conference on Image Processing, Taipei, Taiwan, Sept. 2019,
  IEEE, pp.~3557--3561.

\bibitem{PH16}
{\sc P.~Peter, S.~Hoffmann, F.~Nedwed, L.~Hoeltgen, and J.~Weickert}, {\em
  Evaluating the true potential of diffusion-based inpainting in a compression
  context}, Signal Processing: Image Communication, 46 (2016), pp.~40--53.

\bibitem{PW15}
{\sc P.~Peter and J.~Weickert}, {\em Compressing images with diffusion- and
  exemplar-based inpainting}, in Scale Space and Variational Methods in
  Computer Vision, J.-F. Aujol, M.~Nikolova, and N.~Papadakis, eds., vol.~9087
  of Lecture Notes in Computer Science, Springer, Cham, Switzerland, June 2015,
  pp.~154--165.

\bibitem{RO20}
{\sc L.~Raad, M.~Oliver, C.~Ballester, G.~Haro, and E.~Meinhardt}, {\em On
  anisotropic optical flow inpainting algorithms}, Image Processing On Line, 10
  (2020), pp.~78--104.

\bibitem{RC11}
{\sc T.~Ru{\v{z}}i{\'{c}}, B.~Cornelis, L.~Plati{\v{s}}a, A.~Pi{\v{z}}urica,
  A.~Dooms, W.~Philips, M.~Martens, M.~De~Mey, and I.~Daubechies}, {\em Virtual
  restoration of the {G}hent altarpiece using crack detection and inpainting},
  in Advanced Concepts for Intelligent Vision Systems, J.~Blanc-Talon,
  R.~Kleihorst, W.~Philips, D.~Popescu, and P.~Scheunders, eds., vol.~6915 of
  Lecture Notes in Computer Science, Springer, Berlin, Aug. 2011, pp.~417--428.

\bibitem{Sa03}
{\sc Y.~Saad}, {\em Iterative Methods for Sparse Linear Systems}, Society for
  Industrial and Applied Mathematics, Philadelphia, PA, 2003.

\bibitem{SP14}
{\sc C.~Schmaltz, P.~Peter, M.~Mainberger, F.~Ebel, J.~Weickert, and A.~Bruhn},
  {\em Understanding, optimising, and extending data compression with
  anisotropic diffusion}, International Journal of Computer Vision, 108 (2014),
  pp.~222--240.

\bibitem{Sc15}
{\sc C.-B. Sch\"onlieb}, {\em Partial Differential Equation Methods for Image
  Inpainting}, Cambridge University Press, New York, 2015.

\bibitem{SB00}
{\sc S.~Schussman, M.~Bertram, B.~Hamann, and K.~I. Joy}, {\em Hierarchical
  data representations based on planar {V}oronoi diagrams}, in Data
  Visualization 2000, W.~C. de~Leeuw and R.~van Liere, eds., vol.~31 of
  Eurographics, Springer, Vienna, Austria, 2000, pp.~63--72.

\bibitem{SM96}
{\sc P.~R. Shapiro, H.~Martel, J.~V. Villumsen, and J.~M. Owen}, {\em Adaptive
  smoothed particle hydrodynamics, with application to cosmology:
  {M}ethodology}, The Astrophysical Journal Supplement Series, 103 (1996),
  pp.~269--330.

\bibitem{Sh68}
{\sc D.~S. Shepard}, {\em A two-dimensional interpolation function for
  irregularly-spaced data}, in Proc.~23rd ACM National Conference, New York,
  NY, Jan. 1968, Association for Computing Machinery, pp.~517--524.

\bibitem{SA17}
{\sc G.~Shobeyri and R.~R. Ardakani}, {\em Improving accuracy of {SPH} method
  using {V}oronoi diagram}, Iranian Journal of Science and Technology,
  Transactions of Civil Engineering, 41 (2017), pp.~345--350.

\bibitem{SE05}
{\sc J.-L. Starck, M.~Elad, and D.~L. Donoho}, {\em Image decomposition via the
  combination of sparse representations and a variational approach}, IEEE
  Transactions on Image Processing, 14 (2005), pp.~1570--1582.

\bibitem{TF07}
{\sc H.~Takeda, S.~Farsiu, and P.~Milanfar}, {\em Kernel regression for image
  processing and reconstruction}, IEEE Transactions on Image Processing, 16
  (2007), pp.~349--366.

\bibitem{TM02}
{\sc D.~S. Taubman and M.~W. Marcellin}, {\em {JPEG 2000}: Image Compression
  Fundamentals, Standards and Practice}, vol.~642 of The Springer International
  Series in Engineering and Computer Science, Springer, New York, NY, 2002.
\newblock originally published by Kluwer Academic Publishers, 2002.

\bibitem{TB19}
{\sc R.~Tovey, M.~Benning, C.~Brune, M.~J. Jagerwerf, S.~M. Collins, R.~K.
  Leary, P.~A. Midgley, and C.-B. Sch{\"{o}}nlieb}, {\em Directional sinogram
  impainting for limited angle tomography}, Inverse Problems, 35 (2019),
  pp.~024004:1--29.

\bibitem{US06}
{\sc K.~Uhlir and V.~Skala}, {\em Radial basis function use for the restoration
  of damaged images}, in Computer Vision and Graphics, K.~Wojciechowski,
  B.~Smolka, H.~Palus, R.~S. Kozera, W.~Skarbek, and L.~Noakes, eds., vol.~32
  of Computational Imaging and Vision, Springer, Dordrecht, Netherlands, 2006,
  pp.~839--844.

\bibitem{UVL18}
{\sc D.~Ulyanov, A.~Vedaldi, and V.~Lempitsky}, {\em Deep image prior}, in
  Proc.~2018 IEEE Computer Society Conference on Computer Vision and Pattern
  Recognition, Salt Lake City, UT, June 2018, IEEE Computer Society Press,
  pp.~9446--9454.

\bibitem{Bo92}
{\sc R.~van~den Boomgaard}, {\em The morphological equivalent of the {G}auss
  convolution}, Nieuw Archief voor Wiskunde, 10 (1992), pp.~219--236.

\bibitem{We98}
{\sc J.~Weickert}, {\em Anisotropic Diffusion in Image Processing}, Teubner,
  Stuttgart, 1998.

\bibitem{WW06}
{\sc J.~Weickert and M.~Welk}, {\em Tensor field interpolation with {PDEs}}, in
  Visualization and Processing of Tensor Fields, J.~Weickert and H.~Hagen,
  eds., Springer, Berlin, 2006, pp.~315--325.

\bibitem{WW13}
{\sc J.~Weickert, M.~Welk, and M.~Wickert}, {\em {$L^{2}$}-stable nonstandard
  finite differences for anisotropic diffusion}, in Scale Space and Variational
  Methods in Computer Vision, A.~Kuijper, K.~Bredies, T.~Pock, and H.~Bischof,
  eds., vol.~7893 of Lecture Notes in Computer Science, Springer, Berlin, 2013,
  pp.~380--391.

\bibitem{We05}
{\sc H.~Wendland}, {\em Scattered Data Approximation}, Cambridge Monographs on
  Applied and Computational Mathematics, Cambridge University Press, Cambridge,
  UK, 2005.

\bibitem{YT13}
{\sc J.~Yu and G.~Turk}, {\em Reconstructing surfaces of particle-based fluids
  using anisotropic kernels}, ACM Transactions on Graphics, 32 (2013),
  pp.~5:1--12.

\bibitem{ZB09}
{\sc G.~M. Zhang and R.~C. Batra}, {\em Symmetric smoothed particle
  hydrodynamics ({SSPH}) method and its application to elastic problems},
  Computational Mechanics, 43 (2009), pp.~321--340.

\end{thebibliography}

\end{document}


\maketitle

\begin{figure}[htb]
 \captionsetup[subfigure]{justification=centering}
 \centering
 \begin{subfigure}[t]{0.29\textwidth}
  \centering
  \includegraphics[width=\textwidth]{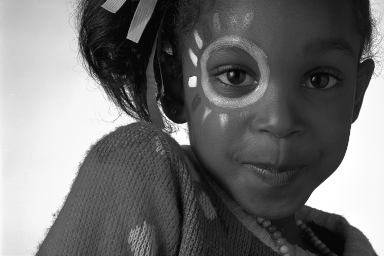}
  \caption{Girl}
 \end{subfigure}
 \quad
 \begin{subfigure}[t]{0.29\textwidth}
  \centering
  \includegraphics[width=\textwidth]{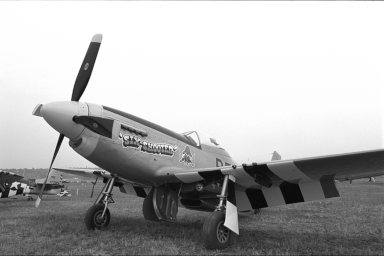}
  \caption{Plane}
 \end{subfigure}
 \quad
 \begin{subfigure}[t]{0.29\textwidth}
  \centering
  \includegraphics[width=\textwidth]{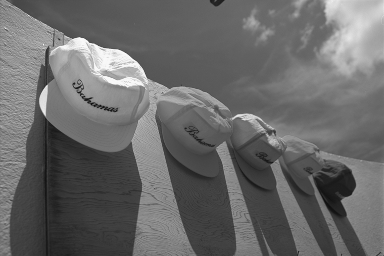}
  \caption{Hats}
 \end{subfigure}
 \caption[Further test images]{The $384 \times 256$ test 
  images ``girl'' (\textbf{a}), ``plane'' (\textbf{b}), and ``hats'' 
  (\textbf{c}).}
 \label{fig:Kodak}
\end{figure}

The purpose of this supplement is to present some further 
results for the various inpainting tasks included in the main article. Besides
the images already included there, we also consider some more examples from the
Kodak database, namely ``girl'', ``plane'', and ``hats'' rescaled to size 
$384 \times 256$; see \cref{fig:Kodak}. The presentation here follows the same
order as in the main article.

\section{Inpainting on Regular and Random Masks}

\begin{figure}[htb]
 \captionsetup[subfigure]{justification=centering}
 \centering
 \begin{subfigure}[t]{0.29\textwidth}
  \centering
  \includegraphics[width=\textwidth]{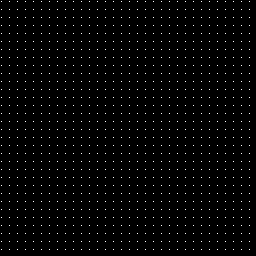}
  \caption{1.5625 \% mask}
 \end{subfigure}
 \quad 
 \begin{subfigure}[t]{0.29\textwidth}
  \centering
  \includegraphics[width=\textwidth]{Figures_art/mask_reg_256_256_D00625}
  \caption{6.25 \% mask}
 \end{subfigure}
 \quad 
 \begin{subfigure}[t]{0.29\textwidth}
  \centering
  \includegraphics[width=\textwidth]{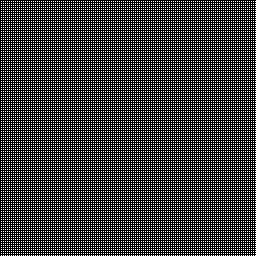}
  \caption{25 \% mask}
 \end{subfigure}

 \begin{subfigure}[t]{0.29\textwidth}
  \centering
  \includegraphics[width=\textwidth]{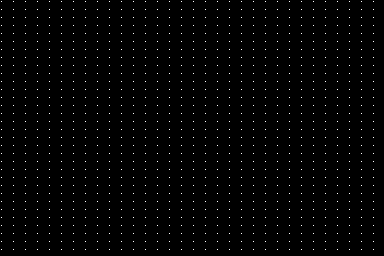}
  \caption{1.04 \% mask}
 \end{subfigure}
 \quad 
 \begin{subfigure}[t]{0.29\textwidth}
  \centering
  \includegraphics[width=\textwidth]{Figures_art/mask_reg_384_256_D00416}
  \caption{4.16 \% mask}
 \end{subfigure}
 \quad 
 \begin{subfigure}[t]{0.29\textwidth}
  \centering
  \includegraphics[width=\textwidth]{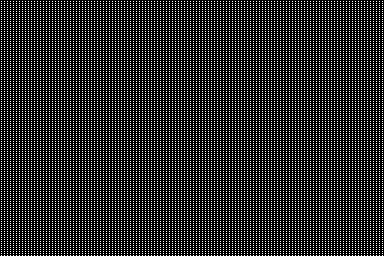}
  \caption{16.66 \% mask}
 \end{subfigure}
 \caption[Regular masks]{Regular masks of different densities 
  for images of size $256 \times 256$ (top row) and size $384 \times 256$ 
  (bottom row).}
 \label{fig:reg_masks}
\end{figure}

In our first batch of examples, we compare the performance of 
SPH inpainting to that of diffusion- and exemplar-based inpainting for a couple 
of regular masks. For the test images of size $256 \times 256$, we consider the
6.25 \% mask (grid size 4 pixels) from the main article, but also a mask of 
density 1.5625 \% (grid size of 8 pixels) and of density 25 \% (grid size of 
2 pixels). 
For the images from the Kodak database, the same grid sizes results in 
densities of 1.04 \%, 4.16 \%, and 16.66 \%, respectively. For convenience, we 
include all six masks in \cref{fig:reg_masks} as well as some of the results
already presented for ``trui'' and ``parrots''.

\begin{figure}[htb]
 \captionsetup[subfigure]{justification=centering}
 \centering
 \begin{subfigure}[t]{0.29\textwidth}
  \centering
  \includegraphics[width=\textwidth]{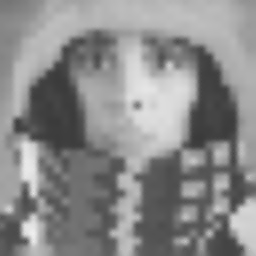}
  \caption{$\textrm{MSE} = 282.58$}
 \end{subfigure}
 \quad 
 \begin{subfigure}[t]{0.29\textwidth}
  \centering
  \includegraphics[width=\textwidth]{Figures_art/trui_reg_D00625_gauss_zero}
  \caption{$\textrm{MSE} = 83.28$}
 \end{subfigure} 
 \quad 
 \begin{subfigure}[t]{0.29\textwidth}
  \centering
  \includegraphics[width=\textwidth]{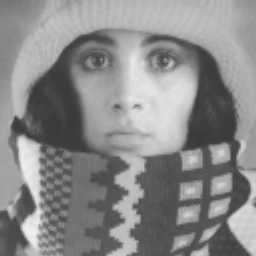}
  \caption{$\textrm{MSE} = 18.61$}
 \end{subfigure}

 \begin{subfigure}[t]{0.29\textwidth}
  \centering
  \includegraphics[width=\textwidth]{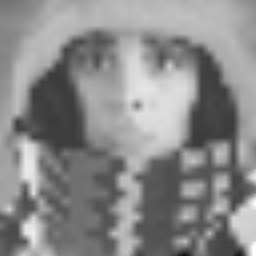}
  \caption{$\textrm{MSE} = 288.33$}
 \end{subfigure}
 \quad 
 \begin{subfigure}[t]{0.29\textwidth}
  \centering
  \includegraphics[width=\textwidth]{Figures_art/trui_reg_D00625_gauss_first}
  \caption{$\textrm{MSE} = 85.01$}
 \end{subfigure}
 \quad 
 \begin{subfigure}[t]{0.29\textwidth}
  \centering
  \includegraphics[width=\textwidth]{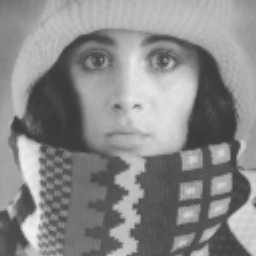}
  \caption{$\textrm{MSE} = 18.47$}
 \end{subfigure}
 \caption[SPH inpainting of ``trui'' with regular masks]{
  Inpainting of 
  ``trui'' for regular masks with isotropic Gaussian kernels. 
  Densities from left to right are 1.5625 \%, 6.25 \%, and 25 \%.
  Top row: Zero order consistency SPH inpainting. 
  Bottom row: First order consistency SPH inpainting.}
\end{figure}

\begin{figure}[htb]
 \captionsetup[subfigure]{justification=centering}
 \centering
 \begin{subfigure}[t]{0.29\textwidth}
  \centering
  \includegraphics[width=\textwidth]{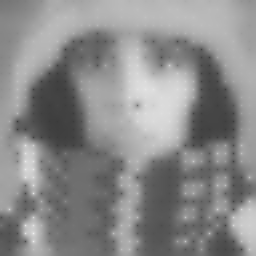}
  \caption{$\textrm{MSE} = 400.14$}
 \end{subfigure}
 \quad 
 \begin{subfigure}[t]{0.29\textwidth}
  \centering
  \includegraphics[width=\textwidth]{Figures_art/trui_reg_D00625_harm}
  \caption{$\textrm{MSE} = 121.96$}
 \end{subfigure}
 \quad 
 \begin{subfigure}[t]{0.29\textwidth}
  \centering
  \includegraphics[width=\textwidth]{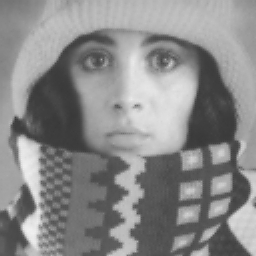}
  \caption{$\textrm{MSE} = 23.29$}
 \end{subfigure}

 \begin{subfigure}[t]{0.29\textwidth}
  \centering
  \includegraphics[width=\textwidth]{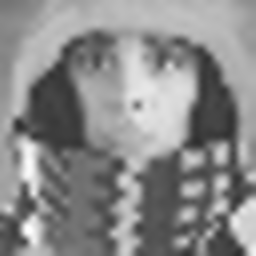}
  \caption{$\textrm{MSE} = 270.70$}
 \end{subfigure}
 \quad
 \begin{subfigure}[t]{0.29\textwidth}
  \centering
  \includegraphics[width=\textwidth]{Figures_art/trui_reg_D00625_biharm}
  \caption{$\textrm{MSE} = 67.95$}
 \end{subfigure}
 \quad
 \begin{subfigure}[t]{0.29\textwidth}
  \centering
  \includegraphics[width=\textwidth]{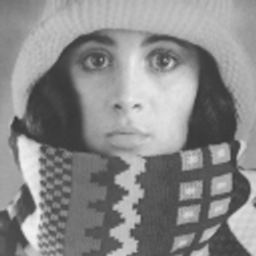}
  \caption{$\textrm{MSE} = 11.10$}
 \end{subfigure}

 \begin{subfigure}[t]{0.29\textwidth}
  \centering
  \includegraphics[
   width=\textwidth]{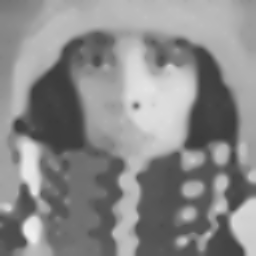}
  \caption{$\textrm{MSE} = 265.70$}
 \end{subfigure}
 \quad 
 \begin{subfigure}[t]{0.29\textwidth}
  \centering
  \includegraphics[
   width=\textwidth]{Figures_art/trui_reg_D00625_eed_lambda_0_2_sigma_0_8}
  \caption{$\textrm{MSE} = 60.68$}
 \end{subfigure}
 \quad 
 \begin{subfigure}[t]{0.29\textwidth}
  \centering
  \includegraphics[
   width=\textwidth]{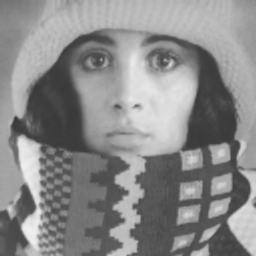}
  \caption{$\textrm{MSE} = 10.56$}
 \end{subfigure}
 \caption[Diffusion inpainting of ``trui'' with regular masks]{
  Inpainting of 
  ``trui'' for regular masks. Densities from left to right are 
  1.5625 \%, 6.25 \%, and 25 \%.
  Top row: Harmonic inpainting. 
  Middle row: Biharmonic inpainting. 
  Bottom row: Inpainting with EED. 
  Parameters are from left to right 
  $\lambda=0.3$ and $\sigma=1.3$, $\lambda=0.2$ and $\sigma=0.8$, and
  $\lambda=0.3$ and $\sigma=0.6$.}
\end{figure}

\begin{figure}[htb]
 \captionsetup[subfigure]{justification=centering}
 \centering
 \begin{subfigure}[t]{0.29\textwidth}
  \centering
  \includegraphics[
   width=\textwidth]{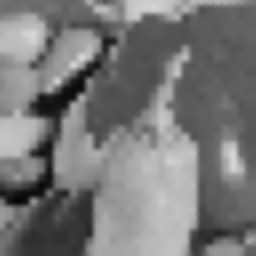}
  \caption{$\textrm{MSE} = 261.05$}
 \end{subfigure}
 \quad 
 \begin{subfigure}[t]{0.29\textwidth}
  \centering
  \includegraphics[width=\textwidth]{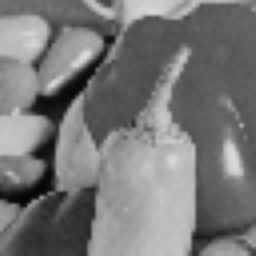}
  \caption{$\textrm{MSE} = 95.31$}
 \end{subfigure} 
 \quad 
 \begin{subfigure}[t]{0.29\textwidth}
  \centering
  \includegraphics[width=\textwidth]{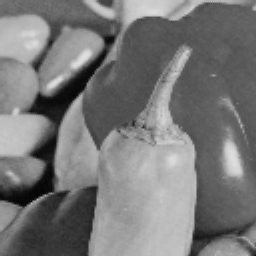}
  \caption{$\textrm{MSE} = 33.81$}
 \end{subfigure}

 \begin{subfigure}[t]{0.29\textwidth}
  \centering
  \includegraphics[
   width=\textwidth]{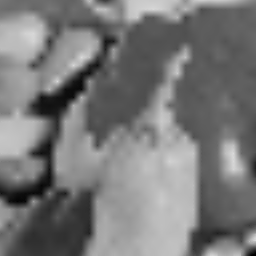}
  \caption{$\textrm{MSE} = 264.02$}
 \end{subfigure}
 \quad 
 \begin{subfigure}[t]{0.29\textwidth}
  \centering
  \includegraphics[width=\textwidth]{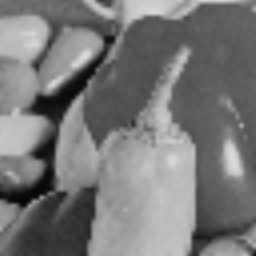}
  \caption{$\textrm{MSE} = 95.39$}
 \end{subfigure}
 \quad 
 \begin{subfigure}[t]{0.29\textwidth}
  \centering
  \includegraphics[width=\textwidth]{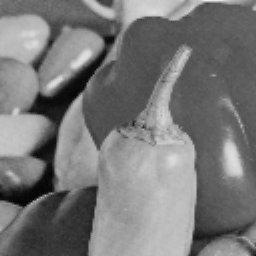}
  \caption{$\textrm{MSE} = 34.19$}
 \end{subfigure}
 \caption[SPH inpainting of ``peppers'' with regular masks]{
  Inpainting of 
  ``peppers'' for regular masks with isotropic Gaussian kernels. 
  Densities from left to right are 1.5625 \%, 6.25 \%, and 25 \%.
  Top row: Zero order consistency SPH inpainting. 
  Bottom row: First order consistency SPH inpainting.}
\end{figure}

\begin{figure}[htb]
 \captionsetup[subfigure]{justification=centering}
 \centering
 \begin{subfigure}[t]{0.29\textwidth}
  \centering
  \includegraphics[width=\textwidth]{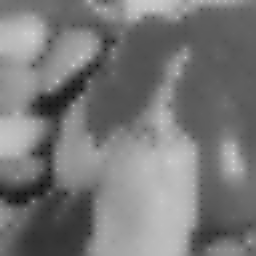}
  \caption{$\textrm{MSE} = 381.26$}
 \end{subfigure}
 \quad 
 \begin{subfigure}[t]{0.29\textwidth}
  \centering
  \includegraphics[width=\textwidth]{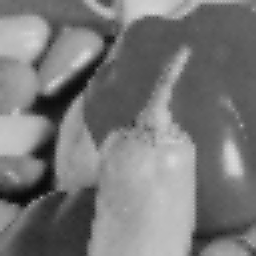}
  \caption{$\textrm{MSE} = 127.12$}
 \end{subfigure}
 \quad 
 \begin{subfigure}[t]{0.29\textwidth}
  \centering
  \includegraphics[width=\textwidth]{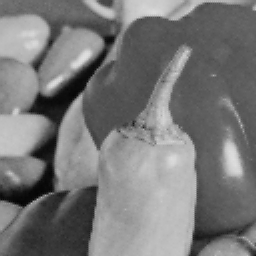}
  \caption{$\textrm{MSE} = 37.00$}
 \end{subfigure}

 \begin{subfigure}[t]{0.29\textwidth}
  \centering
  \includegraphics[width=\textwidth]{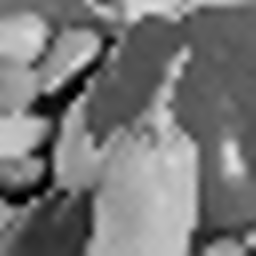}
  \caption{$\textrm{MSE} = 244.80$}
 \end{subfigure}
 \quad
 \begin{subfigure}[t]{0.29\textwidth}
  \centering
  \includegraphics[width=\textwidth]{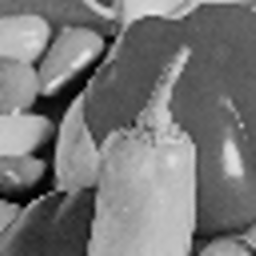}
  \caption{$\textrm{MSE} = 86.40$}
 \end{subfigure}
 \quad
 \begin{subfigure}[t]{0.29\textwidth}
  \centering
  \includegraphics[width=\textwidth]{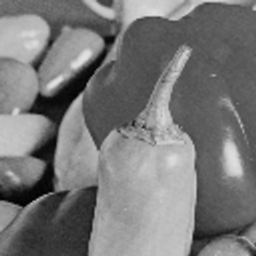}
  \caption{$\textrm{MSE} = 30.01$}
 \end{subfigure}

 \begin{subfigure}[t]{0.29\textwidth}
  \centering
  \includegraphics[
   width=\textwidth]{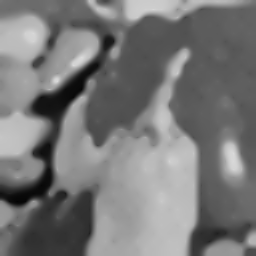}
  \caption{$\textrm{MSE} = 216.62$}
 \end{subfigure}
 \quad 
 \begin{subfigure}[t]{0.29\textwidth}
  \centering
  \includegraphics[
   width=\textwidth]{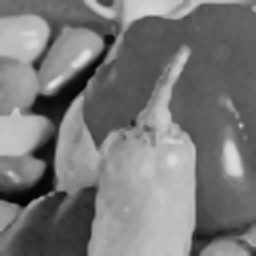}
  \caption{$\textrm{MSE} = 72.63$}
 \end{subfigure}
 \quad 
 \begin{subfigure}[t]{0.29\textwidth}
  \centering
  \includegraphics[
   width=\textwidth]{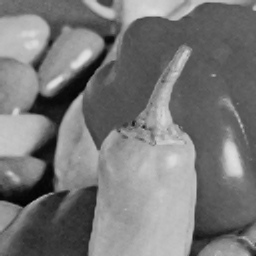}
  \caption{$\textrm{MSE} = 25.94$}
 \end{subfigure}
 \caption[Diffusion inpainting of ``peppers'' with regular masks]{
  Inpainting of 
  ``peppers'' for regular masks. Densities from left to right are 
  1.5625 \%, 6.25 \%, and 25 \%.
  Top row: Harmonic inpainting. 
  Middle row: Biharmonic inpainting. 
  Bottom row: Inpainting with EED. 
  Parameters are from left to right 
  $\lambda=0.3$ and $\sigma=1.9$, $\lambda=0.3$ and $\sigma=1.8$, and
  $\lambda=0.6$ and $\sigma=1.3$.}
\end{figure}

\begin{figure}[htb]
 \captionsetup[subfigure]{justification=centering}
 \centering
 \begin{subfigure}[t]{0.29\textwidth}
  \centering
  \includegraphics[width=\textwidth]{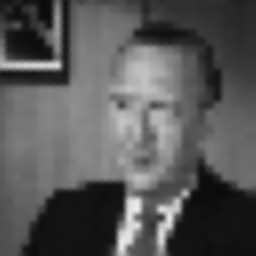}
  \caption{$\textrm{MSE} = 252.35$}
 \end{subfigure}
 \quad 
 \begin{subfigure}[t]{0.29\textwidth}
  \centering
  \includegraphics[width=\textwidth]{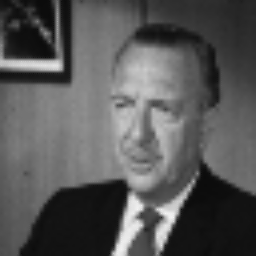}
  \caption{$\textrm{MSE} = 69.13$}
 \end{subfigure} 
 \quad 
 \begin{subfigure}[t]{0.29\textwidth}
  \centering
  \includegraphics[width=\textwidth]{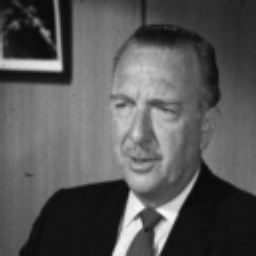}
  \caption{$\textrm{MSE} = 12.47$}
 \end{subfigure}

 \begin{subfigure}[t]{0.29\textwidth}
  \centering
  \includegraphics[
   width=\textwidth]{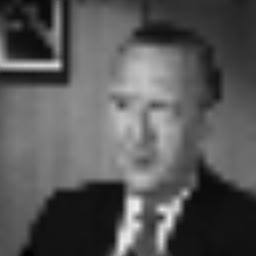}
  \caption{$\textrm{MSE} = 254.53$}
 \end{subfigure}
 \quad 
 \begin{subfigure}[t]{0.29\textwidth}
  \centering
  \includegraphics[width=\textwidth]{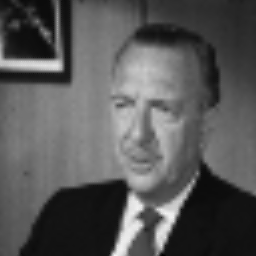}
  \caption{$\textrm{MSE} = 69.28$}
 \end{subfigure}
 \quad 
 \begin{subfigure}[t]{0.29\textwidth}
  \centering
  \includegraphics[width=\textwidth]{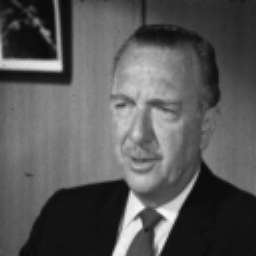}
  \caption{$\textrm{MSE} = 11.66$}
 \end{subfigure}
 \caption[SPH inpainting of ``walter'' with regular masks]{
  Inpainting of 
  ``walter'' for regular masks with isotropic Gaussian kernels. 
  Densities from left to right are 1.5625~\%, 6.25~\%, and 25~\%.
  Top row: Zero order consistency SPH inpainting. 
  Bottom row: First order consistency SPH inpainting.}
\end{figure}

\begin{figure}[htb]
 \captionsetup[subfigure]{justification=centering}
 \centering 
 \begin{subfigure}[t]{0.29\textwidth}
  \centering
  \includegraphics[width=\textwidth]{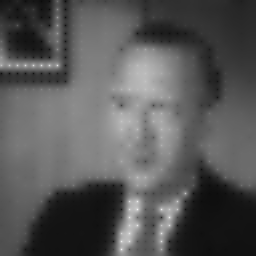}
  \caption{$\textrm{MSE} = 423.03$}
 \end{subfigure}
 \quad 
 \begin{subfigure}[t]{0.29\textwidth}
  \centering
  \includegraphics[width=\textwidth]{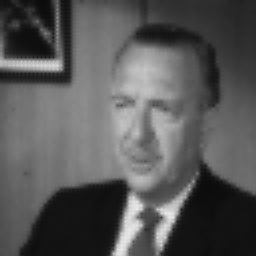}
  \caption{$\textrm{MSE} = 115.18$}
 \end{subfigure}
 \quad 
 \begin{subfigure}[t]{0.29\textwidth}
  \centering
  \includegraphics[width=\textwidth]{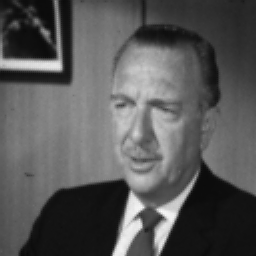}
  \caption{$\textrm{MSE} = 17.02$}
 \end{subfigure}

 \begin{subfigure}[t]{0.29\textwidth}
  \centering
  \includegraphics[width=\textwidth]{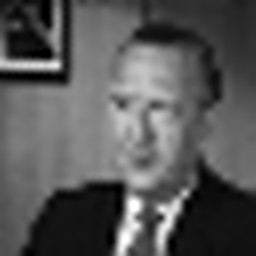}
  \caption{$\textrm{MSE} = 229.23$}
 \end{subfigure}
 \quad
 \begin{subfigure}[t]{0.29\textwidth}
  \centering
  \includegraphics[width=\textwidth]{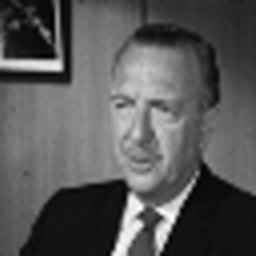}
  \caption{$\textrm{MSE} = 48.54$}
 \end{subfigure}
 \quad
 \begin{subfigure}[t]{0.29\textwidth}
  \centering
  \includegraphics[width=\textwidth]{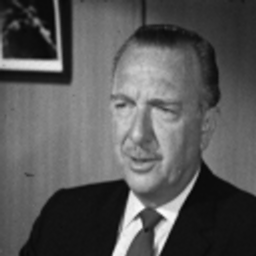}
  \caption{$\textrm{MSE} = 4.73$}
 \end{subfigure}

 \begin{subfigure}[t]{0.29\textwidth}
  \centering
  \includegraphics[
   width=\textwidth]{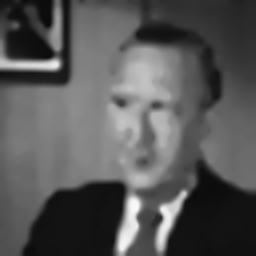}
  \caption{$\textrm{MSE} = 221.08$}
 \end{subfigure}
 \quad 
 \begin{subfigure}[t]{0.29\textwidth}
  \centering
  \includegraphics[
   width=\textwidth]{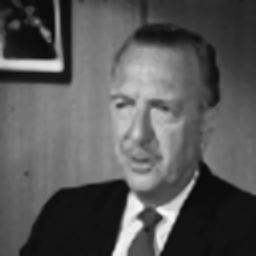}
  \caption{$\textrm{MSE} = 38.20$}
 \end{subfigure}
 \quad 
 \begin{subfigure}[t]{0.29\textwidth}
  \centering
  \includegraphics[
   width=\textwidth]{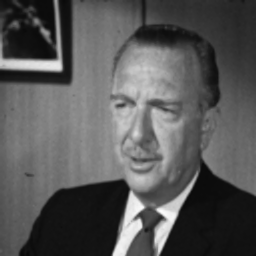}
  \caption{$\textrm{MSE} = 4.25$}
 \end{subfigure}
 \caption[Diffusion inpainting of ``walter'' with regular masks]{
  Inpainting of 
  ``walter'' for regular masks. Densities from left to right are 
  1.5625 \%, 6.25 \%, and 25 \%.
  Top row: Harmonic inpainting. 
  Middle row: Biharmonic inpainting. 
  Bottom row: Inpainting with EED. 
  Parameters are from left to right 
  $\lambda=0.2$ and $\sigma=1.2$, $\lambda=0.1$ and $\sigma=1.1$, and
  $\lambda=0.1$ and $\sigma=0.7$.}
\end{figure}

\begin{figure}[htb]
 \captionsetup[subfigure]{justification=centering}
 \centering
 \begin{subfigure}[t]{0.29\textwidth}
  \centering
  \includegraphics[width=\textwidth]{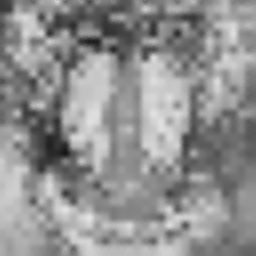}
  \caption{$\textrm{MSE} = 1082.97$}
 \end{subfigure}
 \quad 
 \begin{subfigure}[t]{0.29\textwidth}
  \centering
  \includegraphics[width=\textwidth]{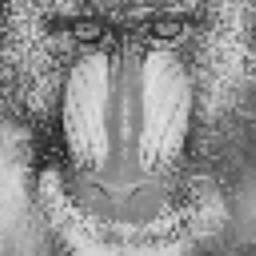}
  \caption{$\textrm{MSE} = 786.14$}
 \end{subfigure} 
 \quad 
 \begin{subfigure}[t]{0.29\textwidth}
  \centering
  \includegraphics[width=\textwidth]{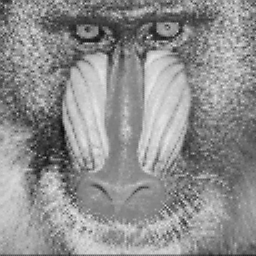}
  \caption{$\textrm{MSE} = 484.12$}
 \end{subfigure}

 \begin{subfigure}[t]{0.29\textwidth}
  \centering
  \includegraphics[
   width=\textwidth]{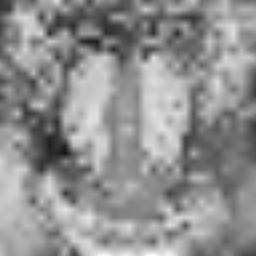}
  \caption{$\textrm{MSE} = 1069.05$}
 \end{subfigure}
 \quad 
 \begin{subfigure}[t]{0.29\textwidth}
  \centering
  \includegraphics[width=\textwidth]{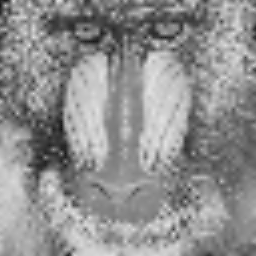}
  \caption{$\textrm{MSE} = 787.46$}
 \end{subfigure}
 \quad 
 \begin{subfigure}[t]{0.29\textwidth}
  \centering
  \includegraphics[width=\textwidth]{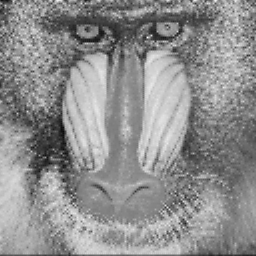}
  \caption{$\textrm{MSE} = 488.93$}
 \end{subfigure}
 \caption[SPH inpainting of ``baboon'' with regular masks]{
  Inpainting of 
  ``baboon'' for regular masks with isotropic Gaussian kernels. 
  Densities from left to right are 1.5625 \%, 6.25 \%, and 25 \%.
  Top row: Zero order consistency SPH inpainting. 
  Bottom row: First order consistency SPH inpainting.}
\end{figure}

\begin{figure}[htb]
 \captionsetup[subfigure]{justification=centering}
 \centering 
 \begin{subfigure}[t]{0.29\textwidth}
  \centering
  \includegraphics[width=\textwidth]{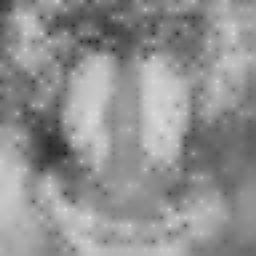}
  \caption{$\textrm{MSE} = 943.91$}
 \end{subfigure}
 \quad 
 \begin{subfigure}[t]{0.29\textwidth}
  \centering
  \includegraphics[width=\textwidth]{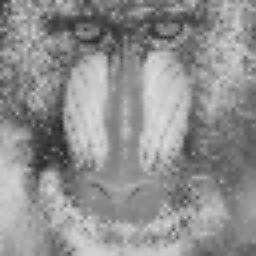}
  \caption{$\textrm{MSE} = 738.12$}
 \end{subfigure}
 \quad 
 \begin{subfigure}[t]{0.29\textwidth}
  \centering
  \includegraphics[width=\textwidth]{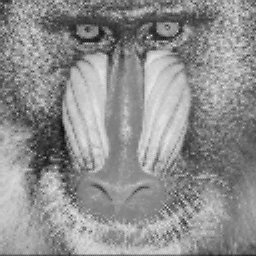}
  \caption{$\textrm{MSE} = 473.74$}
 \end{subfigure}

 \begin{subfigure}[t]{0.29\textwidth}
  \centering
  \includegraphics[width=\textwidth]{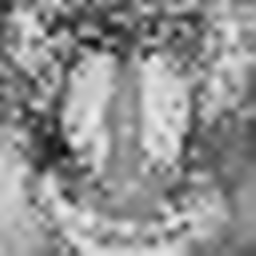}
  \caption{$\textrm{MSE} = 1203.91$}
 \end{subfigure}
 \quad
 \begin{subfigure}[t]{0.29\textwidth}
  \centering
  \includegraphics[width=\textwidth]{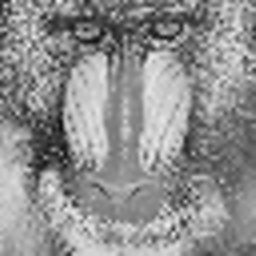}
  \caption{$\textrm{MSE} = 890.13$}
 \end{subfigure}
 \quad
 \begin{subfigure}[t]{0.29\textwidth}
  \centering
  \includegraphics[width=\textwidth]{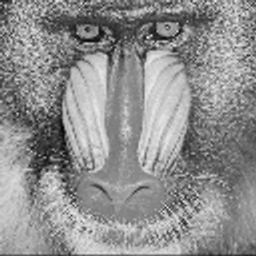}
  \caption{$\textrm{MSE} = 540.67$}
 \end{subfigure}

 \begin{subfigure}[t]{0.29\textwidth}
  \centering
  \includegraphics[
   width=\textwidth]{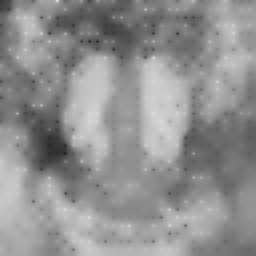}
  \caption{$\textrm{MSE} = 940.73$}
 \end{subfigure}
 \quad 
 \begin{subfigure}[t]{0.29\textwidth}
  \centering
  \includegraphics[
   width=\textwidth]{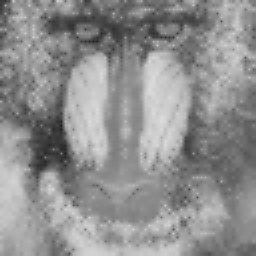}
  \caption{$\textrm{MSE} = 733.31$}
 \end{subfigure}
 \quad 
 \begin{subfigure}[t]{0.29\textwidth}
  \centering
  \includegraphics[
   width=\textwidth]{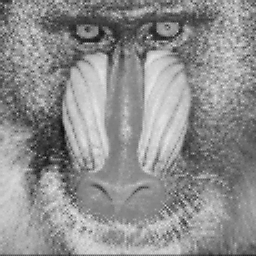}
  \caption{$\textrm{MSE} = 473.66$}
 \end{subfigure}
 \caption[Diffusion inpainting of ``baboon'' with regular masks]{
  Inpainting of 
  ``baboon'' for regular masks. Densities from left to right are 
  1.5625 \%, 6.25 \%, and 25 \%.
  Top row: Harmonic inpainting. 
  Middle row: Biharmonic inpainting. 
  Bottom row: Inpainting with EED. 
  Parameters are from left to right 
  $\lambda=5.2$ and $\sigma=0.6$, $\lambda=11.4$ and $\sigma=0.1$, and
  $\lambda=9.6$ and $\sigma=3.0$.}
\end{figure}

\begin{figure}[htb]
 \captionsetup[subfigure]{justification=centering}
 \centering
 \begin{subfigure}[t]{0.29\textwidth}
  \centering
  \includegraphics[width=\textwidth]{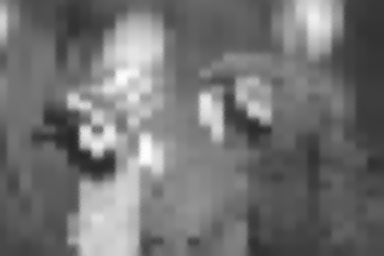}
  \caption{$\textrm{MSE} = 277.88$}
 \end{subfigure}
 \quad 
 \begin{subfigure}[t]{0.29\textwidth}
  \centering
  \includegraphics[width=\textwidth]{Figures_art/parrots_reg_D00416_gauss_zero}
  \caption{$\textrm{MSE} = 132.79$}
 \end{subfigure}
 \quad 
 \begin{subfigure}[t]{0.29\textwidth}
  \centering
  \includegraphics[width=\textwidth]{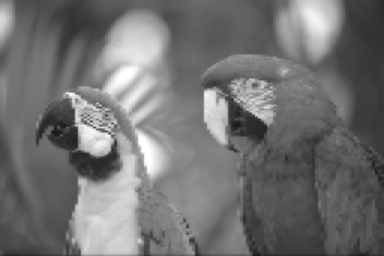}
  \caption{$\textrm{MSE} = 57.52$}
 \end{subfigure}
 
  \begin{subfigure}[t]{0.29\textwidth}
  \centering
  \includegraphics[width=\textwidth]{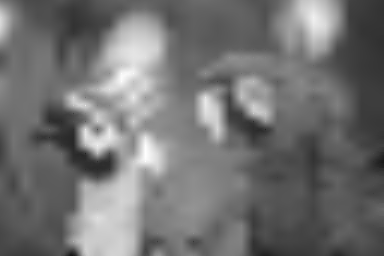}
  \caption{$\textrm{MSE} = 261.11$}
 \end{subfigure}
 \quad 
 \begin{subfigure}[t]{0.29\textwidth}
  \centering
  \includegraphics[width=\textwidth]{Figures_art/parrots_reg_D00416_gauss_first}
  \caption{$\textrm{MSE} = 124.81$}
 \end{subfigure}
 \quad 
 \begin{subfigure}[t]{0.29\textwidth}
  \centering
  \includegraphics[width=\textwidth]{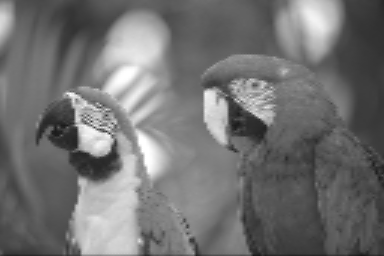}
  \caption{$\textrm{MSE} = 55.87$}
 \end{subfigure}
 \caption[SPH inpainting of ``parrots'' with regular masks]{
  Inpainting of 
  ``parrots'' for regular masks with isotropic Gaussian kernels. 
  Densities from left to right are 1.04 \%, 4.16 \%, and 16.66 \%.
  Top row: Zero order consistency SPH inpainting. 
  Bottom row: First order consistency SPH inpainting.}
\end{figure}

\begin{figure}[htb]
 \captionsetup[subfigure]{justification=centering}
 \centering
 \begin{subfigure}[t]{0.29\textwidth}
  \centering
  \includegraphics[width=\textwidth]{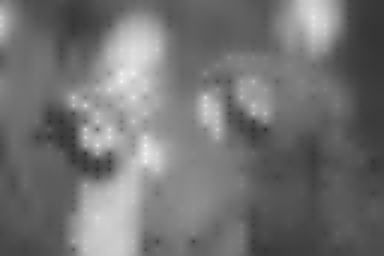}
  \caption{$\textrm{MSE} = 313.94$}
 \end{subfigure}
 \quad 
 \begin{subfigure}[t]{0.29\textwidth}
  \centering
  \includegraphics[width=\textwidth]{Figures_art/parrots_reg_D00416_harm}
  \caption{$\textrm{MSE} = 139.10$}
 \end{subfigure}
 \quad 
 \begin{subfigure}[t]{0.29\textwidth}
  \centering
  \includegraphics[width=\textwidth]{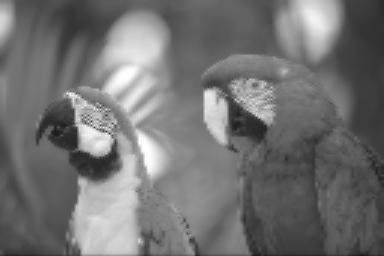}
  \caption{$\textrm{MSE} = 60.39$}
 \end{subfigure}
 
 \begin{subfigure}[t]{0.29\textwidth}
  \centering
  \includegraphics[width=\textwidth]{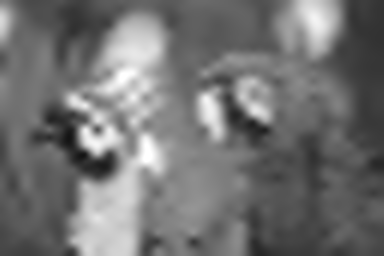}
  \caption{$\textrm{MSE} = 266.33$}
 \end{subfigure}
 \quad 
 \begin{subfigure}[t]{0.29\textwidth}
  \centering
  \includegraphics[width=\textwidth]{Figures_art/parrots_reg_D00416_biharm}
  \caption{$\textrm{MSE} = 128.18$}
 \end{subfigure}
 \quad 
 \begin{subfigure}[t]{0.29\textwidth}
  \centering
  \includegraphics[width=\textwidth]{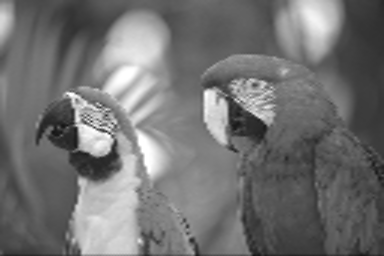}
  \caption{$\textrm{MSE} = 54.70$}
 \end{subfigure}
 
 \begin{subfigure}[t]{0.29\textwidth}
  \centering
  \includegraphics[
   width=\textwidth]{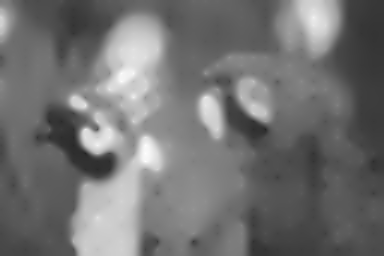}
  \caption{$\textrm{MSE} = 258.32$}
 \end{subfigure}
 \quad 
 \begin{subfigure}[t]{0.29\textwidth}
  \centering
  \includegraphics[
   width=\textwidth]{Figures_art/parrots_reg_D00416_eed_lambda_1_5_sigma_2}
  \caption{$\textrm{MSE} = 118.79$}
 \end{subfigure}
 \quad 
 \begin{subfigure}[t]{0.29\textwidth}
  \centering
  \includegraphics[
   width=\textwidth]{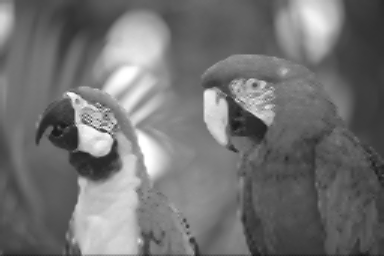}
  \caption{$\textrm{MSE} = 52.08$}
 \end{subfigure}
 \caption[Diffusion inpainting of ``parrots'' with regular masks]{
  Inpainting of 
  ``parrots'' for regular masks. Densities from left to right are 
  1.04 \%, 4.16 \%, and 16.66 \%.
  Top row: Harmonic inpainting. 
  Middle row: Biharmonic inpainting. 
  Bottom row: Inpainting with EED. 
  Parameters are from left to right 
  $\lambda=0.7$ and $\sigma=2.0$, $\lambda=1.5$ and $\sigma=2.0$, and
  $\lambda=1.6$ and $\sigma=1.5$.}
\end{figure}

\begin{figure}[htb]
 \captionsetup[subfigure]{justification=centering}
 \centering
 \begin{subfigure}[t]{0.29\textwidth}
  \centering
  \includegraphics[width=\textwidth]{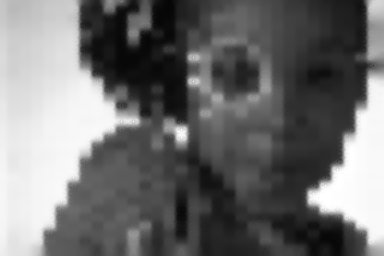}
  \caption{$\textrm{MSE} = 502.30$}
 \end{subfigure}
 \quad 
 \begin{subfigure}[t]{0.29\textwidth}
  \centering
  \includegraphics[width=\textwidth]{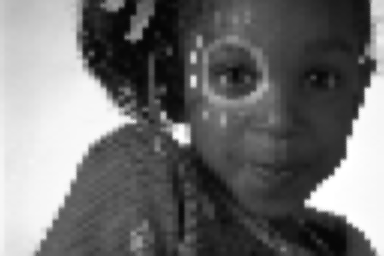}
  \caption{$\textrm{MSE} = 225.16$}
 \end{subfigure}
 \quad 
 \begin{subfigure}[t]{0.29\textwidth}
  \centering
  \includegraphics[width=\textwidth]{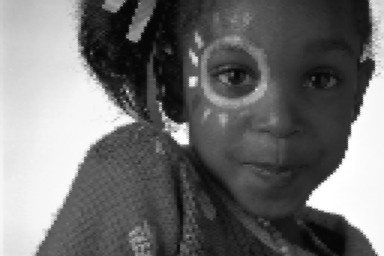}
  \caption{$\textrm{MSE} = 104.52$}
 \end{subfigure}
 
  \begin{subfigure}[t]{0.29\textwidth}
  \centering
  \includegraphics[width=\textwidth]{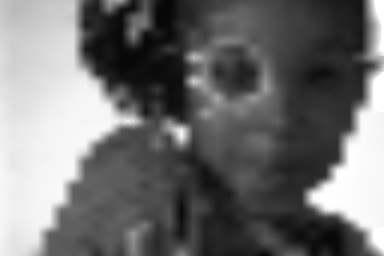}
  \caption{$\textrm{MSE} = 501.80$}
 \end{subfigure}
 \quad 
 \begin{subfigure}[t]{0.29\textwidth}
  \centering
  \includegraphics[width=\textwidth]{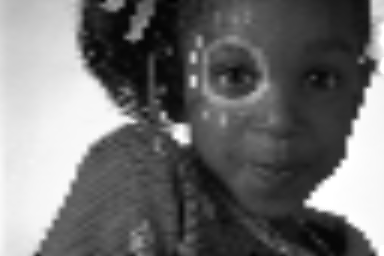}
  \caption{$\textrm{MSE} = 218.19$}
 \end{subfigure}
 \quad 
 \begin{subfigure}[t]{0.29\textwidth}
  \centering
  \includegraphics[width=\textwidth]{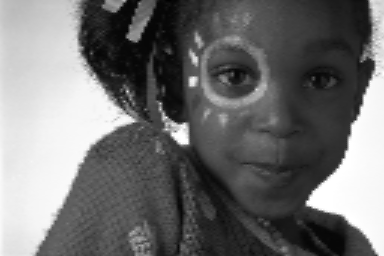}
  \caption{$\textrm{MSE} = 99.95$}
 \end{subfigure}
 \caption[SPH inpainting of ``girl'' with regular masks]{
  Inpainting of 
  ``girl'' for regular masks with isotropic Gaussian kernels. 
  Densities from left to right are 1.04 \%, 4.16 \%, and 16.66 \%.
  Top row: Zero order consistency SPH inpainting. 
  Bottom row: First order consistency SPH inpainting.}
\end{figure}

\begin{figure}[htb]
 \captionsetup[subfigure]{justification=centering}
 \centering
 \begin{subfigure}[t]{0.29\textwidth}
  \centering
  \includegraphics[width=\textwidth]{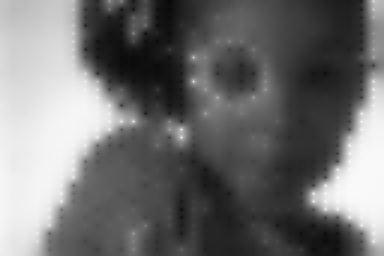}
  \caption{$\textrm{MSE} = 593.92$}
 \end{subfigure}
 \quad 
 \begin{subfigure}[t]{0.29\textwidth}
  \centering
  \includegraphics[width=\textwidth]{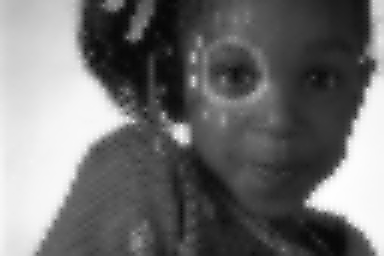}
  \caption{$\textrm{MSE} = 247.97$}
 \end{subfigure}
 \quad 
 \begin{subfigure}[t]{0.29\textwidth}
  \centering
  \includegraphics[width=\textwidth]{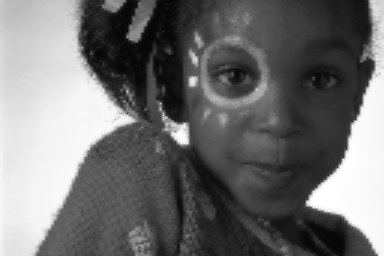}
  \caption{$\textrm{MSE} = 104.55$}
 \end{subfigure}
 
 \begin{subfigure}[t]{0.29\textwidth}
  \centering
  \includegraphics[width=\textwidth]{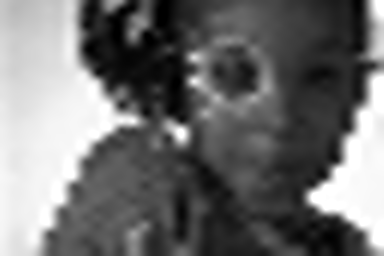}
  \caption{$\textrm{MSE} = 469.80$}
 \end{subfigure}
 \quad 
 \begin{subfigure}[t]{0.29\textwidth}
  \centering
  \includegraphics[width=\textwidth]{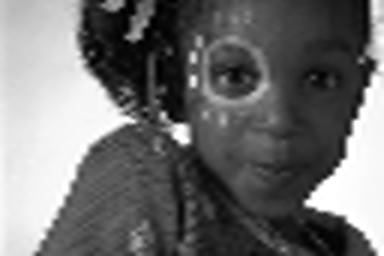}
  \caption{$\textrm{MSE} = 215.31$}
 \end{subfigure}
 \quad 
 \begin{subfigure}[t]{0.29\textwidth}
  \centering
  \includegraphics[width=\textwidth]{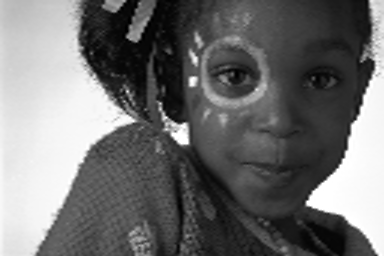}
  \caption{$\textrm{MSE} = 99.54$}
 \end{subfigure}
 
 \begin{subfigure}[t]{0.29\textwidth}
  \centering
  \includegraphics[
   width=\textwidth]{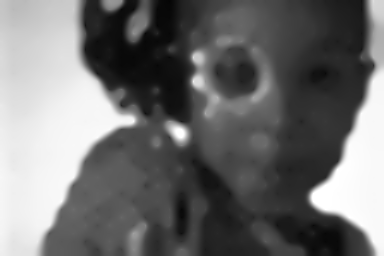}
  \caption{$\textrm{MSE} = 371.32$}
 \end{subfigure}
 \quad 
 \begin{subfigure}[t]{0.29\textwidth}
  \centering
  \includegraphics[
   width=\textwidth]{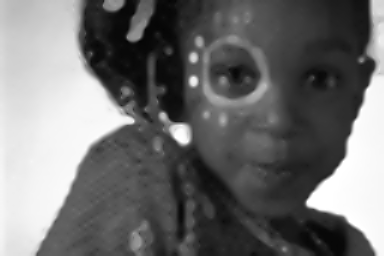}
  \caption{$\textrm{MSE} = 166.89$}
 \end{subfigure}
 \quad 
 \begin{subfigure}[t]{0.29\textwidth}
  \centering
  \includegraphics[
   width=\textwidth]{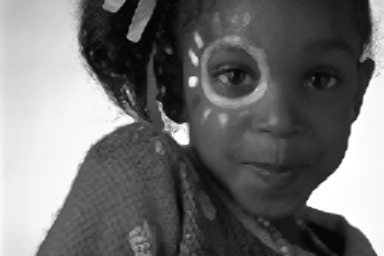}
  \caption{$\textrm{MSE} = 78.41$}
 \end{subfigure}
 \caption[Diffusion inpainting of ``girl'' with regular masks]{
  Inpainting of 
  ``girl'' for regular masks. Densities from left to right are 
  1.04 \%, 4.16 \%, and 16.66 \%.
  Top row: Harmonic inpainting. 
  Middle row: Biharmonic inpainting. 
  Bottom row: Inpainting with EED. 
  Parameters are from left to right 
  $\lambda=0.3$ and $\sigma=2.0$, $\lambda=0.6$ and $\sigma=2.0$, and
  $\lambda=0.7$ and $\sigma=2.0$.}
\end{figure}

\begin{figure}[htb]
 \captionsetup[subfigure]{justification=centering}
 \centering
 \begin{subfigure}[t]{0.29\textwidth}
  \centering
  \includegraphics[width=\textwidth]{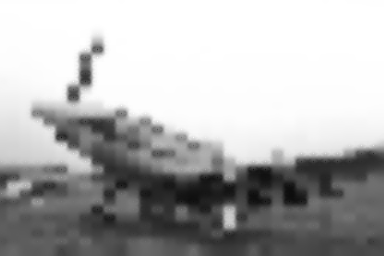}
  \caption{$\textrm{MSE} = 489.67$}
 \end{subfigure}
 \quad 
 \begin{subfigure}[t]{0.29\textwidth}
  \centering
  \includegraphics[width=\textwidth]{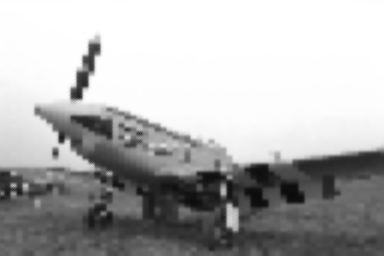}
  \caption{$\textrm{MSE} = 247.10$}
 \end{subfigure}
 \quad 
 \begin{subfigure}[t]{0.29\textwidth}
  \centering
  \includegraphics[width=\textwidth]{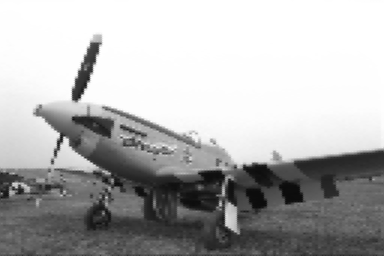}
  \caption{$\textrm{MSE} = 108.35$}
 \end{subfigure}
 
  \begin{subfigure}[t]{0.29\textwidth}
  \centering
  \includegraphics[width=\textwidth]{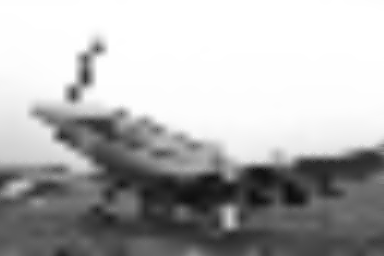}
  \caption{$\textrm{MSE} = 477.34$}
 \end{subfigure}
 \quad 
 \begin{subfigure}[t]{0.29\textwidth}
  \centering
  \includegraphics[width=\textwidth]{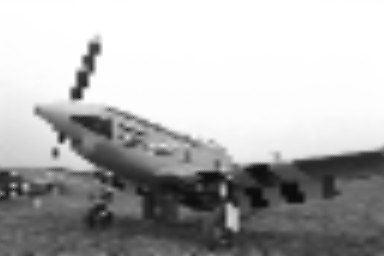}
  \caption{$\textrm{MSE} = 244.17$}
 \end{subfigure}
 \quad 
 \begin{subfigure}[t]{0.29\textwidth}
  \centering
  \includegraphics[width=\textwidth]{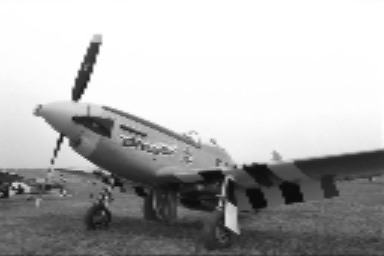}
  \caption{$\textrm{MSE} = 105.77$}
 \end{subfigure}
 \caption[SPH inpainting of ``plane'' with regular masks]{
  Inpainting of 
  ``plane'' for regular masks with isotropic Gaussian kernels. 
  Densities from left to right are 1.04 \%, 4.16 \%, and 16.66 \%.
  Top row: Zero order consistency SPH inpainting. 
  Bottom row: First order consistency SPH inpainting.}
\end{figure}

\begin{figure}[htb]
 \captionsetup[subfigure]{justification=centering}
 \centering
 \begin{subfigure}[t]{0.29\textwidth}
  \centering
  \includegraphics[width=\textwidth]{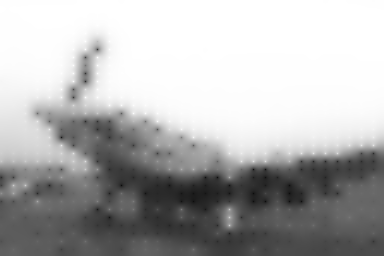}
  \caption{$\textrm{MSE} = 551.02$}
 \end{subfigure}
 \quad 
 \begin{subfigure}[t]{0.29\textwidth}
  \centering
  \includegraphics[width=\textwidth]{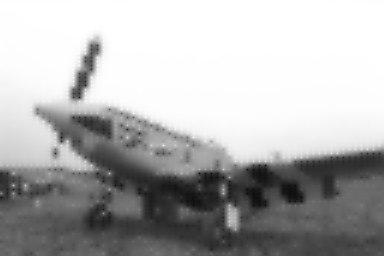}
  \caption{$\textrm{MSE} = 277.22$}
 \end{subfigure}
 \quad 
 \begin{subfigure}[t]{0.29\textwidth}
  \centering
  \includegraphics[width=\textwidth]{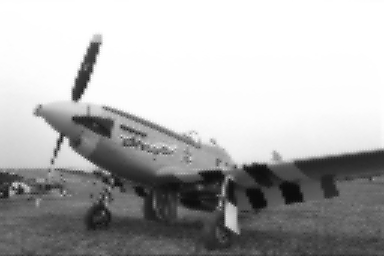}
  \caption{$\textrm{MSE} = 116.11$}
 \end{subfigure}
 
 \begin{subfigure}[t]{0.29\textwidth}
  \centering
  \includegraphics[width=\textwidth]{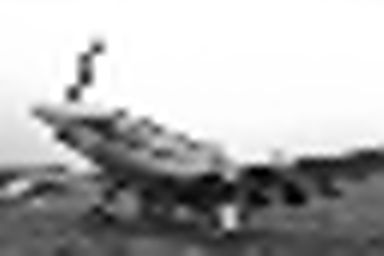}
  \caption{$\textrm{MSE} = 497.11$}
 \end{subfigure}
 \quad 
 \begin{subfigure}[t]{0.29\textwidth}
  \centering
  \includegraphics[width=\textwidth]{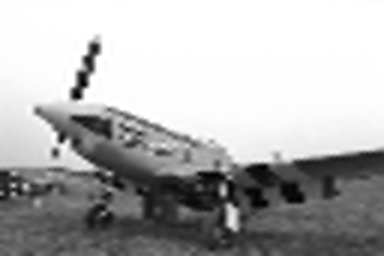}
  \caption{$\textrm{MSE} = 250.55$}
 \end{subfigure}
 \quad 
 \begin{subfigure}[t]{0.29\textwidth}
  \centering
  \includegraphics[width=\textwidth]{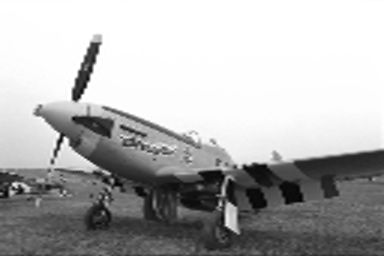}
  \caption{$\textrm{MSE} = 103.77$}
 \end{subfigure}
 
 \begin{subfigure}[t]{0.29\textwidth}
  \centering
  \includegraphics[
   width=\textwidth]{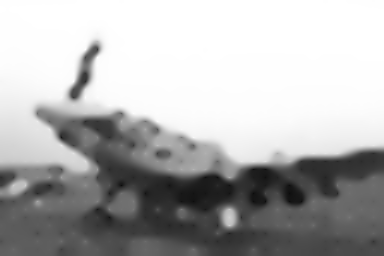}
  \caption{$\textrm{MSE} = 426.88$}
 \end{subfigure}
 \quad 
 \begin{subfigure}[t]{0.29\textwidth}
  \centering
  \includegraphics[
   width=\textwidth]{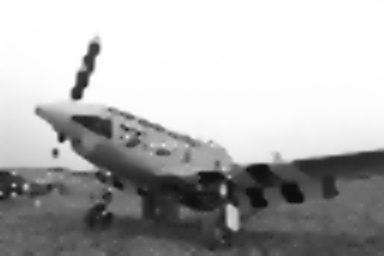}
  \caption{$\textrm{MSE} = 225.21$}
 \end{subfigure}
 \quad 
 \begin{subfigure}[t]{0.29\textwidth}
  \centering
  \includegraphics[
   width=\textwidth]{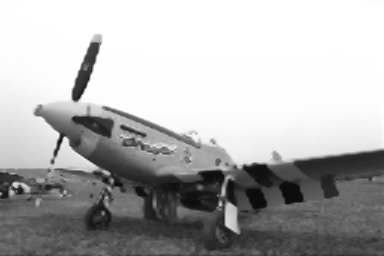}
  \caption{$\textrm{MSE} = 92.32$}
 \end{subfigure}
 \caption[Diffusion inpainting of ``plane'' with regular masks]{
  Inpainting of 
  ``plane'' for regular masks. Densities from left to right are 
  1.04 \%, 4.16 \%, and 16.66 \%.
  Top row: Harmonic inpainting. 
  Middle row: Biharmonic inpainting. 
  Bottom row: Inpainting with EED. 
  Parameters are from left to right 
  $\lambda=0.8$ and $\sigma=2.0$, $\lambda=1.8$ and $\sigma=0.9$, and
  $\lambda=1.4$ and $\sigma=1.6$.}
\end{figure}

\begin{figure}[htb]
 \captionsetup[subfigure]{justification=centering}
 \centering
 \begin{subfigure}[t]{0.29\textwidth}
  \centering
  \includegraphics[width=\textwidth]{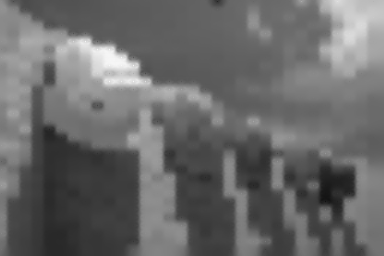}
  \caption{$\textrm{MSE} = 284.43$}
 \end{subfigure}
 \quad 
 \begin{subfigure}[t]{0.29\textwidth}
  \centering
  \includegraphics[width=\textwidth]{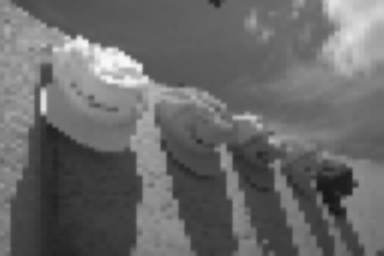}
  \caption{$\textrm{MSE} = 143.04$}
 \end{subfigure}
 \quad 
 \begin{subfigure}[t]{0.29\textwidth}
  \centering
  \includegraphics[width=\textwidth]{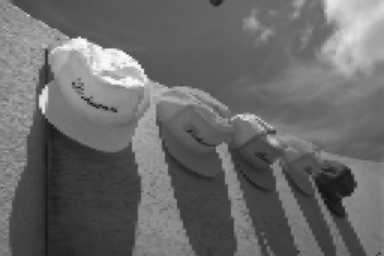}
  \caption{$\textrm{MSE} = 67.49$}
 \end{subfigure}
 
  \begin{subfigure}[t]{0.29\textwidth}
  \centering
  \includegraphics[width=\textwidth]{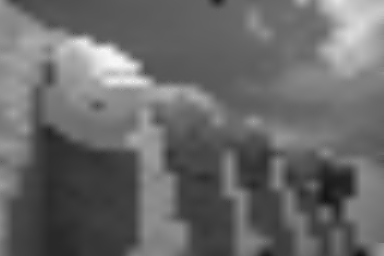}
  \caption{$\textrm{MSE} = 274.08$}
 \end{subfigure}
 \quad 
 \begin{subfigure}[t]{0.29\textwidth}
  \centering
  \includegraphics[width=\textwidth]{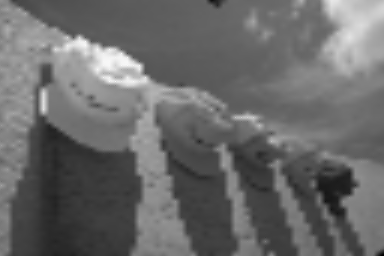}
  \caption{$\textrm{MSE} = 140.25$}
 \end{subfigure}
 \quad 
 \begin{subfigure}[t]{0.29\textwidth}
  \centering
  \includegraphics[width=\textwidth]{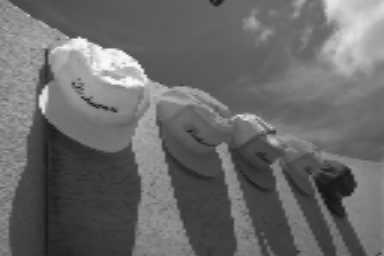}
  \caption{$\textrm{MSE} = 65.32$}
 \end{subfigure}
 \caption[Inpainting of ``hats'' with regular masks]{
  Inpainting of 
  ``hats'' for regular masks with isotropic Gaussian kernels. 
  Densities from left to right are 1.04 \%, 4.16 \%, and 16.66 \%.
  Top row: Zero order consistency SPH inpainting. 
  Bottom row: First order consistency SPH inpainting.}
\end{figure}

\begin{figure}[htb]
 \captionsetup[subfigure]{justification=centering}
 \centering
 \begin{subfigure}[t]{0.29\textwidth}
  \centering
  \includegraphics[width=\textwidth]{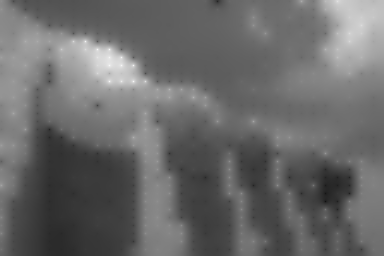}
  \caption{$\textrm{MSE} = 305.05$}
 \end{subfigure}
 \quad 
 \begin{subfigure}[t]{0.29\textwidth}
  \centering
  \includegraphics[width=\textwidth]{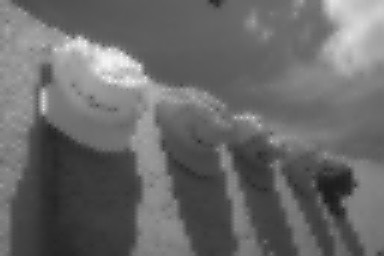}
  \caption{$\textrm{MSE} = 155.78$}
 \end{subfigure}
 \quad 
 \begin{subfigure}[t]{0.29\textwidth}
  \centering
  \includegraphics[width=\textwidth]{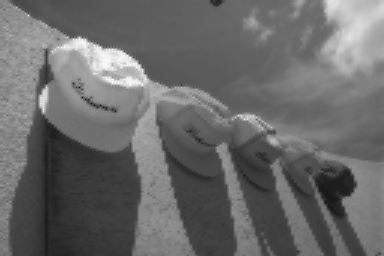}
  \caption{$\textrm{MSE} = 69.45$}
 \end{subfigure}
 
 \begin{subfigure}[t]{0.29\textwidth}
  \centering
  \includegraphics[width=\textwidth]{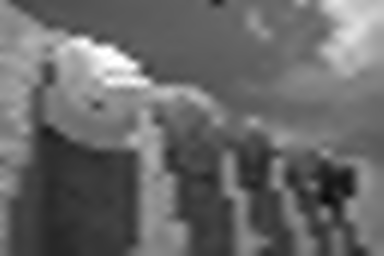}
  \caption{$\textrm{MSE} = 283.68$}
 \end{subfigure}
 \quad 
 \begin{subfigure}[t]{0.29\textwidth}
  \centering
  \includegraphics[width=\textwidth]{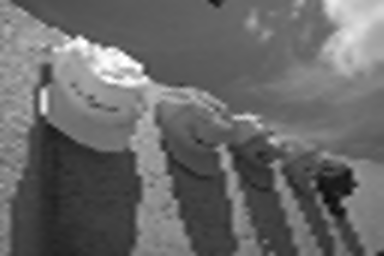}
  \caption{$\textrm{MSE} = 140.67$}
 \end{subfigure}
 \quad 
 \begin{subfigure}[t]{0.29\textwidth}
  \centering
  \includegraphics[width=\textwidth]{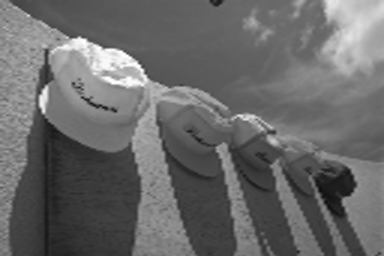}
  \caption{$\textrm{MSE} = 65.96$}
 \end{subfigure}
 
 \begin{subfigure}[t]{0.29\textwidth}
  \centering
  \includegraphics[
   width=\textwidth]{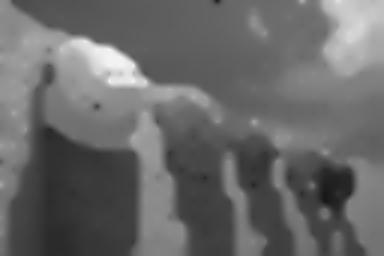}
  \caption{$\textrm{MSE} = 254.01$}
 \end{subfigure}
 \quad 
 \begin{subfigure}[t]{0.29\textwidth}
  \centering
  \includegraphics[
   width=\textwidth]{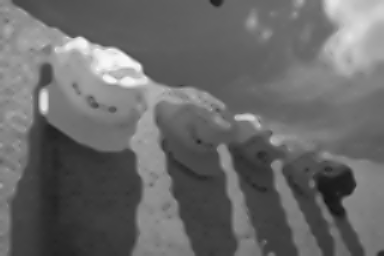}
  \caption{$\textrm{MSE} = 119.05$}
 \end{subfigure}
 \quad 
 \begin{subfigure}[t]{0.29\textwidth}
  \centering
  \includegraphics[
   width=\textwidth]{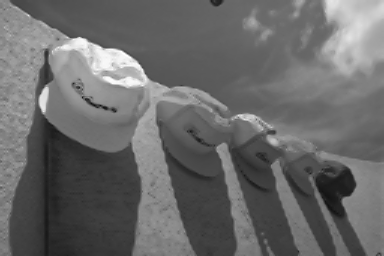}
  \caption{$\textrm{MSE} = 55.24$}
 \end{subfigure}
 \caption[Diffusion inpainting of ``hats'' with regular masks]{
  Inpainting of 
  ``hats'' for regular masks. Densities from left to right are 
  1.04 \%, 4.16 \%, and 16.66 \%.
  Top row: Harmonic inpainting. 
  Middle row: Biharmonic inpainting. 
  Bottom row: Inpainting with EED. 
  Parameters are from left to right 
  $\lambda=0.6$ and $\sigma=2.0$, $\lambda=0.5$ and $\sigma=2.0$, and
  $\lambda=0.6$ and $\sigma=2.0$.}
\end{figure}
Over all images and densities, the results of SPH inpainting 
are somewhere between the results achieved by harmonic and biharmonic
diffusion, respectively. However, SPH can, in general, not achieve the same 
quality as EED inpainting.

To investigate further, we consider the same images, but now 
equipped with masks of randomly chosen mask pixels instead of regular masks.
For each image, we have created random masks of densities 1 \%, 5 \%, and 
10 \%. 
For each of these masks, we perform SPH inpainting with an isotropic Gaussian
kernel either for zero or first order consistency, harmonic inpainting, 
biharmonic inpainting, and inpainting with EED.

\begin{figure}[htb]
 \captionsetup[subfigure]{justification=centering}
 \centering
 \begin{subfigure}[t]{0.29\textwidth}
  \centering
  \includegraphics[width=\textwidth]{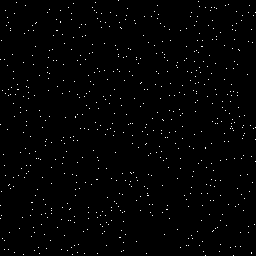}
  \caption{1 \% mask}
 \end{subfigure}
 \quad 
 \begin{subfigure}[t]{0.29\textwidth}
  \centering
  \includegraphics[width=\textwidth]{Figures_art/mask_ran_D005_trui}
  \caption{5 \% mask}
 \end{subfigure}
 \quad 
 \begin{subfigure}[t]{0.29\textwidth}
  \centering
  \includegraphics[width=\textwidth]{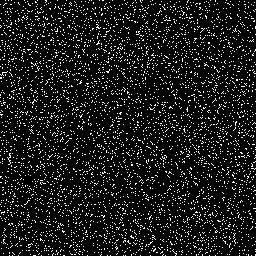}
  \caption{10 \% mask}
 \end{subfigure}

 \begin{subfigure}[t]{0.29\textwidth}
  \centering
  \includegraphics[width=\textwidth]{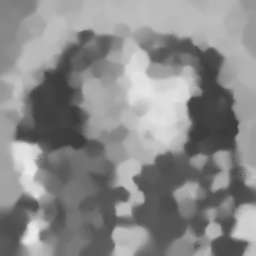}
  \caption{$\textrm{MSE} = 649.00$}
 \end{subfigure}
 \quad 
 \begin{subfigure}[t]{0.29\textwidth}
  \centering
  \includegraphics[width=\textwidth]{Figures_art/trui_ran_D005_gauss_zero}
  \caption{$\textrm{MSE} = 208.48$}
 \end{subfigure}
 \quad 
 \begin{subfigure}[t]{0.29\textwidth}
  \centering
  \includegraphics[width=\textwidth]{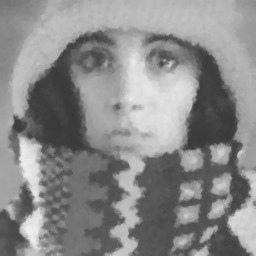}
  \caption{$\textrm{MSE} = 110.48$}
 \end{subfigure}

 \begin{subfigure}[t]{0.29\textwidth}
  \centering
  \includegraphics[width=\textwidth]{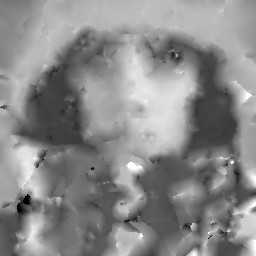}
  \caption{$\textrm{MSE} = 693.58$}
 \end{subfigure}
 \quad 
 \begin{subfigure}[t]{0.29\textwidth}
  \centering
  \includegraphics[width=\textwidth]{Figures_art/trui_ran_D005_gauss_first}
  \caption{$\textrm{MSE} = 197.58$}
 \end{subfigure}
 \quad 
 \begin{subfigure}[t]{0.29\textwidth}
  \centering
  \includegraphics[width=\textwidth]{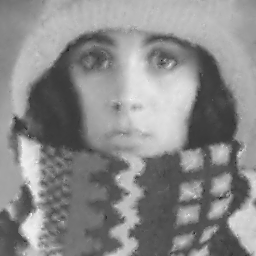}
  \caption{$\textrm{MSE} = 99.23$}
 \end{subfigure}
 \caption[SPH inpainting of ``trui'' with random masks]{
  Inpainting of ``trui'' 
  for random masks of different densities.
  Top row: Masks with densities of 1 \%, 5 \%, and 10 \%. 
  Middle row: Zero order consistency SPH inpainting with isotropic 
  Gaussian kernel. 
  Bottom row: First order consistency SPH inpainting with isotropic 
  Gaussian kernel.}
\end{figure}

\begin{figure}[htb]
 \captionsetup[subfigure]{justification=centering}
 \centering
 \begin{subfigure}[t]{0.29\textwidth}
  \centering
  \includegraphics[width=\textwidth]{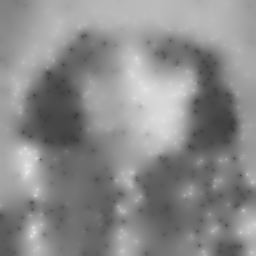}
  \caption{$\textrm{MSE} = 655.65$}
 \end{subfigure}
 \quad 
 \begin{subfigure}[t]{0.29\textwidth}
  \centering
  \includegraphics[width=\textwidth]{Figures_art/trui_ran_D005_harm}
  \caption{$\textrm{MSE} = 226.06$}
 \end{subfigure}
 \quad 
 \begin{subfigure}[t]{0.29\textwidth}
  \centering
  \includegraphics[width=\textwidth]{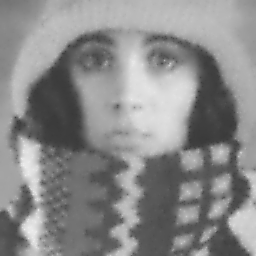}
  \caption{$\textrm{MSE} = 121.66$}
 \end{subfigure}

 \begin{subfigure}[t]{0.29\textwidth}
  \centering
  \includegraphics[width=\textwidth]{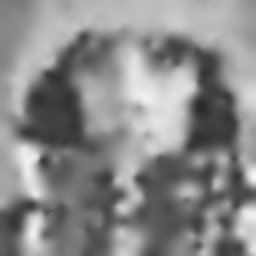}
  \caption{$\textrm{MSE} = 579.71$}
 \end{subfigure}
 \quad 
 \begin{subfigure}[t]{0.29\textwidth}
  \centering
  \includegraphics[width=\textwidth]{Figures_art/trui_ran_D005_biharm}
  \caption{$\textrm{MSE} = 146.46$}
 \end{subfigure}
 \quad 
 \begin{subfigure}[t]{0.29\textwidth}
  \centering
  \includegraphics[width=\textwidth]{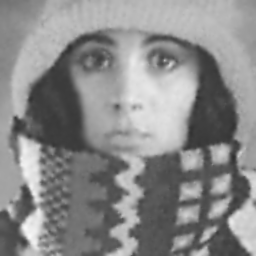}
  \caption{$\textrm{MSE} = 70.99$}
 \end{subfigure}

 \begin{subfigure}[t]{0.29\textwidth}
  \centering
  \includegraphics[
   width=\textwidth]{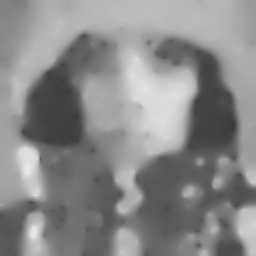}
  \caption{$\textrm{MSE} = 554.98$}
 \end{subfigure}
 \quad 
 \begin{subfigure}[t]{0.29\textwidth}
  \centering
  \includegraphics[
   width=\textwidth]{Figures_art/trui_ran_D005_eed_lambda_0_2_sigma_0_8}
  \caption{$\textrm{MSE} = 134.92$}
 \end{subfigure}
 \quad 
 \begin{subfigure}[t]{0.29\textwidth}
  \centering
  \includegraphics[
   width=\textwidth]{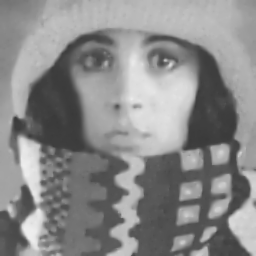}
  \caption{$\textrm{MSE} = 62.51$}
 \end{subfigure}
 \caption[Diffusion inpainting of ``trui'' with random masks]{
  Inpainting of
  ``trui'' for random masks of densities 1 \%, 5 \%, and 10 \% (left to right).
  Top row: Harmonic inpainting. 
  Middle row: Biharmonic inpainting. 
  Bottom row: Inpainting with EED. 
  Parameters are from left to right 
  $\lambda=1.0$ and $\sigma=2.0$, $\lambda=0.2$ and $\sigma=0.8$, and
  $\lambda=0.4$ and $\sigma=0.8$.}
\end{figure}

\begin{figure}[htb]
 \captionsetup[subfigure]{justification=centering}
 \centering
 \begin{subfigure}[t]{0.29\textwidth}
  \centering
  \includegraphics[width=\textwidth]{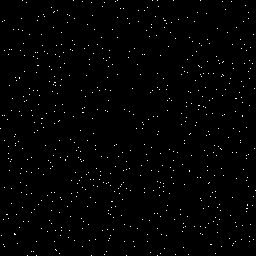}
  \caption{1 \% mask}
 \end{subfigure}
 \quad 
 \begin{subfigure}[t]{0.29\textwidth}
  \centering
  \includegraphics[width=\textwidth]{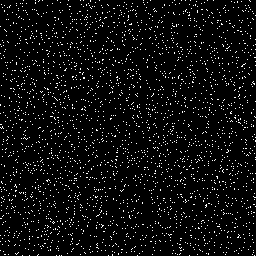}
  \caption{5 \% mask}
 \end{subfigure}
 \quad 
 \begin{subfigure}[t]{0.29\textwidth}
  \centering
  \includegraphics[width=\textwidth]{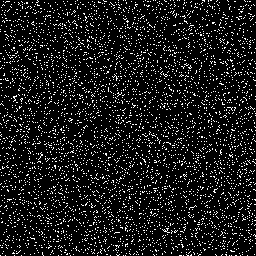}
  \caption{10 \% mask}
 \end{subfigure}

 \begin{subfigure}[t]{0.29\textwidth}
  \centering
  \includegraphics[width=\textwidth]{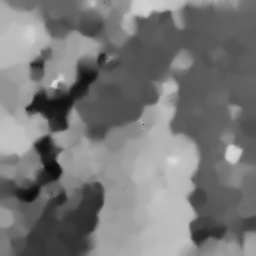}
  \caption{$\textrm{MSE} = 643.74$}
 \end{subfigure}
 \quad 
 \begin{subfigure}[t]{0.29\textwidth}
  \centering
  \includegraphics[width=\textwidth]{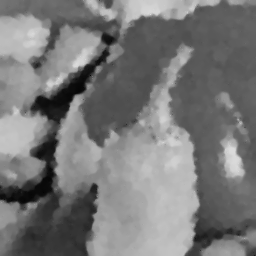}
  \caption{$\textrm{MSE} = 196.64$}
 \end{subfigure}
 \quad 
 \begin{subfigure}[t]{0.29\textwidth}
  \centering
  \includegraphics[width=\textwidth]{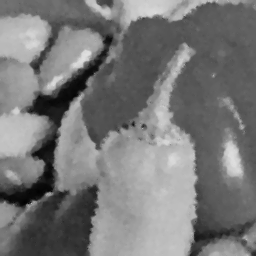}
  \caption{$\textrm{MSE} = 115.62$}
 \end{subfigure}

 \begin{subfigure}[t]{0.29\textwidth}
  \centering
  \includegraphics[width=\textwidth]{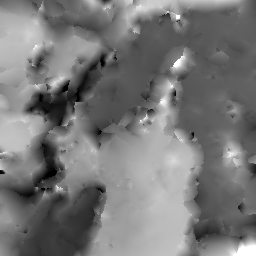}
  \caption{$\textrm{MSE} = 807.45$}
 \end{subfigure}
 \quad 
 \begin{subfigure}[t]{0.29\textwidth}
  \centering
  \includegraphics[width=\textwidth]{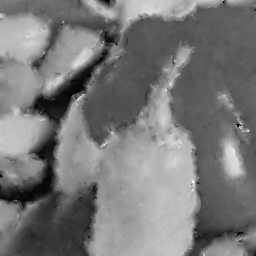}
  \caption{$\textrm{MSE} = 201.50$}
 \end{subfigure}
 \quad 
 \begin{subfigure}[t]{0.29\textwidth}
  \centering
  \includegraphics[width=\textwidth]{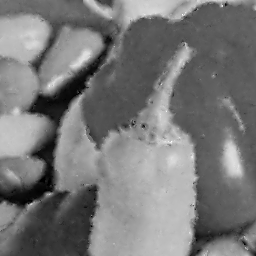}
  \caption{$\textrm{MSE} = 109.25$}
 \end{subfigure}
 \caption[SPH inpainting of ``peppers'' with random masks]{
  Inpainting of 
  ``peppers''  for random masks of different densities.
  Top row: Masks with densities of 1 \%, 5 \%, and 10 \%. 
  Middle row: Zero order consistency SPH inpainting with isotropic 
  Gaussian kernel. 
  Bottom row: First order consistency SPH inpainting with isotropic 
  Gaussian kernel.}
\end{figure}

\begin{figure}[htb]
 \captionsetup[subfigure]{justification=centering}
 \centering
 \begin{subfigure}[t]{0.29\textwidth}
  \centering
  \includegraphics[width=\textwidth]{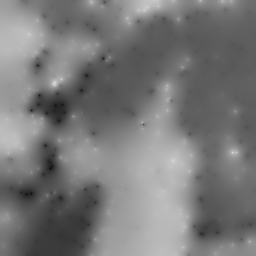}
  \caption{$\textrm{MSE} = 712.59$}
 \end{subfigure}
 \quad 
 \begin{subfigure}[t]{0.29\textwidth}
  \centering
  \includegraphics[width=\textwidth]{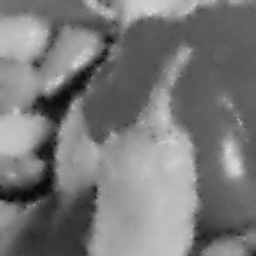}
  \caption{$\textrm{MSE} = 217.79$}
 \end{subfigure}
 \quad 
 \begin{subfigure}[t]{0.29\textwidth}
  \centering
  \includegraphics[width=\textwidth]{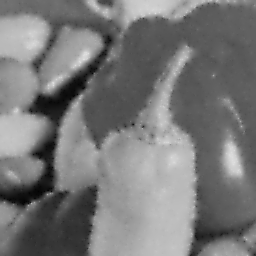}
  \caption{$\textrm{MSE} = 119.63$}
 \end{subfigure}

 \begin{subfigure}[t]{0.29\textwidth}
  \centering
  \includegraphics[width=\textwidth]{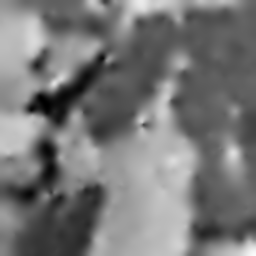}
  \caption{$\textrm{MSE} = 553.56$}
 \end{subfigure}
 \quad 
 \begin{subfigure}[t]{0.29\textwidth}
  \centering
  \includegraphics[width=\textwidth]{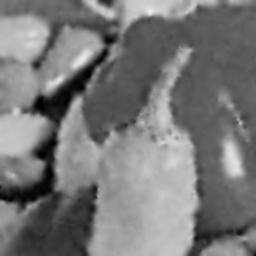}
  \caption{$\textrm{MSE} = 151.51$}
 \end{subfigure}
 \quad 
 \begin{subfigure}[t]{0.29\textwidth}
  \centering
  \includegraphics[width=\textwidth]{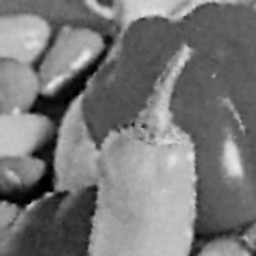}
  \caption{$\textrm{MSE} = 88.45$}
 \end{subfigure}

 \begin{subfigure}[t]{0.29\textwidth}
  \centering
  \includegraphics[
   width=\textwidth]{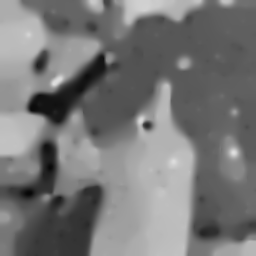}
  \caption{$\textrm{MSE} = 542.87$}
 \end{subfigure}
 \quad 
 \begin{subfigure}[t]{0.29\textwidth}
  \centering
  \includegraphics[
   width=\textwidth]{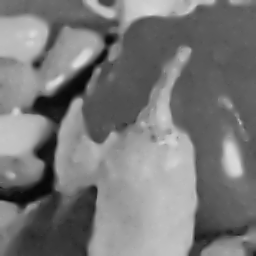}
  \caption{$\textrm{MSE} = 135.32$}
 \end{subfigure}
 \quad 
 \begin{subfigure}[t]{0.29\textwidth}
  \centering
  \includegraphics[
   width=\textwidth]{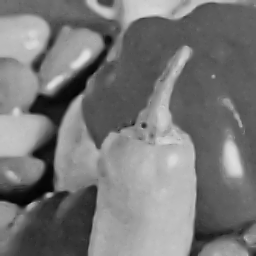}
  \caption{$\textrm{MSE} = 68.56$}
 \end{subfigure}
 \caption[Diffusion inpainting of ``peppers'' with random masks]{
  Inpainting of
  ``peppers'' for random masks of densities 1 \%, 5 \%, and 10
  \% (left to right).
  Top row: Harmonic inpainting. 
  Middle row: Biharmonic inpainting. 
  Bottom row: Inpainting with EED. 
  Parameters are from left to right 
  $\lambda=0.1$ and $\sigma=0.9$, $\lambda=0.5$ and $\sigma=1.5$, and
  $\lambda=0.4$ and $\sigma=1.8$.}
\end{figure}

\begin{figure}[htb]
 \captionsetup[subfigure]{justification=centering}
 \centering
 \begin{subfigure}[t]{0.29\textwidth}
  \centering
  \includegraphics[width=\textwidth]{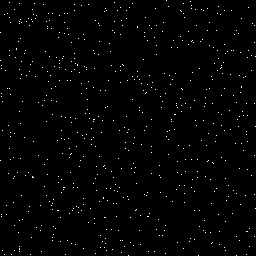}
  \caption{1 \% mask}
 \end{subfigure}
 \quad 
 \begin{subfigure}[t]{0.29\textwidth}
  \centering
  \includegraphics[width=\textwidth]{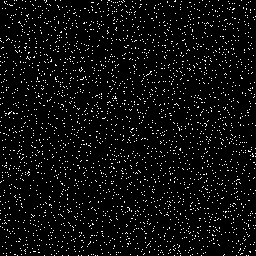}
  \caption{5 \% mask}
 \end{subfigure}
 \quad 
 \begin{subfigure}[t]{0.29\textwidth}
  \centering
  \includegraphics[width=\textwidth]{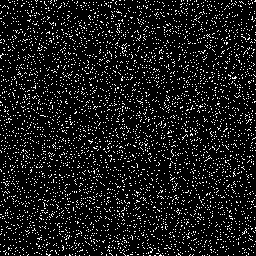}
  \caption{10 \% mask}
 \end{subfigure}

 \begin{subfigure}[t]{0.29\textwidth}
  \centering
  \includegraphics[width=\textwidth]{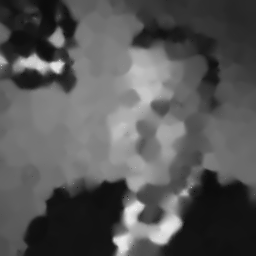}
  \caption{$\textrm{MSE} = 600.02$}
 \end{subfigure}
 \quad 
 \begin{subfigure}[t]{0.29\textwidth}
  \centering
  \includegraphics[width=\textwidth]{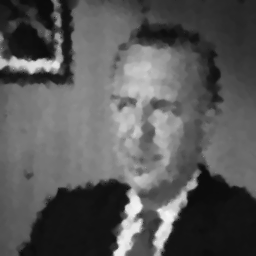}
  \caption{$\textrm{MSE} = 180.58$}
 \end{subfigure}
 \quad 
 \begin{subfigure}[t]{0.29\textwidth}
  \centering
  \includegraphics[width=\textwidth]{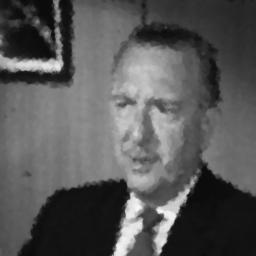}
  \caption{$\textrm{MSE} = 93.79$}
 \end{subfigure}

 \begin{subfigure}[t]{0.29\textwidth}
  \centering
  \includegraphics[width=\textwidth]{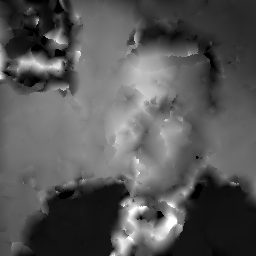}
  \caption{$\textrm{MSE} = 745.89$}
 \end{subfigure}
 \quad 
 \begin{subfigure}[t]{0.29\textwidth}
  \centering
  \includegraphics[width=\textwidth]{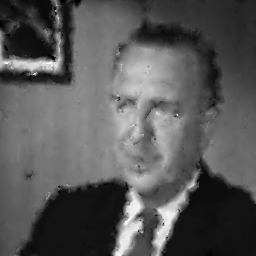}
  \caption{$\textrm{MSE} = 167.00$}
 \end{subfigure}
 \quad 
 \begin{subfigure}[t]{0.29\textwidth}
  \centering
  \includegraphics[width=\textwidth]{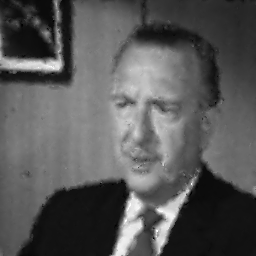}
  \caption{$\textrm{MSE} = 80.22$}
 \end{subfigure}
 \caption[SPH inpainting of ``walter'' with random masks]{
  Inpainting of 
  ``walter''  for random masks of different densities.
  Top row: Masks with densities of 1 \%, 5 \%, and 10 \%. 
  Middle row: Zero order consistency SPH inpainting with isotropic 
  Gaussian kernel. 
  Bottom row: First order consistency SPH inpainting with isotropic 
  Gaussian kernel.}
\end{figure}

\begin{figure}[htb]
 \captionsetup[subfigure]{justification=centering}
 \centering
 \begin{subfigure}[t]{0.29\textwidth}
  \centering
  \includegraphics[width=\textwidth]{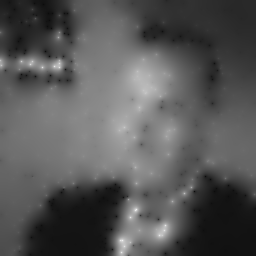}
  \caption{$\textrm{MSE} = 672.50$}
 \end{subfigure}
 \quad 
 \begin{subfigure}[t]{0.29\textwidth}
  \centering
  \includegraphics[width=\textwidth]{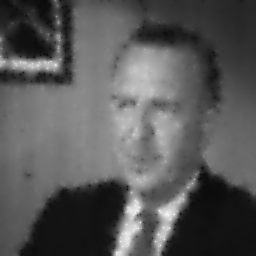}
  \caption{$\textrm{MSE} = 212.61$}
 \end{subfigure}
 \quad 
 \begin{subfigure}[t]{0.29\textwidth}
  \centering
  \includegraphics[width=\textwidth]{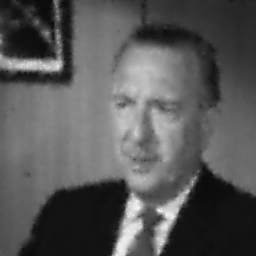}
  \caption{$\textrm{MSE} = 113.88$}
 \end{subfigure}

 \begin{subfigure}[t]{0.29\textwidth}
  \centering
  \includegraphics[width=\textwidth]{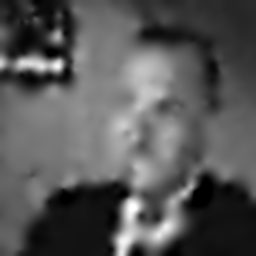}
  \caption{$\textrm{MSE} = 526.65$}
 \end{subfigure}
 \quad 
 \begin{subfigure}[t]{0.29\textwidth}
  \centering
  \includegraphics[width=\textwidth]{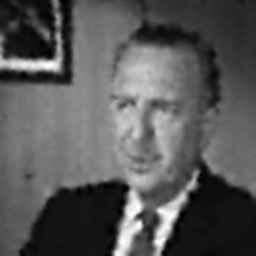}
  \caption{$\textrm{MSE} = 115.72$}
 \end{subfigure}
 \quad 
 \begin{subfigure}[t]{0.29\textwidth}
  \centering
  \includegraphics[width=\textwidth]{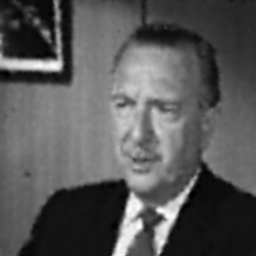}
  \caption{$\textrm{MSE} = 50.82$}
 \end{subfigure}

 \begin{subfigure}[t]{0.29\textwidth}
  \centering
  \includegraphics[
   width=\textwidth]{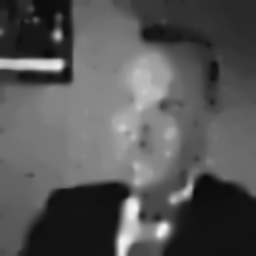}
  \caption{$\textrm{MSE} = 466.66$}
 \end{subfigure}
 \quad 
 \begin{subfigure}[t]{0.29\textwidth}
  \centering
  \includegraphics[
   width=\textwidth]{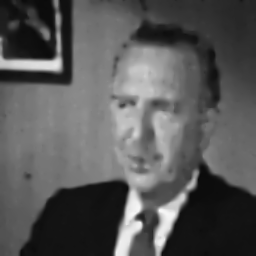}
  \caption{$\textrm{MSE} = 91.90$}
 \end{subfigure}
 \quad 
 \begin{subfigure}[t]{0.29\textwidth}
  \centering
  \includegraphics[
   width=\textwidth]{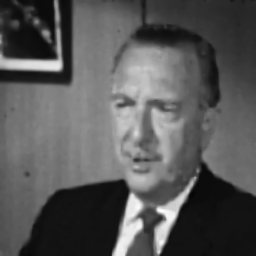}
  \caption{$\textrm{MSE} = 37.92$}
 \end{subfigure}
 \caption[Diffusion inpainting of ``walter'' with random masks]{
  Inpainting of
  ``walter'' for random masks of densities 1 \%, 5 \%, and 10 \% 
  (left to right).
  Top row: Harmonic inpainting. 
  Middle row: Biharmonic inpainting. 
  Bottom row: Inpainting with EED. 
  Parameters are from left to right 
  $\lambda=0.1$ and $\sigma=0.6$, $\lambda=0.2$ and $\sigma=0.9$, and
  $\lambda=0.2$ and $\sigma=0.8$.}
\end{figure}

\begin{figure}[htb]
 \captionsetup[subfigure]{justification=centering}
 \centering
 \begin{subfigure}[t]{0.29\textwidth}
  \centering
  \includegraphics[width=\textwidth]{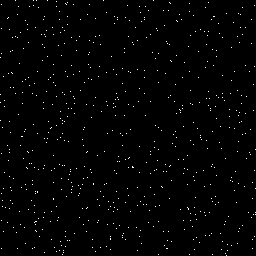}
  \caption{1 \% mask}
 \end{subfigure}
 \quad 
 \begin{subfigure}[t]{0.29\textwidth}
  \centering
  \includegraphics[width=\textwidth]{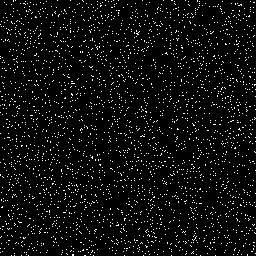}
  \caption{5 \% mask}
 \end{subfigure}
 \quad 
 \begin{subfigure}[t]{0.29\textwidth}
  \centering
  \includegraphics[width=\textwidth]{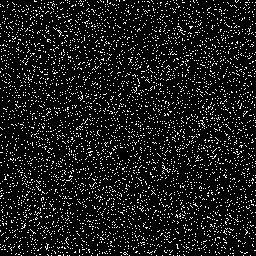}
  \caption{10 \% mask}
 \end{subfigure}

 \begin{subfigure}[t]{0.29\textwidth}
  \centering
  \includegraphics[width=\textwidth]{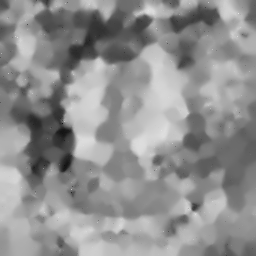}
  \caption{$\textrm{MSE} = 1253.29$}
 \end{subfigure}
 \quad 
 \begin{subfigure}[t]{0.29\textwidth}
  \centering
  \includegraphics[width=\textwidth]{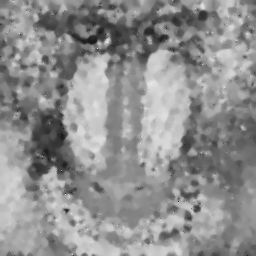}
  \caption{$\textrm{MSE} = 939.62$}
 \end{subfigure}
 \quad 
 \begin{subfigure}[t]{0.29\textwidth}
  \centering
  \includegraphics[width=\textwidth]{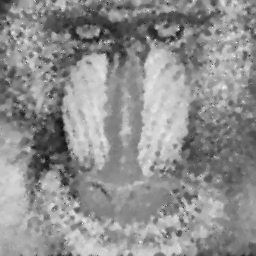}
  \caption{$\textrm{MSE} = 778.42$}
 \end{subfigure}

 \begin{subfigure}[t]{0.29\textwidth}
  \centering
  \includegraphics[width=\textwidth]{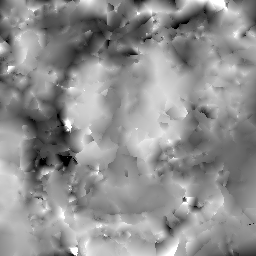}
  \caption{$\textrm{MSE} = 1481.98$}
 \end{subfigure}
 \quad 
 \begin{subfigure}[t]{0.29\textwidth}
  \centering
  \includegraphics[width=\textwidth]{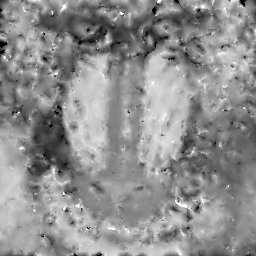}
  \caption{$\textrm{MSE} = 1124.76$}
 \end{subfigure}
 \quad 
 \begin{subfigure}[t]{0.29\textwidth}
  \centering
  \includegraphics[width=\textwidth]{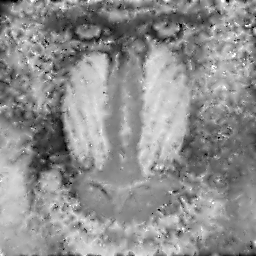}
  \caption{$\textrm{MSE} = 873.64$}
 \end{subfigure}
 \caption[SPH inpainting of ``baboon'' with random masks]{
  Inpainting of 
  ``baboon''  for random masks of different densities.
  Top row: Masks with densities of 1 \%, 5 \%, and 10 \%. 
  Middle row: Zero order consistency SPH inpainting with isotropic 
  Gaussian kernel. 
  Bottom row: First order consistency SPH inpainting with isotropic 
  Gaussian kernel.}
  \label{fig:baboon_ran_sph}
\end{figure}

\begin{figure}[htb]
 \captionsetup[subfigure]{justification=centering}
 \centering
 \begin{subfigure}[t]{0.29\textwidth}
  \centering
  \includegraphics[width=\textwidth]{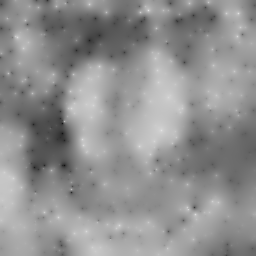}
  \caption{$\textrm{MSE} = 1038.79$}
 \end{subfigure}
 \quad 
 \begin{subfigure}[t]{0.29\textwidth}
  \centering
  \includegraphics[width=\textwidth]{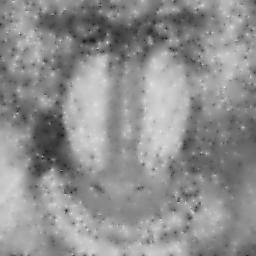}
  \caption{$\textrm{MSE} = 794.68$}
 \end{subfigure}
 \quad 
 \begin{subfigure}[t]{0.29\textwidth}
  \centering
  \includegraphics[width=\textwidth]{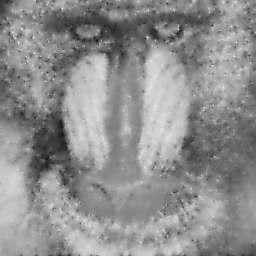}
  \caption{$\textrm{MSE} = 688.23$}
 \end{subfigure}

 \begin{subfigure}[t]{0.29\textwidth}
  \centering
  \includegraphics[width=\textwidth]{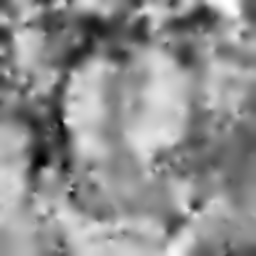}
  \caption{$\textrm{MSE} = 1441.11$}
 \end{subfigure}
 \quad 
 \begin{subfigure}[t]{0.29\textwidth}
  \centering
  \includegraphics[width=\textwidth]{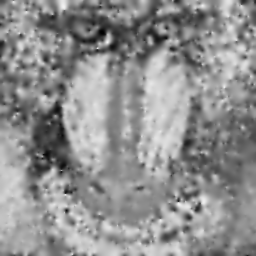}
  \caption{$\textrm{MSE} = 1049.28$}
 \end{subfigure}
 \quad 
 \begin{subfigure}[t]{0.29\textwidth}
  \centering
  \includegraphics[width=\textwidth]{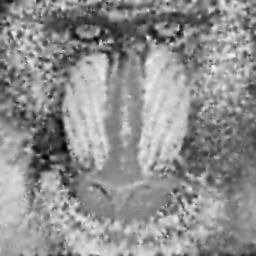}
  \caption{$\textrm{MSE} = 864.59$}
 \end{subfigure}

 \begin{subfigure}[t]{0.29\textwidth}
  \centering
  \includegraphics[
   width=\textwidth]{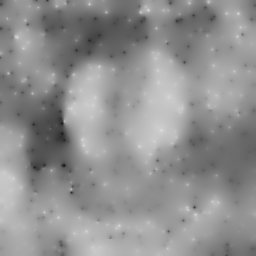}
  \caption{$\textrm{MSE} = 1028.85$}
 \end{subfigure}
 \quad 
 \begin{subfigure}[t]{0.29\textwidth}
  \centering
  \includegraphics[
   width=\textwidth]{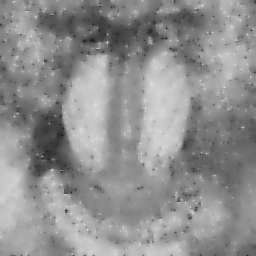}
  \caption{$\textrm{MSE} = 789.49$}
 \end{subfigure}
 \quad 
 \begin{subfigure}[t]{0.29\textwidth}
  \centering
  \includegraphics[
   width=\textwidth]{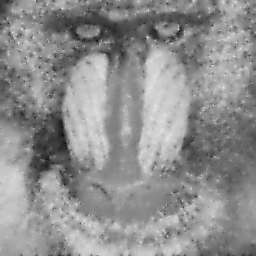}
  \caption{$\textrm{MSE} = 686.94$}
 \end{subfigure}
 \caption[Diffusion inpainting of ``baboon'' with random masks]{
  Inpainting of
  ``baboon'' for random masks of densities 1 \%, 5 \%, and 10 \% 
  (left to right).
  Top row: Harmonic inpainting. 
  Middle row: Biharmonic inpainting. 
  Bottom row: Inpainting with EED. 
  Parameters are from left to right 
  $\lambda=5.0$ and $\sigma=0.7$, $\lambda=6.7$ and $\sigma=0.3$, and
  $\lambda=5.8$ and $\sigma=3.0$.}
  \label{fig:baboon_ran_diff}
\end{figure}

\begin{figure}[htb]
 \captionsetup[subfigure]{justification=centering}
 \centering
 \begin{subfigure}[t]{0.29\textwidth}
  \centering
  \includegraphics[width=\textwidth]{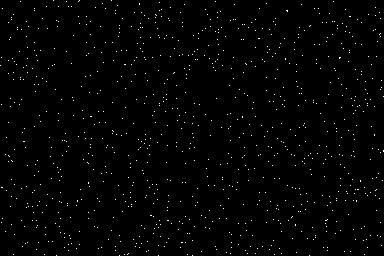}
  \caption{1 \% mask}
 \end{subfigure}
 \quad 
 \begin{subfigure}[t]{0.29\textwidth}
  \centering
  \includegraphics[width=\textwidth]{Figures_art/mask_ran_D005_parrots}
  \caption{5 \% mask}
 \end{subfigure}
 \quad 
 \begin{subfigure}[t]{0.29\textwidth}
  \centering
  \includegraphics[width=\textwidth]{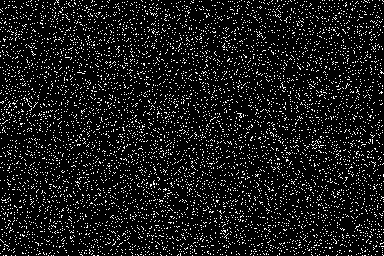}
  \caption{10 \% mask}
 \end{subfigure}
 
 \begin{subfigure}[t]{0.29\textwidth}
  \centering
  \includegraphics[width=\textwidth]{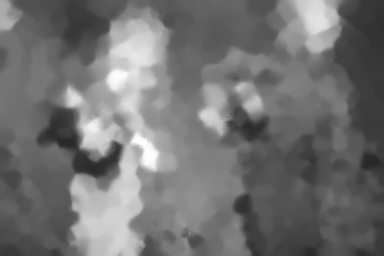}
  \caption{$\textrm{MSE} = 354.73$}
 \end{subfigure}
 \quad 
 \begin{subfigure}[t]{0.29\textwidth}
  \centering
  \includegraphics[width=\textwidth]{Figures_art/parrots_ran_D005_gauss_zero}
  \caption{$\textrm{MSE} = 169.62$}
 \end{subfigure}
 \quad 
 \begin{subfigure}[t]{0.29\textwidth}
  \centering
  \includegraphics[width=\textwidth]{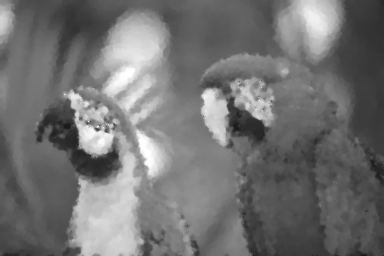}
  \caption{$\textrm{MSE} = 111.69$}
 \end{subfigure}
 
  \begin{subfigure}[t]{0.29\textwidth}
  \centering
  \includegraphics[width=\textwidth]{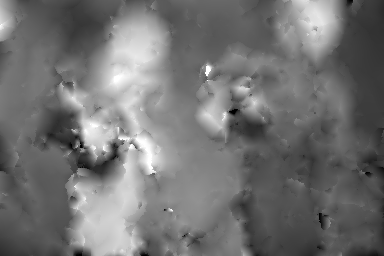}
  \caption{$\textrm{MSE} = 394.41$}
 \end{subfigure}
 \quad 
 \begin{subfigure}[t]{0.29\textwidth}
  \centering
  \includegraphics[width=\textwidth]{Figures_art/parrots_ran_D005_gauss_first}
  \caption{$\textrm{MSE} = 173.71$}
 \end{subfigure}
 \quad 
 \begin{subfigure}[t]{0.29\textwidth}
  \centering
  \includegraphics[width=\textwidth]{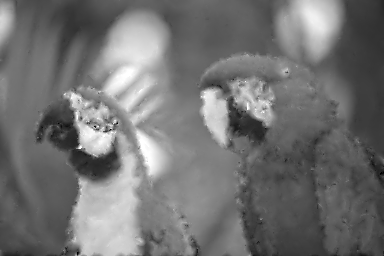}
  \caption{$\textrm{MSE} = 115.51$}
 \end{subfigure}
 \caption[SPH inpainting of ``parrots'' with random masks]{
  Inpainting of 
  ``parrots''  for random masks of different densities.
  Top row: Masks with densities of 1 \%, 5 \%, and 10 \%. 
  Middle row: Zero order consistency SPH inpainting with isotropic 
  Gaussian kernel. 
  Bottom row: First order consistency SPH inpainting with isotropic 
  Gaussian kernel.}
\end{figure}

\begin{figure}[htb]
 \captionsetup[subfigure]{justification=centering}
 \centering
 \begin{subfigure}[t]{0.29\textwidth}
  \centering
  \includegraphics[width=\textwidth]{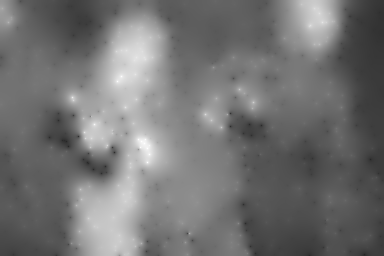}
  \caption{$\textrm{MSE} = 387.24$}
 \end{subfigure}
 \quad 
 \begin{subfigure}[t]{0.29\textwidth}
  \centering
  \includegraphics[width=\textwidth]{Figures_art/parrots_ran_D005_harm}
  \caption{$\textrm{MSE} = 162.53$}
 \end{subfigure}
 \quad 
 \begin{subfigure}[t]{0.29\textwidth}
  \centering
  \includegraphics[width=\textwidth]{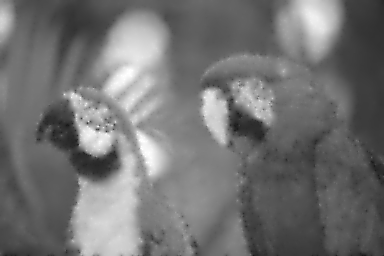}
  \caption{$\textrm{MSE} = 106.41$}
 \end{subfigure}
 
 \begin{subfigure}[t]{0.29\textwidth}
  \centering
  \includegraphics[width=\textwidth]{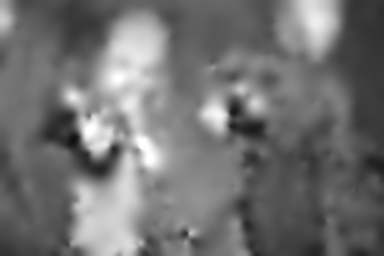}
  \caption{$\textrm{MSE} = 304.72$}
 \end{subfigure}
 \quad 
 \begin{subfigure}[t]{0.29\textwidth}
  \centering
  \includegraphics[width=\textwidth]{Figures_art/parrots_ran_D005_biharm}
  \caption{$\textrm{MSE} = 147.75$}
 \end{subfigure}
 \quad 
 \begin{subfigure}[t]{0.29\textwidth}
  \centering
  \includegraphics[width=\textwidth]{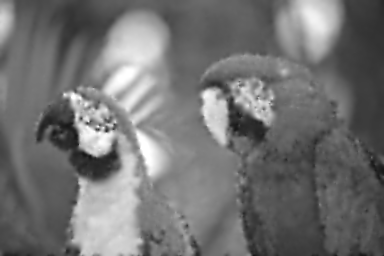}
  \caption{$\textrm{MSE} = 102.72$}
 \end{subfigure}
 
 \begin{subfigure}[t]{0.29\textwidth}
  \centering
  \includegraphics[
   width=\textwidth]{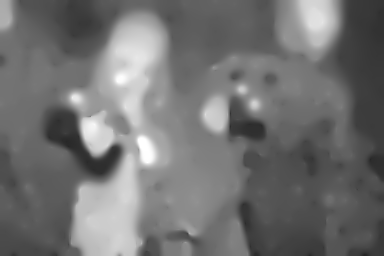}
  \caption{$\textrm{MSE} = 294.05$}
 \end{subfigure}
 \quad 
 \begin{subfigure}[t]{0.29\textwidth}
  \centering
  \includegraphics[
   width=\textwidth]{Figures_art/parrots_ran_D005_eed_lambda_1_2_sigma_2}
  \caption{$\textrm{MSE} = 137.04$}
 \end{subfigure}
 \quad 
 \begin{subfigure}[t]{0.29\textwidth}
  \centering
  \includegraphics[
   width=\textwidth]{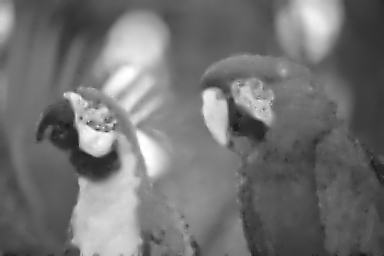}
  \caption{$\textrm{MSE} = 92.75$}
 \end{subfigure}
 \caption[Diffusion inpainting of ``parrots'' with random masks]{
  Inpainting of
  ``parrots'' for random masks of densities 1 \%, 5 \%, and 10 \% 
  (left to right).
  Top row: Harmonic inpainting. 
  Middle row: Biharmonic inpainting. 
  Bottom row: Inpainting with EED. 
  Parameters are from left to right 
  $\lambda=0.2$ and $\sigma=1.8$, $\lambda=1.2$ and $\sigma=2.0$, and
  $\lambda=1.7$ and $\sigma=2.0$.}
\end{figure}

\begin{figure}[htb]
 \captionsetup[subfigure]{justification=centering}
 \centering
 \begin{subfigure}[t]{0.29\textwidth}
  \centering
  \includegraphics[width=\textwidth]{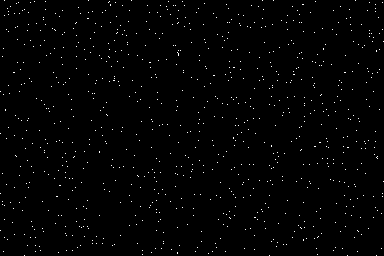}
  \caption{1 \% mask}
 \end{subfigure}
 \quad 
 \begin{subfigure}[t]{0.29\textwidth}
  \centering
  \includegraphics[width=\textwidth]{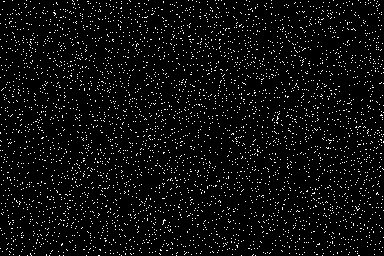}
  \caption{5 \% mask}
 \end{subfigure}
 \quad 
 \begin{subfigure}[t]{0.29\textwidth}
  \centering
  \includegraphics[width=\textwidth]{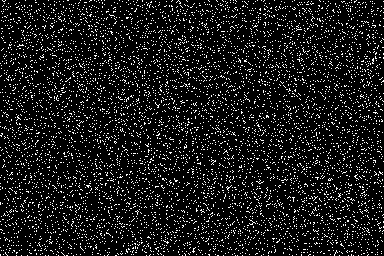}
  \caption{10 \% mask}
 \end{subfigure}
 
 \begin{subfigure}[t]{0.29\textwidth}
  \centering
  \includegraphics[width=\textwidth]{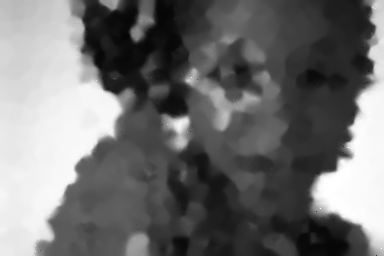}
  \caption{$\textrm{MSE} = 662.11$}
 \end{subfigure}
 \quad 
 \begin{subfigure}[t]{0.29\textwidth}
  \centering
  \includegraphics[width=\textwidth]{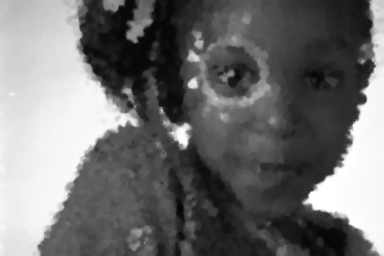}
  \caption{$\textrm{MSE} = 271.68$}
 \end{subfigure}
 \quad 
 \begin{subfigure}[t]{0.29\textwidth}
  \centering
  \includegraphics[width=\textwidth]{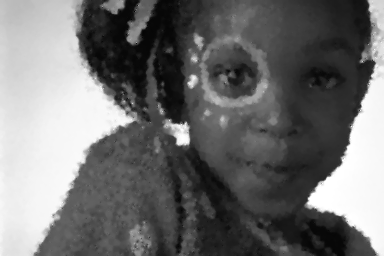}
  \caption{$\textrm{MSE} = 175.94$}
 \end{subfigure}
 
  \begin{subfigure}[t]{0.29\textwidth}
  \centering
  \includegraphics[width=\textwidth]{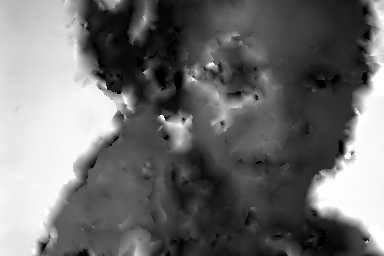}
  \caption{$\textrm{MSE} = 735.35$}
 \end{subfigure}
 \quad 
 \begin{subfigure}[t]{0.29\textwidth}
  \centering
  \includegraphics[width=\textwidth]{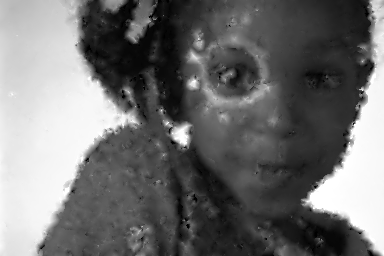}
  \caption{$\textrm{MSE} = 273.61$}
 \end{subfigure}
 \quad 
 \begin{subfigure}[t]{0.29\textwidth}
  \centering
  \includegraphics[width=\textwidth]{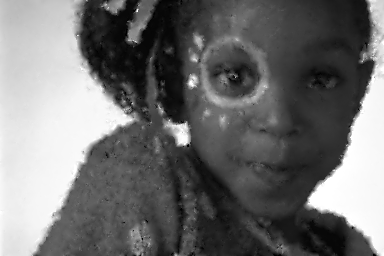}
  \caption{$\textrm{MSE} = 184.10$}
 \end{subfigure}
 \caption[SPH inpainting of ``girl'' with random masks]{
  Inpainting of 
  ``girl''  for random masks of different densities.
  Top row: Masks with densities of 1 \%, 5 \%, and 10 \%. 
  Middle row: Zero order consistency SPH inpainting with isotropic 
  Gaussian kernel. 
  Bottom row: First order consistency SPH inpainting with isotropic 
  Gaussian kernel.}
\end{figure}

\begin{figure}[htb]
 \captionsetup[subfigure]{justification=centering}
 \centering
 \begin{subfigure}[t]{0.29\textwidth}
  \centering
  \includegraphics[width=\textwidth]{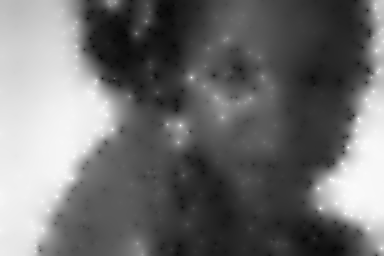}
  \caption{$\textrm{MSE} = 710.72$}
 \end{subfigure}
 \quad 
 \begin{subfigure}[t]{0.29\textwidth}
  \centering
  \includegraphics[width=\textwidth]{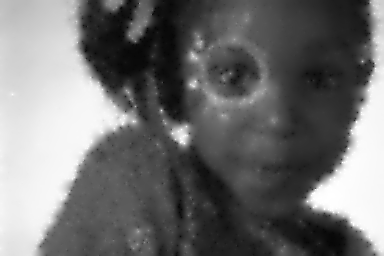}
  \caption{$\textrm{MSE} = 265.84$}
 \end{subfigure}
 \quad 
 \begin{subfigure}[t]{0.29\textwidth}
  \centering
  \includegraphics[width=\textwidth]{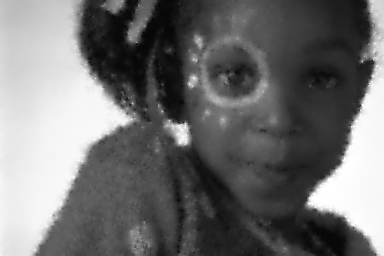}
  \caption{$\textrm{MSE} = 177.32$}
 \end{subfigure}
 
 \begin{subfigure}[t]{0.29\textwidth}
  \centering
  \includegraphics[width=\textwidth]{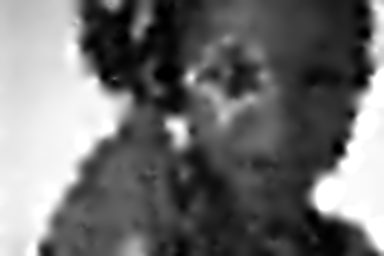}
  \caption{$\textrm{MSE} = 612.65$}
 \end{subfigure}
 \quad 
 \begin{subfigure}[t]{0.29\textwidth}
  \centering
  \includegraphics[width=\textwidth]{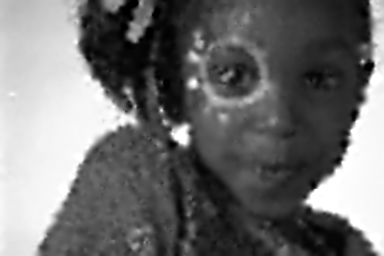}
  \caption{$\textrm{MSE} = 233.97$}
 \end{subfigure}
 \quad 
 \begin{subfigure}[t]{0.29\textwidth}
  \centering
  \includegraphics[width=\textwidth]{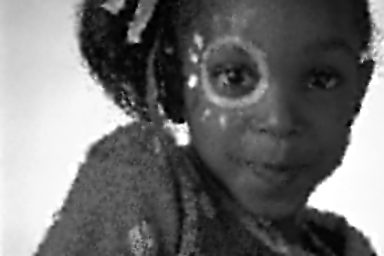}
  \caption{$\textrm{MSE} = 159.95$}
 \end{subfigure}
 
 \begin{subfigure}[t]{0.29\textwidth}
  \centering
  \includegraphics[
   width=\textwidth]{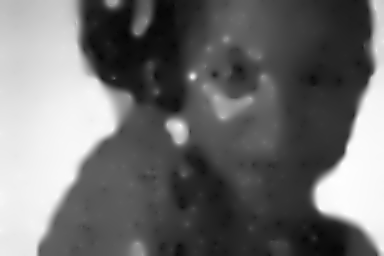}
  \caption{$\textrm{MSE} = 492.03$}
 \end{subfigure}
 \quad 
 \begin{subfigure}[t]{0.29\textwidth}
  \centering
  \includegraphics[
   width=\textwidth]{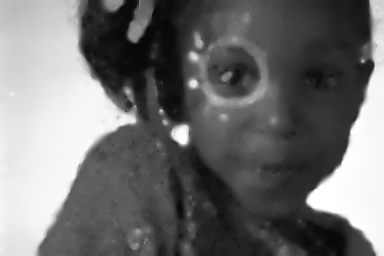}
  \caption{$\textrm{MSE} = 182.14$}
 \end{subfigure}
 \quad 
 \begin{subfigure}[t]{0.29\textwidth}
  \centering
  \includegraphics[
   width=\textwidth]{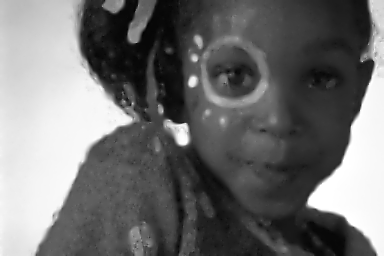}
  \caption{$\textrm{MSE} = 126.27$}
 \end{subfigure}
 \caption[Diffusion inpainting of ``girl'' with random masks]{
  Inpainting of
  ``girl'' for random masks of densities 1 \%, 5 \%, and 10 \% (left to right).
  Top row: Harmonic inpainting. 
  Middle row: Biharmonic inpainting. 
  Bottom row: Inpainting with EED. 
  Parameters are from left to right 
  $\lambda=0.8$ and $\sigma=2.0$, $\lambda=1.4$ and $\sigma=2.0$, and
  $\lambda=1.0$ and $\sigma=2.0$.}
\end{figure}

\begin{figure}[htb]
 \captionsetup[subfigure]{justification=centering}
 \centering
 \begin{subfigure}[t]{0.29\textwidth}
  \centering
  \includegraphics[width=\textwidth]{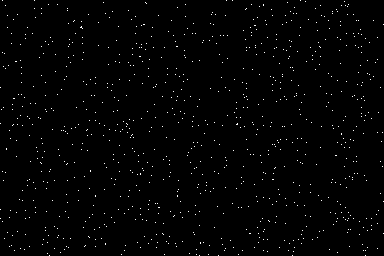}
  \caption{1 \% mask}
 \end{subfigure}
 \quad 
 \begin{subfigure}[t]{0.29\textwidth}
  \centering
  \includegraphics[width=\textwidth]{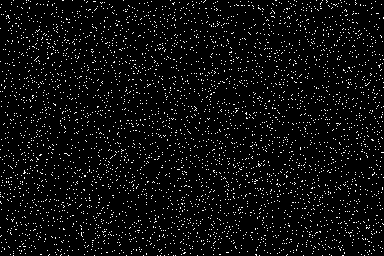}
  \caption{5 \% mask}
 \end{subfigure}
 \quad 
 \begin{subfigure}[t]{0.29\textwidth}
  \centering
  \includegraphics[width=\textwidth]{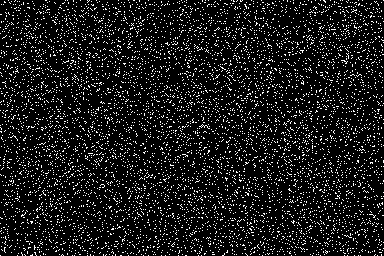}
  \caption{10 \% mask}
 \end{subfigure}
 
 \begin{subfigure}[t]{0.29\textwidth}
  \centering
  \includegraphics[width=\textwidth]{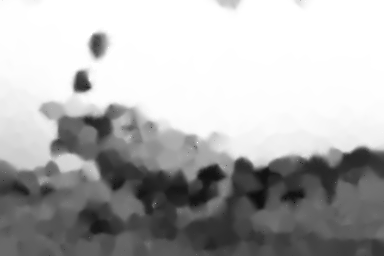}
  \caption{$\textrm{MSE} = 717.64$}
 \end{subfigure}
 \quad 
 \begin{subfigure}[t]{0.29\textwidth}
  \centering
  \includegraphics[width=\textwidth]{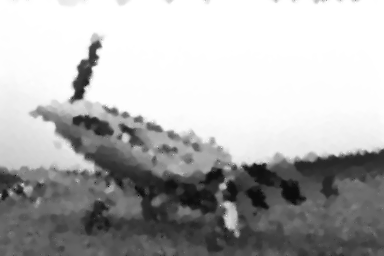}
  \caption{$\textrm{MSE} = 313.08$}
 \end{subfigure}
 \quad 
 \begin{subfigure}[t]{0.29\textwidth}
  \centering
  \includegraphics[width=\textwidth]{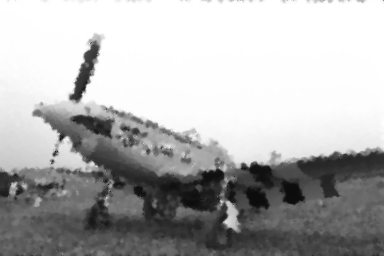}
  \caption{$\textrm{MSE} = 209.96$}
 \end{subfigure}
 
  \begin{subfigure}[t]{0.29\textwidth}
  \centering
  \includegraphics[width=\textwidth]{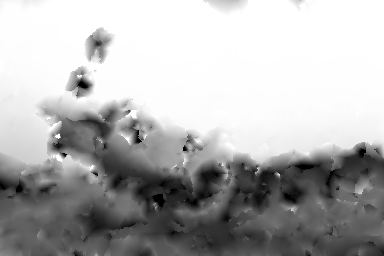}
  \caption{$\textrm{MSE} = 845.86$}
 \end{subfigure}
 \quad 
 \begin{subfigure}[t]{0.29\textwidth}
  \centering
  \includegraphics[width=\textwidth]{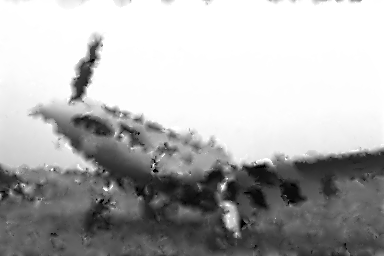}
  \caption{$\textrm{MSE} = 341.57$}
 \end{subfigure}
 \quad 
 \begin{subfigure}[t]{0.29\textwidth}
  \centering
  \includegraphics[width=\textwidth]{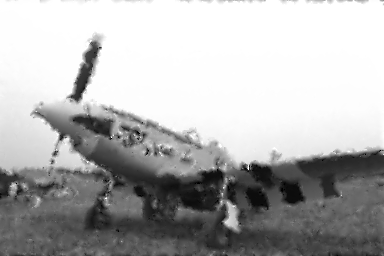}
  \caption{$\textrm{MSE} = 213.02$}
 \end{subfigure}
 \caption[SPH inpainting of ``plane'' with random masks]{
  Inpainting of 
  ``plane''  for random masks of different densities.
  Top row: Masks with densities of 1 \%, 5 \%, and 10 \%. 
  Middle row: Zero order consistency SPH inpainting with isotropic 
  Gaussian kernel. 
  Bottom row: First order consistency SPH inpainting with isotropic 
  Gaussian kernel.}
\end{figure}

\begin{figure}[htb]
 \captionsetup[subfigure]{justification=centering}
 \centering
 \begin{subfigure}[t]{0.29\textwidth}
  \centering
  \includegraphics[width=\textwidth]{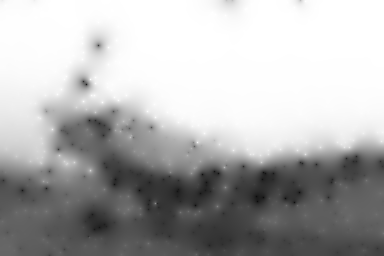}
  \caption{$\textrm{MSE} = 705.55$}
 \end{subfigure}
 \quad 
 \begin{subfigure}[t]{0.29\textwidth}
  \centering
  \includegraphics[width=\textwidth]{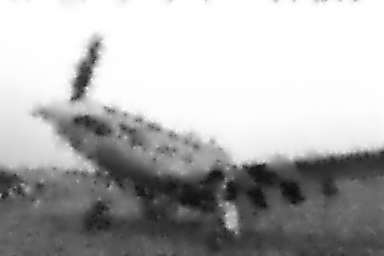}
  \caption{$\textrm{MSE} = 309.14$}
 \end{subfigure}
 \quad 
 \begin{subfigure}[t]{0.29\textwidth}
  \centering
  \includegraphics[width=\textwidth]{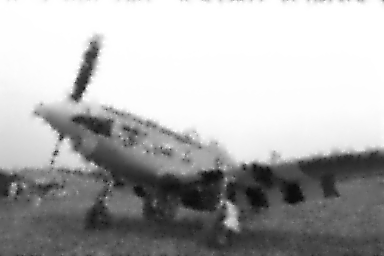}
  \caption{$\textrm{MSE} = 209.28$}
 \end{subfigure}
 
 \begin{subfigure}[t]{0.29\textwidth}
  \centering
  \includegraphics[width=\textwidth]{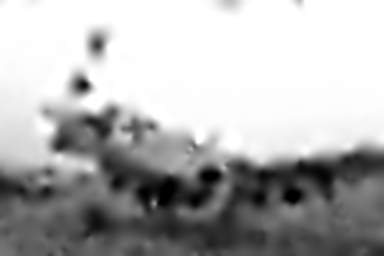}
  \caption{$\textrm{MSE} = 718.31$}
 \end{subfigure}
 \quad 
 \begin{subfigure}[t]{0.29\textwidth}
  \centering
  \includegraphics[width=\textwidth]{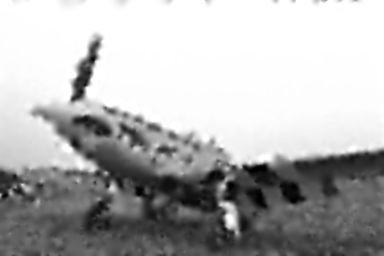}
  \caption{$\textrm{MSE} = 315.87$}
 \end{subfigure}
 \quad 
 \begin{subfigure}[t]{0.29\textwidth}
  \centering
  \includegraphics[width=\textwidth]{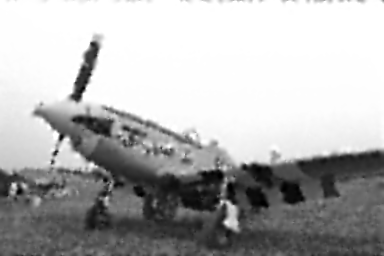}
  \caption{$\textrm{MSE} = 182.87$}
 \end{subfigure}
 
 \begin{subfigure}[t]{0.29\textwidth}
  \centering
  \includegraphics[
   width=\textwidth]{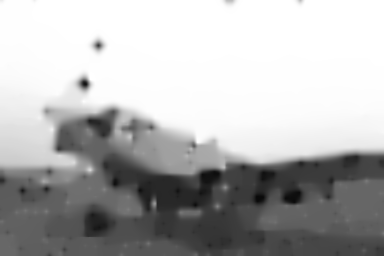}
  \caption{$\textrm{MSE} = 640.09$}
 \end{subfigure}
 \quad 
 \begin{subfigure}[t]{0.29\textwidth}
  \centering
  \includegraphics[
   width=\textwidth]{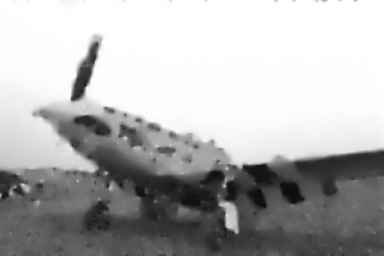}
  \caption{$\textrm{MSE} = 255.48$}
 \end{subfigure}
 \quad 
 \begin{subfigure}[t]{0.29\textwidth}
  \centering
  \includegraphics[
   width=\textwidth]{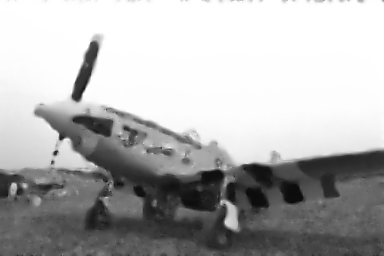}
  \caption{$\textrm{MSE} = 162.08$}
 \end{subfigure}
 \caption[Diffusion inpainting of ``plane'' with random masks]{
  Inpainting of
  ``plane'' for random masks of densities 1 \%, 5 \%, and 10 \% 
  (left to right).
  Top row: Harmonic inpainting. 
  Middle row: Biharmonic inpainting. 
  Bottom row: Inpainting with EED. 
  Parameters are from left to right 
  $\lambda=0.1$ and $\sigma=0.4$, $\lambda=1.9$ and $\sigma=0.6$, and
  $\lambda=1.2$ and $\sigma=2.0$.}
\end{figure}

\begin{figure}[htb]
 \captionsetup[subfigure]{justification=centering}
 \centering
 \begin{subfigure}[t]{0.29\textwidth}
  \centering
  \includegraphics[width=\textwidth]{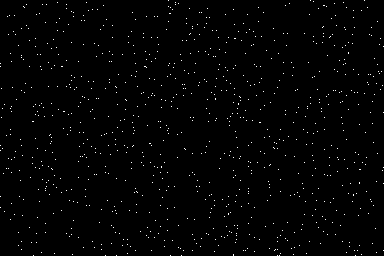}
  \caption{1 \% mask}
 \end{subfigure}
 \quad 
 \begin{subfigure}[t]{0.29\textwidth}
  \centering
  \includegraphics[width=\textwidth]{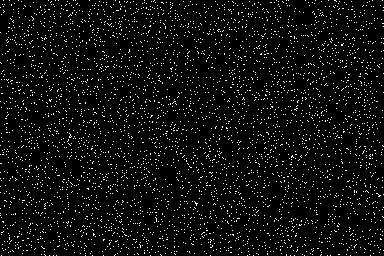}
  \caption{5 \% mask}
 \end{subfigure}
 \quad 
 \begin{subfigure}[t]{0.29\textwidth}
  \centering
  \includegraphics[width=\textwidth]{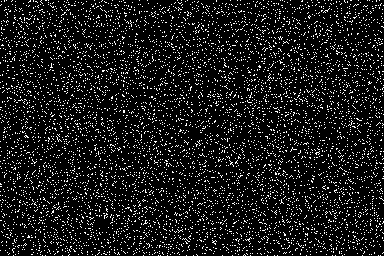}
  \caption{10 \% mask}
 \end{subfigure}
 
 \begin{subfigure}[t]{0.29\textwidth}
  \centering
  \includegraphics[width=\textwidth]{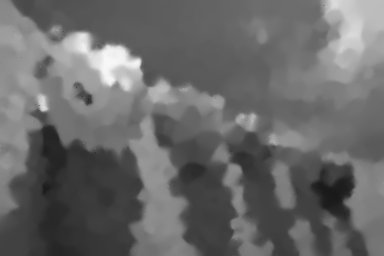}
  \caption{$\textrm{MSE} = 353.40$}
 \end{subfigure}
 \quad 
 \begin{subfigure}[t]{0.29\textwidth}
  \centering
  \includegraphics[width=\textwidth]{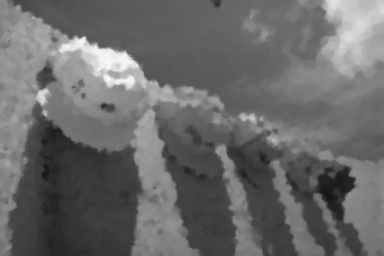}
  \caption{$\textrm{MSE} = 171.43$}
 \end{subfigure}
 \quad 
 \begin{subfigure}[t]{0.29\textwidth}
  \centering
  \includegraphics[width=\textwidth]{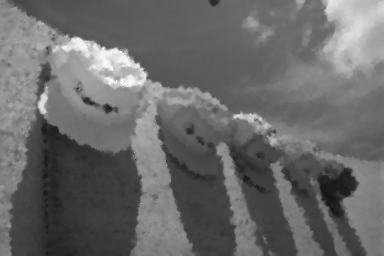}
  \caption{$\textrm{MSE} = 123.37$}
 \end{subfigure}
 
  \begin{subfigure}[t]{0.29\textwidth}
  \centering
  \includegraphics[width=\textwidth]{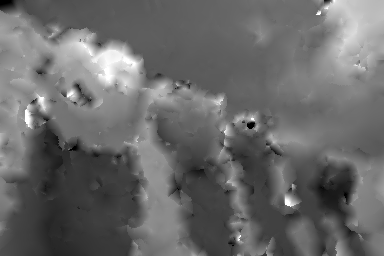}
  \caption{$\textrm{MSE} = 470.24$}
 \end{subfigure}
 \quad 
 \begin{subfigure}[t]{0.29\textwidth}
  \centering
  \includegraphics[width=\textwidth]{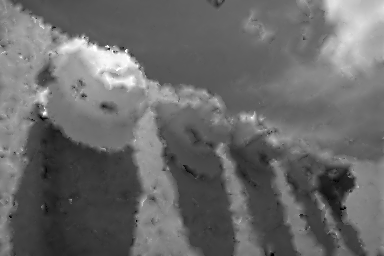}
  \caption{$\textrm{MSE} = 191.77$}
 \end{subfigure}
 \quad 
 \begin{subfigure}[t]{0.29\textwidth}
  \centering
  \includegraphics[width=\textwidth]{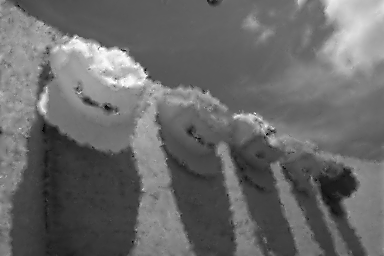}
  \caption{$\textrm{MSE} = 128.98$}
 \end{subfigure}
 \caption[SPH inpainting of ``hats'' with random masks]{
  Inpainting of 
  ``hats''  for random masks of different densities.
  Top row: Masks with densities of 1 \%, 5 \%, and 10 \%. 
  Middle row: Zero order consistency SPH inpainting with isotropic 
  Gaussian kernel. 
  Bottom row: First order consistency SPH inpainting with isotropic 
  Gaussian kernel.}
\end{figure}

\begin{figure}[htb]
 \captionsetup[subfigure]{justification=centering}
 \centering
 \begin{subfigure}[t]{0.29\textwidth}
  \centering
  \includegraphics[width=\textwidth]{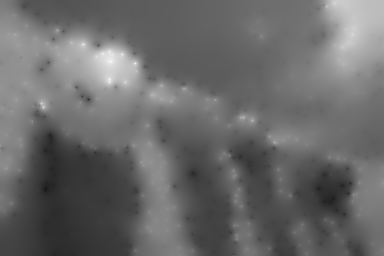}
  \caption{$\textrm{MSE} = 349.80$}
 \end{subfigure}
 \quad 
 \begin{subfigure}[t]{0.29\textwidth}
  \centering
  \includegraphics[width=\textwidth]{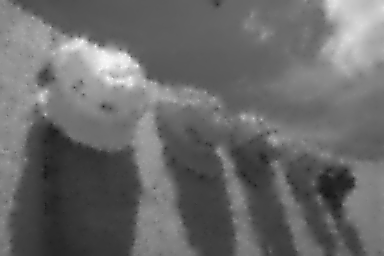}
  \caption{$\textrm{MSE} = 175.68$}
 \end{subfigure}
 \quad 
 \begin{subfigure}[t]{0.29\textwidth}
  \centering
  \includegraphics[width=\textwidth]{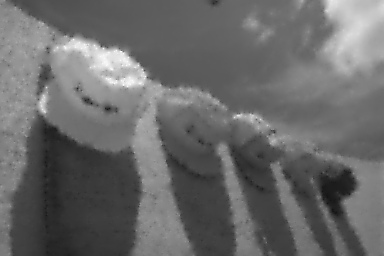}
  \caption{$\textrm{MSE} = 119.83$}
 \end{subfigure}
 
 \begin{subfigure}[t]{0.29\textwidth}
  \centering
  \includegraphics[width=\textwidth]{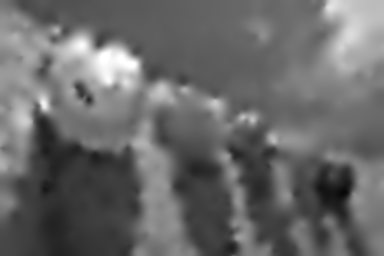}
  \caption{$\textrm{MSE} = 344.37$}
 \end{subfigure}
 \quad 
 \begin{subfigure}[t]{0.29\textwidth}
  \centering
  \includegraphics[width=\textwidth]{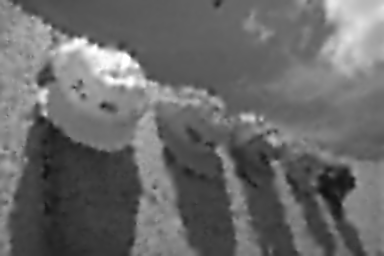}
  \caption{$\textrm{MSE} = 165.10$}
 \end{subfigure}
 \quad 
 \begin{subfigure}[t]{0.29\textwidth}
  \centering
  \includegraphics[width=\textwidth]{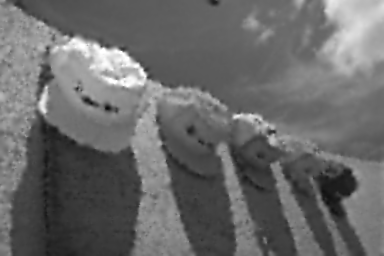}
  \caption{$\textrm{MSE} = 112.54$}
 \end{subfigure}
 
 \begin{subfigure}[t]{0.29\textwidth}
  \centering
  \includegraphics[
   width=\textwidth]{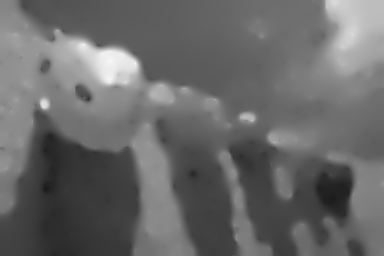}
  \caption{$\textrm{MSE} = 289.27$}
 \end{subfigure}
 \quad 
 \begin{subfigure}[t]{0.29\textwidth}
  \centering
  \includegraphics[
   width=\textwidth]{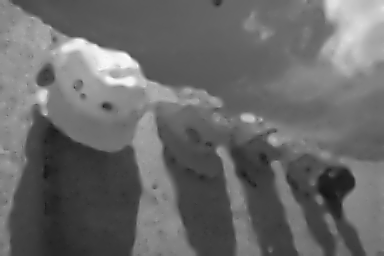}
  \caption{$\textrm{MSE} = 142.61$}
 \end{subfigure}
 \quad 
 \begin{subfigure}[t]{0.29\textwidth}
  \centering
  \includegraphics[
   width=\textwidth]{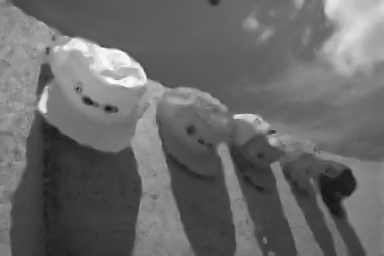}
  \caption{$\textrm{MSE} = 91.07$}
 \end{subfigure}
 \caption[Diffusion inpainting of ``hats'' with random masks]{
  Inpainting of
  ``hats'' for random masks of densities 1 \%, 5 \%, and 10 \% (left to right).
  Top row: Harmonic inpainting. 
  Middle row: Biharmonic inpainting. 
  Bottom row: Inpainting with EED. 
  Parameters are from left to right 
  $\lambda=0.6$ and $\sigma=2.0$, $\lambda=0.7$ and $\sigma=2.0$, and
  $\lambda=0.6$ and $\sigma=2.0$.}
\end{figure}

Once again, we observe that first order consistency SPH 
inpainting is prone to producing artifacts in the form of under- and overshoots.
These are in particular visible for lower densities. Zero order consistency
SPH inpainting is more stable and achieves better MSEs than harmonic inpainting. 
Biharmonic inpainting is more suited to the distribution of mask points in most 
cases and EED can once more benefit from its nonlinear and anisotropic nature.

\section{Scratch and Text Removal}

\begin{figure}[htb]
 \captionsetup[subfigure]{justification=centering}
 \centering
 \begin{subfigure}[t]{0.29\textwidth}
  \centering
  \includegraphics[width=\textwidth]{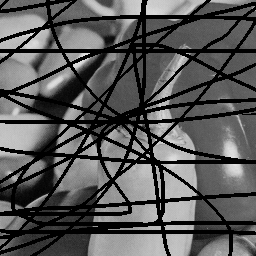}
 \end{subfigure}
 \quad
 \begin{subfigure}[t]{0.29\textwidth}
  \centering
  \includegraphics[width=\textwidth]{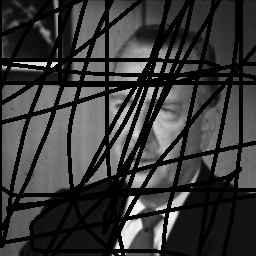}
 \end{subfigure}
 \quad
 \begin{subfigure}[t]{0.29\textwidth}
  \centering
  \includegraphics[width=\textwidth]{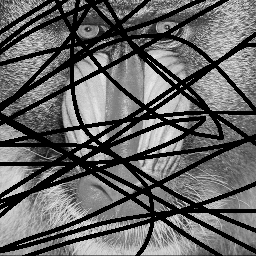}
 \end{subfigure}
 \vspace*{3ex}

 \begin{subfigure}[t]{0.29\textwidth}
  \centering
  \includegraphics[width=\textwidth]{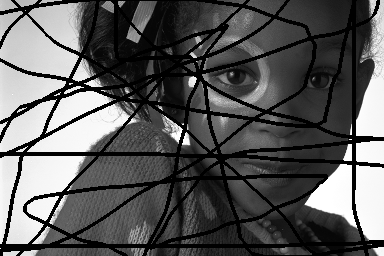}
 \end{subfigure}
 \quad
 \begin{subfigure}[t]{0.29\textwidth}
  \centering
  \includegraphics[width=\textwidth]{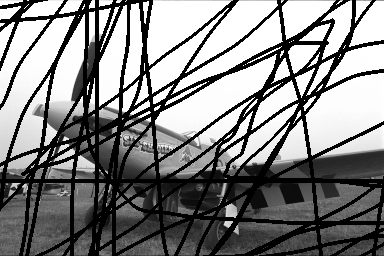}
 \end{subfigure}
 \quad
 \begin{subfigure}[t]{0.29\textwidth}
  \centering
  \includegraphics[width=\textwidth]{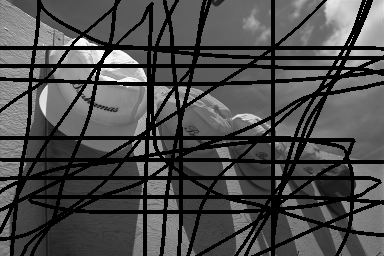}
 \end{subfigure}
 \caption{Images damaged by scratches.}
  \label{fig:images_scratch}
\end{figure}

\begin{figure}[htb]
 \captionsetup[subfigure]{justification=centering}
 \centering
 \begin{subfigure}[t]{0.29\textwidth}
  \centering
  \includegraphics[width=\textwidth]{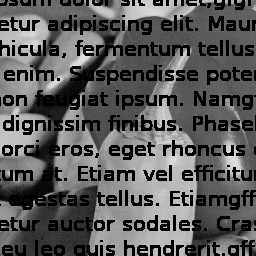}
 \end{subfigure}
 \quad
 \begin{subfigure}[t]{0.29\textwidth}
  \centering
  \includegraphics[width=\textwidth]{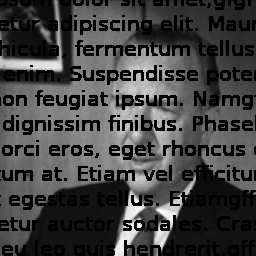}
 \end{subfigure}
 \quad
 \begin{subfigure}[t]{0.29\textwidth}
  \centering
  \includegraphics[width=\textwidth]{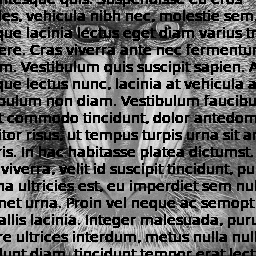}
 \end{subfigure}
 \vspace*{3ex}
 
 \begin{subfigure}[t]{0.29\textwidth}
  \centering
  \includegraphics[width=\textwidth]{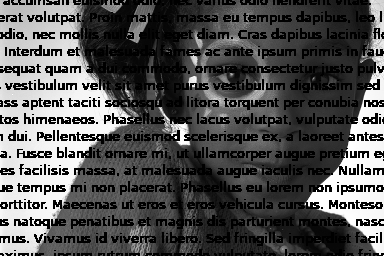}
 \end{subfigure}
 \quad
 \begin{subfigure}[t]{0.29\textwidth}
  \centering
  \includegraphics[width=\textwidth]{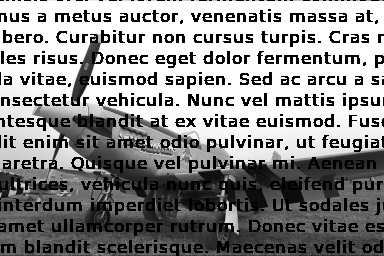}
 \end{subfigure}
 \quad
 \begin{subfigure}[t]{0.29\textwidth}
  \centering
  \includegraphics[width=\textwidth]{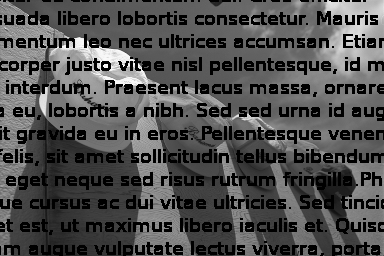}
 \end{subfigure}
 \caption{Images overlaid with text.}
  \label{fig:images_text}
\end{figure}

As classical applications of inpainting, we consider the 
repair of scratches and removal of overlaid text. Images with scratches are 
presented in \cref{fig:images_scratch} whereas \cref{fig:images_text} shows
the images overlaid with text. As competitors, we consider once more harmonic
inpainting, biharmonic inpainting, inpainting with EED, and the exemplar-based
inpainting approach by Criminisi et al. with disc-shaped patches.

\begin{figure}[htb]
 \captionsetup[subfigure]{justification=centering}
 \centering
 \begin{subfigure}[t]{0.29\textwidth}
  \centering
  \includegraphics[width=\textwidth]{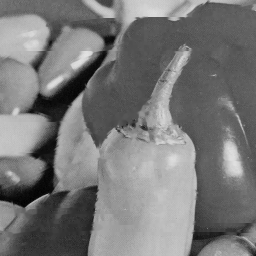}
  \caption{Zero order SPH\\
   $\textrm{MSE} = 30.84$}
 \end{subfigure}
  \quad 
 \begin{subfigure}[t]{0.29\textwidth}
  \centering
  \includegraphics[width=\textwidth]{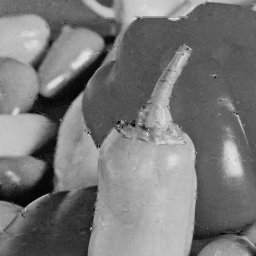}
  \caption{First order SPH\\
   $\textrm{MSE} = 36.90$}
 \end{subfigure}
   \quad 
 \begin{subfigure}[t]{0.29\textwidth}
  \centering
  \includegraphics[width=\textwidth]{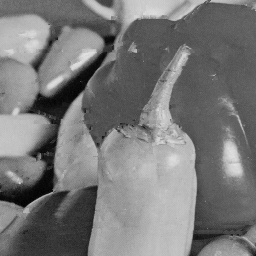}
  \caption{Exemplar-based\\
   $\textrm{MSE} = 59.86$}
 \end{subfigure}

 \begin{subfigure}[t]{0.29\textwidth}
  \centering
  \includegraphics[width=\textwidth]{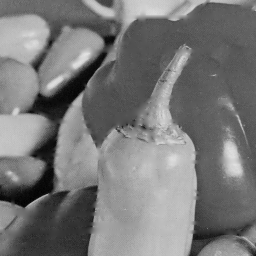}
  \caption{Harmonic\\
   $\textrm{MSE} = 27.76$}
 \end{subfigure}
 \quad 
 \begin{subfigure}[t]{0.29\textwidth}
  \centering
  \includegraphics[width=\textwidth]{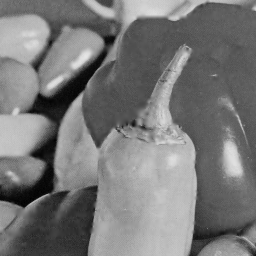}
  \caption{Biharmonic\\
   $\textrm{MSE} = 21.12$}
 \end{subfigure}
   \quad 
 \begin{subfigure}[t]{0.29\textwidth}
  \centering
  \includegraphics[
   width=\textwidth]{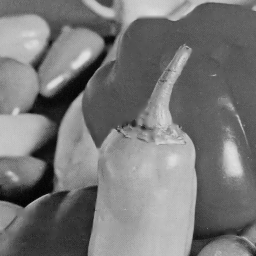}
  \caption{EED\\
   $\textrm{MSE} = 16.90$}
 \end{subfigure}
 \caption[Inpainting of ``peppers'' for scratch removal]{
  Inpainting of 
  damaged image ``peppers'' with different inpainting methods. For SPH 
  inpainting, we used an isotropic Gaussian kernel. 
  Parameters for EED are
  $\lambda=1.1$ and $\sigma=1.1$.}
\end{figure}

\begin{figure}[htb]
 \captionsetup[subfigure]{justification=centering}
 \centering
 \begin{subfigure}[t]{0.29\textwidth}
  \centering
  \includegraphics[width=\textwidth]{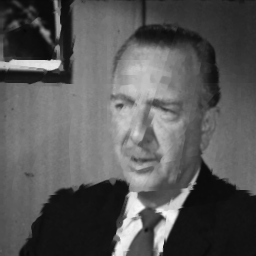}
  \caption{Zero order SPH\\
   $\textrm{MSE} = 83.01$}
 \end{subfigure}
  \quad 
 \begin{subfigure}[t]{0.29\textwidth}
  \centering
  \includegraphics[width=\textwidth]{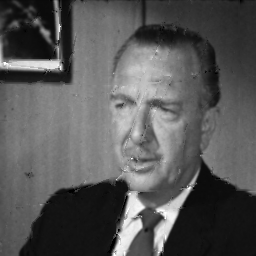}
  \caption{First order SPH\\
   $\textrm{MSE} = 60.14$}
 \end{subfigure}
   \quad 
 \begin{subfigure}[t]{0.29\textwidth}
  \centering
  \includegraphics[width=\textwidth]{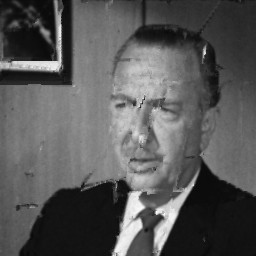}
  \caption{Exemplar-based\\
   $\textrm{MSE} = 132.18$}
 \end{subfigure}

 \begin{subfigure}[t]{0.29\textwidth}
  \centering
  \includegraphics[width=\textwidth]{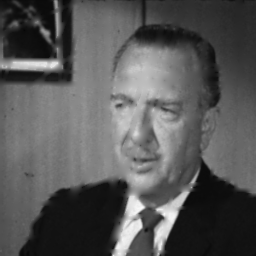}
  \caption{Harmonic\\
   $\textrm{MSE} = 66.73$}
 \end{subfigure}
 \quad 
 \begin{subfigure}[t]{0.29\textwidth}
  \centering
  \includegraphics[width=\textwidth]{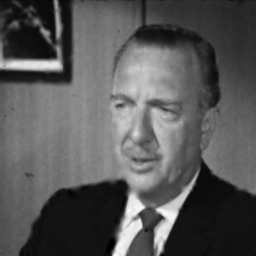}
  \caption{Biharmonic\\
   $\textrm{MSE} = 27.55$}
 \end{subfigure}
   \quad 
 \begin{subfigure}[t]{0.29\textwidth}
  \centering
  \includegraphics[
   width=\textwidth]{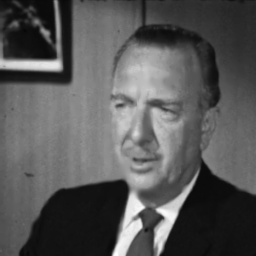}
  \caption{EED\\
   $\textrm{MSE} = 15.63$}
 \end{subfigure}
 \caption[Inpainting of ``walter'' for scratch removal]{
  Inpainting of 
  damaged image ``walter'' with different inpainting methods. For SPH 
  inpainting, we used an isotropic Gaussian kernel. 
  Parameters for EED are
  $\lambda=0.1$ and $\sigma=0.9$.}
\end{figure}

\begin{figure}[htb]
 \captionsetup[subfigure]{justification=centering}
 \centering
 \begin{subfigure}[t]{0.29\textwidth}
  \centering
  \includegraphics[width=\textwidth]{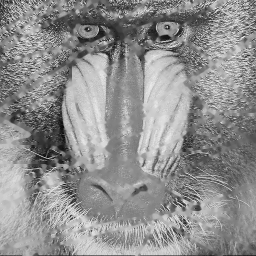}
  \caption{Zero order SPH\\
   $\textrm{MSE} = 283.53$}
 \end{subfigure}
  \quad 
 \begin{subfigure}[t]{0.29\textwidth}
  \centering
  \includegraphics[width=\textwidth]{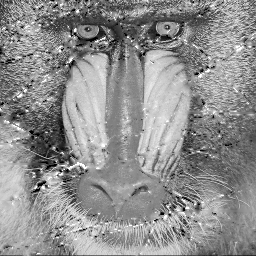}
  \caption{First order SPH\\
   $\textrm{MSE} = 765.51$}
 \end{subfigure}
   \quad 
 \begin{subfigure}[t]{0.29\textwidth}
  \centering
  \includegraphics[width=\textwidth]{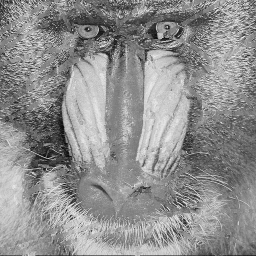}
  \caption{Exemplar-based\\
   $\textrm{MSE} = 363.47$}
 \end{subfigure}
 
 \begin{subfigure}[t]{0.29\textwidth}
  \centering
  \includegraphics[width=\textwidth]{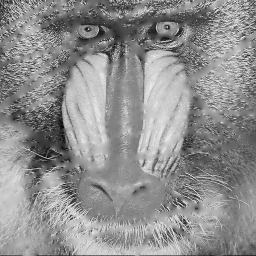}
  \caption{Harmonic\\
   $\textrm{MSE} = 232.43$}
 \end{subfigure}
 \quad 
 \begin{subfigure}[t]{0.29\textwidth}
  \centering
  \includegraphics[width=\textwidth]{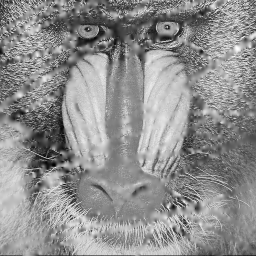}
  \caption{Biharmonic\\
   $\textrm{MSE} = 335.96$}
 \end{subfigure}
   \quad 
 \begin{subfigure}[t]{0.29\textwidth}
  \centering
  \includegraphics[
   width=\textwidth]{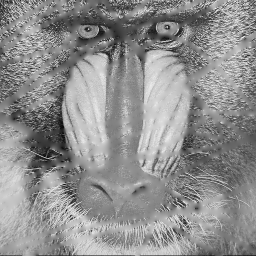}
  \caption{EED\\
   $\textrm{MSE} = 231.40$}
 \end{subfigure}
 \caption[Inpainting of ``baboon'' for scratch removal]{
  Inpainting of 
  damaged image ``baboon'' with different inpainting methods. For SPH 
  inpainting, we used an isotropic Gaussian kernel. 
  Parameters for EED are
  $\lambda=6.0$ and $\sigma=3.0$.}
\end{figure}

\begin{figure}[htb]
 \captionsetup[subfigure]{justification=centering}
 \centering
 \begin{subfigure}[t]{0.29\textwidth}
  \centering
  \includegraphics[width=\textwidth]{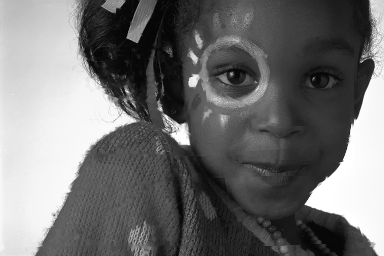}
  \caption{Zero order SPH\\
   $\textrm{MSE} = 40.61$}
 \end{subfigure}
  \quad 
 \begin{subfigure}[t]{0.29\textwidth}
  \centering
  \includegraphics[width=\textwidth]{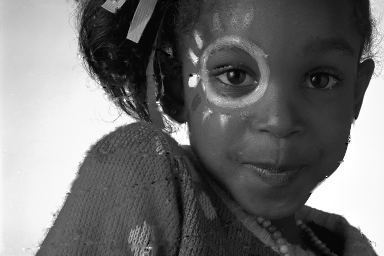}
  \caption{First order SPH\\
   $\textrm{MSE} = 51.99$}
 \end{subfigure}
   \quad 
 \begin{subfigure}[t]{0.29\textwidth}
  \centering
  \includegraphics[width=\textwidth]{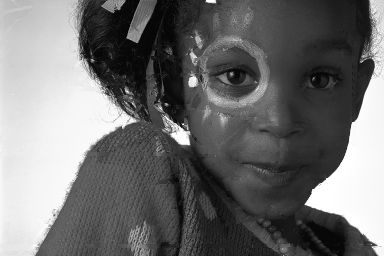}
  \caption{Exemplar-based\\
   $\textrm{MSE} = 86.43$}
 \end{subfigure}
 
 \begin{subfigure}[t]{0.29\textwidth}
  \centering
  \includegraphics[width=\textwidth]{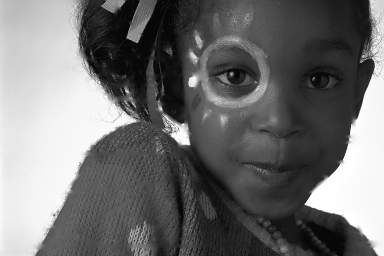}
  \caption{Harmonic\\
   $\textrm{MSE} = 34.97$}
 \end{subfigure}
 \quad 
 \begin{subfigure}[t]{0.29\textwidth}
  \centering
  \includegraphics[width=\textwidth]{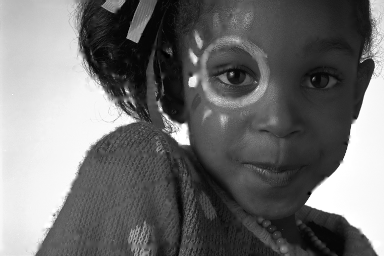}
  \caption{Biharmonic\\
   $\textrm{MSE} = 31.41$}
 \end{subfigure}
   \quad 
 \begin{subfigure}[t]{0.29\textwidth}
  \centering
  \includegraphics[
   width=\textwidth]{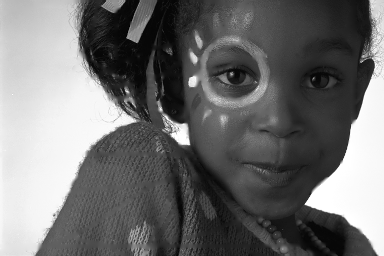}
  \caption{EED\\
   $\textrm{MSE} = 24.81$}
 \end{subfigure}
 \caption[Inpainting of ``girl'' for scratch removal]{
  Inpainting of 
  damaged image ``girl'' with different inpainting methods. For SPH 
  inpainting, we used an isotropic Gaussian kernel. 
  Parameters for EED are
  $\lambda=2.7$ and $\sigma=1.2$.}
\end{figure}

\begin{figure}[htb]
 \captionsetup[subfigure]{justification=centering}
 \centering
 \begin{subfigure}[t]{0.29\textwidth}
  \centering
  \includegraphics[width=\textwidth]{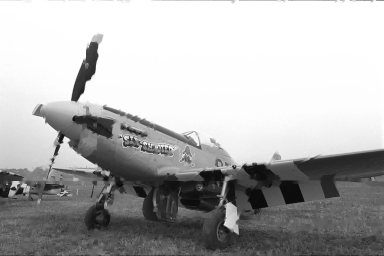}
  \caption{Zero order SPH\\
   $\textrm{MSE} = 92.47$}
 \end{subfigure}
  \quad 
 \begin{subfigure}[t]{0.29\textwidth}
  \centering
  \includegraphics[width=\textwidth]{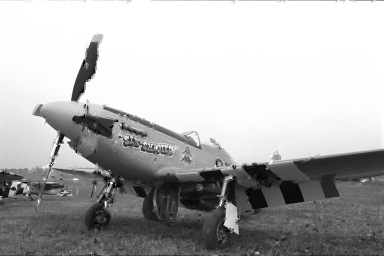}
  \caption{First order SPH\\
   $\textrm{MSE} = 121.49$}
 \end{subfigure}
   \quad 
 \begin{subfigure}[t]{0.29\textwidth}
  \centering
  \includegraphics[width=\textwidth]{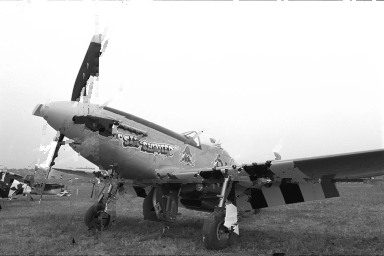}
  \caption{Exemplar-based\\
   $\textrm{MSE} = 190.01$}
 \end{subfigure}

 \begin{subfigure}[t]{0.29\textwidth}
  \centering
  \includegraphics[width=\textwidth]{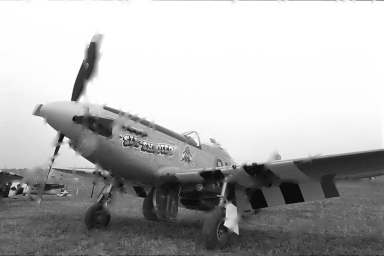}
  \caption{Harmonic\\
   $\textrm{MSE} = 78.81$}
 \end{subfigure}
 \quad 
 \begin{subfigure}[t]{0.29\textwidth}
  \centering
  \includegraphics[width=\textwidth]{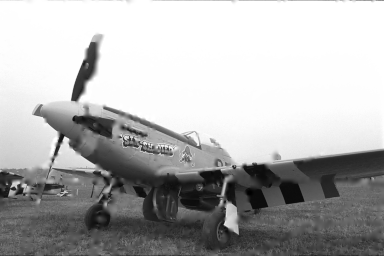}
  \caption{Biharmonic\\
   $\textrm{MSE} = 71.88$}
 \end{subfigure}
   \quad 
 \begin{subfigure}[t]{0.29\textwidth}
  \centering
  \includegraphics[
   width=\textwidth]{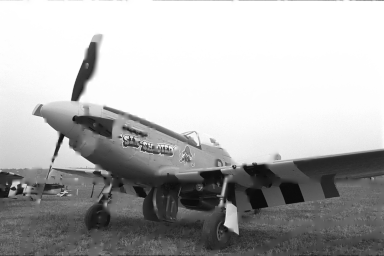}
  \caption{EED\\
   $\textrm{MSE} = 62.87$}
 \end{subfigure}
 \caption[Inpainting of ``plane'' for scratch removal]{
  Inpainting of 
  damaged image ``plane'' with different inpainting methods. For SPH 
  inpainting, we used an isotropic Gaussian kernel. 
  Parameters for EED are
  $\lambda=2.0$ and $\sigma=0.5$.}
\end{figure}

\begin{figure}[htb]
 \captionsetup[subfigure]{justification=centering}
 \centering
 \begin{subfigure}[t]{0.29\textwidth}
  \centering
  \includegraphics[width=\textwidth]{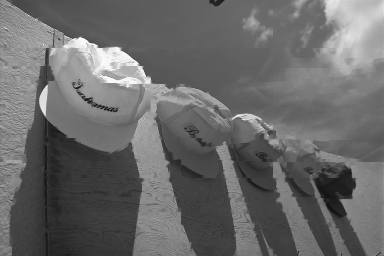}
  \caption{Zero order SPH\\
   $\textrm{MSE} = 47.63$}
 \end{subfigure}
  \quad 
 \begin{subfigure}[t]{0.29\textwidth}
  \centering
  \includegraphics[width=\textwidth]{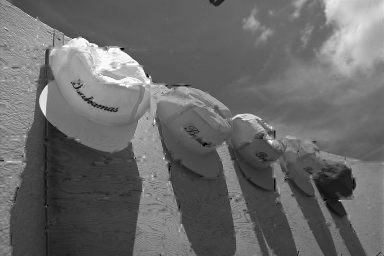}
  \caption{First order SPH\\
   $\textrm{MSE} = 56.53$}
 \end{subfigure}
   \quad 
 \begin{subfigure}[t]{0.29\textwidth}
  \centering
  \includegraphics[width=\textwidth]{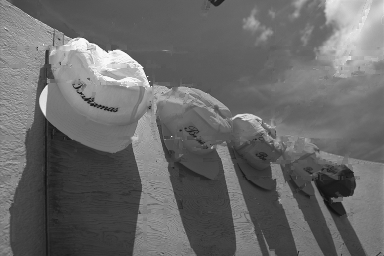}
  \caption{Exemplar-based\\
   $\textrm{MSE} = 66.15$}
 \end{subfigure}
 
 \begin{subfigure}[t]{0.29\textwidth}
  \centering
  \includegraphics[width=\textwidth]{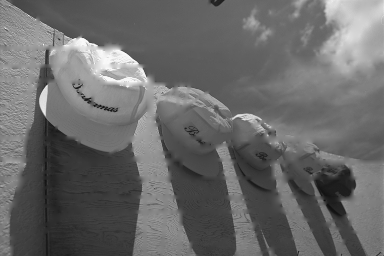}
  \caption{Harmonic\\
   $\textrm{MSE} = 36.97$}
 \end{subfigure}
 \quad 
 \begin{subfigure}[t]{0.29\textwidth}
  \centering
  \includegraphics[width=\textwidth]{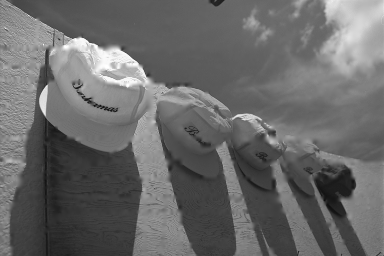}
  \caption{Biharmonic\\
   $\textrm{MSE} = 35.01$}
 \end{subfigure}
   \quad 
 \begin{subfigure}[t]{0.29\textwidth}
  \centering
  \includegraphics[
   width=\textwidth]{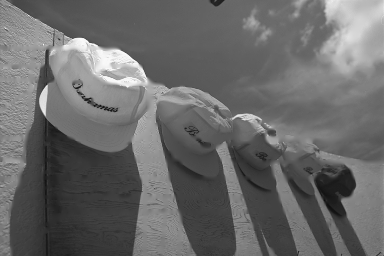}
  \caption{EED\\
   $\textrm{MSE} = 23.99$}
 \end{subfigure}
 \caption[Inpainting of ``hats'' for scratch removal]{
  Inpainting of 
  damaged image ``hats'' with different inpainting methods. For SPH 
  inpainting, we used an isotropic Gaussian kernel. 
  Parameters for EED are
  $\lambda=0.4$ and $\sigma=2.0$.}
\end{figure}

We observe mainly the same results as for the examples ``trui''
and ``parrots'' in the main article: Zero order consistency SPH inpainting
performs better than first order consistency. Both perform better than the 
exemplar-based approach, but cannot reach the same quality as the 
diffusion-based methods. Even when the MSE of the SPH inpainting comes close to 
the MSE of diffusion-based methods, the former often produces some unpleasant
artifacts. However, we remind the reader that the task of repairing such 
damages comes not naturally for SPH inpainting.

\begin{figure}[htb]
 \captionsetup[subfigure]{justification=centering}
 \centering
 \begin{subfigure}[t]{0.29\textwidth}
  \centering
  \includegraphics[width=\textwidth]{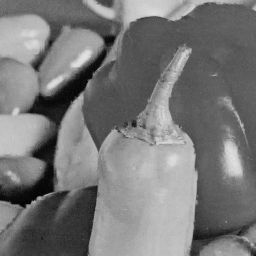}
  \caption{Zero order SPH\\
   $\textrm{MSE} = 24.56$}
 \end{subfigure}
  \quad 
 \begin{subfigure}[t]{0.29\textwidth}
  \centering
  \includegraphics[width=\textwidth]{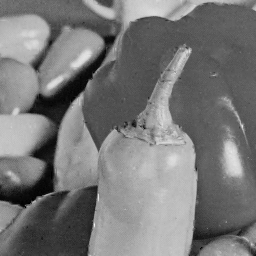}
  \caption{First order SPH\\
   $\textrm{MSE} = 23.76$}
 \end{subfigure}
   \quad 
 \begin{subfigure}[t]{0.29\textwidth}
  \centering
  \includegraphics[width=\textwidth]{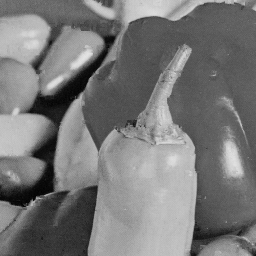}
  \caption{Exemplar-based\\
   $\textrm{MSE} = 39.34$}
 \end{subfigure}
 
 \begin{subfigure}[t]{0.29\textwidth}
  \centering
  \includegraphics[width=\textwidth]{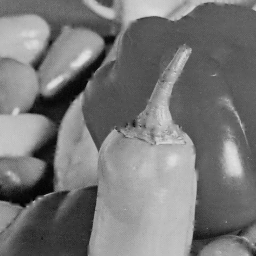}
  \caption{Harmonic\\
   $\textrm{MSE} = 21.12$}
 \end{subfigure}
 \quad 
 \begin{subfigure}[t]{0.29\textwidth}
  \centering
  \includegraphics[width=\textwidth]{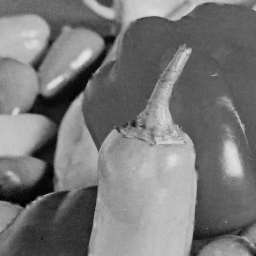}
  \caption{Biharmonic\\
   $\textrm{MSE} = 17.18$}
 \end{subfigure}
   \quad 
 \begin{subfigure}[t]{0.29\textwidth}
  \centering
  \includegraphics[
   width=\textwidth]{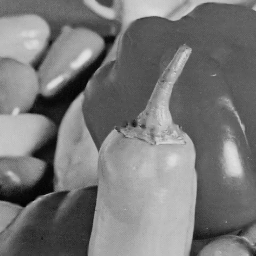}
  \caption{EED\\
   $\textrm{MSE} = 12.44$}
 \end{subfigure}
 \caption[Inpainting of ``peppers'' for text removal]{
  Inpainting of 
  image ``peppers'' overlaid with text for different inpainting methods. For 
  SPH inpainting, we used an isotropic Gaussian kernel. 
  Parameters for EED are
  $\lambda=0.4$ and $\sigma=1.7$.}
\end{figure}

\begin{figure}[htb]
 \captionsetup[subfigure]{justification=centering}
 \centering
 \begin{subfigure}[t]{0.29\textwidth}
  \centering
  \includegraphics[width=\textwidth]{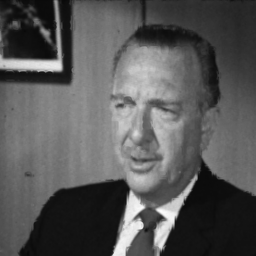}
  \caption{Zero order SPH\\
   $\textrm{MSE} = 20.42$}
 \end{subfigure}
  \quad 
 \begin{subfigure}[t]{0.29\textwidth}
  \centering
  \includegraphics[width=\textwidth]{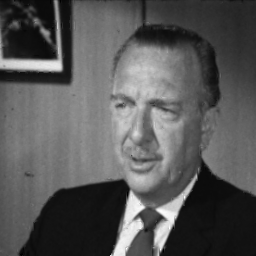}
  \caption{First order SPH\\
   $\textrm{MSE} = 11.82$}
 \end{subfigure}
   \quad 
 \begin{subfigure}[t]{0.29\textwidth}
  \centering
  \includegraphics[width=\textwidth]{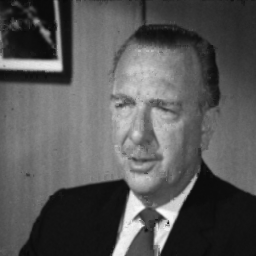}
  \caption{Exemplar-based\\
   $\textrm{MSE} = 24.24$}
 \end{subfigure}
 
 \begin{subfigure}[t]{0.29\textwidth}
  \centering
  \includegraphics[width=\textwidth]{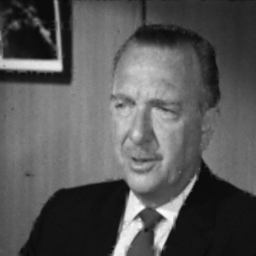}
  \caption{Harmonic\\
   $\textrm{MSE} = 16.35$}
 \end{subfigure}
 \quad 
 \begin{subfigure}[t]{0.29\textwidth}
  \centering
  \includegraphics[width=\textwidth]{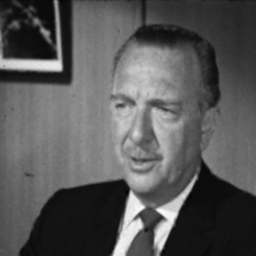}
  \caption{Biharmonic\\
   $\textrm{MSE} = 6.08$}
 \end{subfigure}
   \quad 
 \begin{subfigure}[t]{0.29\textwidth}
  \centering
  \includegraphics[
   width=\textwidth]{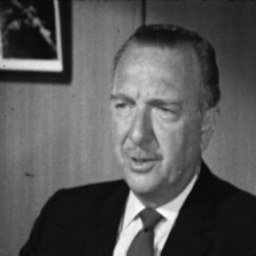}
  \caption{EED\\
   $\textrm{MSE} = 4.37$}
 \end{subfigure}
 \caption[Inpainting of ``walter'' for text removal]{
  Inpainting of 
  image ``walter'' overlaid with text for different inpainting methods. For 
  SPH inpainting, we used an isotropic Gaussian kernel. 
  Parameters for EED are
  $\lambda=0.1$ and $\sigma=0.9$.}
\end{figure}

\begin{figure}[htb]
 \captionsetup[subfigure]{justification=centering}
 \centering
 \begin{subfigure}[t]{0.29\textwidth}
  \centering
  \includegraphics[width=\textwidth]{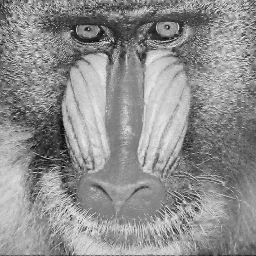}
  \caption{Zero order SPH\\
   $\textrm{MSE} = 183.52$}
 \end{subfigure}
  \quad 
 \begin{subfigure}[t]{0.29\textwidth}
  \centering
  \includegraphics[width=\textwidth]{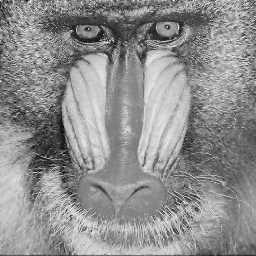}
  \caption{First order SPH\\
   $\textrm{MSE} = 193.30$}
 \end{subfigure}
   \quad 
 \begin{subfigure}[t]{0.29\textwidth}
  \centering
  \includegraphics[width=\textwidth]{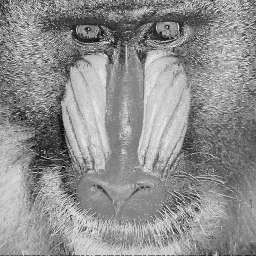}
  \caption{Exemplar-based\\
   $\textrm{MSE} = 273.64$}
 \end{subfigure}
 
 \begin{subfigure}[t]{0.29\textwidth}
  \centering
  \includegraphics[width=\textwidth]{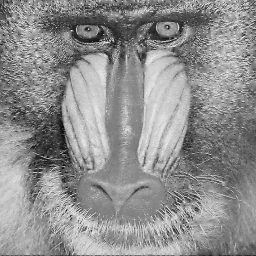}
  \caption{Harmonic\\
   $\textrm{MSE} = 170.17$}
 \end{subfigure}
 \quad 
 \begin{subfigure}[t]{0.29\textwidth}
  \centering
  \includegraphics[width=\textwidth]{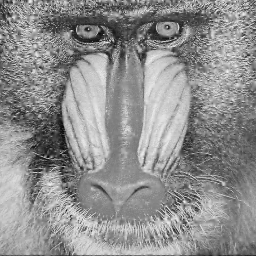}
  \caption{Biharmonic\\
   $\textrm{MSE} = 202.89$}
 \end{subfigure}
   \quad 
 \begin{subfigure}[t]{0.29\textwidth}
  \centering
  \includegraphics[
   width=\textwidth]{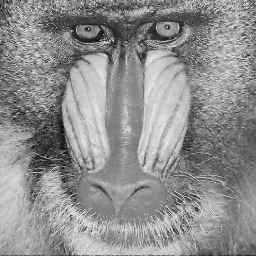}
  \caption{EED\\
   $\textrm{MSE} = 170.69$}
 \end{subfigure}
 \caption[Inpainting of ``baboon'' for text removal]{
  Inpainting of 
  image ``baboon'' overlaid with text for different inpainting methods. For 
  SPH inpainting, we used an isotropic Gaussian kernel. 
  Parameters for EED are
  $\lambda=8.6$ and $\sigma=1.9$.}
\end{figure}

\begin{figure}[htb]
 \captionsetup[subfigure]{justification=centering}
 \centering
 \begin{subfigure}[t]{0.29\textwidth}
  \centering
  \includegraphics[width=\textwidth]{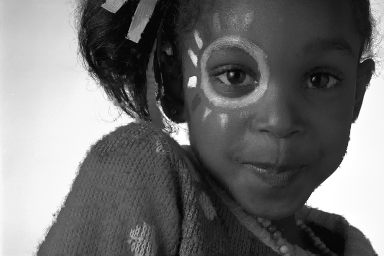}
  \caption{Zero order SPH\\
   $\textrm{MSE} = 31.60$}
 \end{subfigure}
  \quad 
 \begin{subfigure}[t]{0.29\textwidth}
  \centering
  \includegraphics[width=\textwidth]{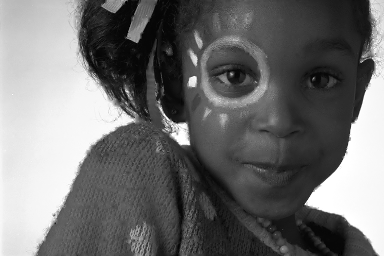}
  \caption{First order SPH\\
   $\textrm{MSE} = 29.88$}
 \end{subfigure}
   \quad 
 \begin{subfigure}[t]{0.29\textwidth}
  \centering
  \includegraphics[width=\textwidth]{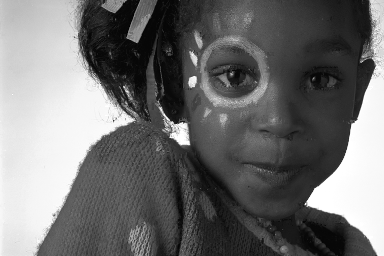}
  \caption{Exemplar-based\\
   $\textrm{MSE} = 45.88$}
 \end{subfigure}
 
 \begin{subfigure}[t]{0.29\textwidth}
  \centering
  \includegraphics[width=\textwidth]{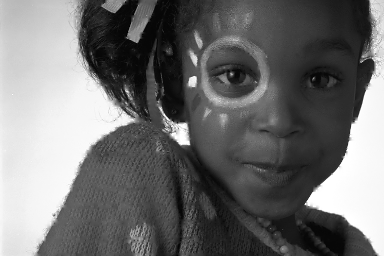}
  \caption{Harmonic\\
   $\textrm{MSE} = 29.08$}
 \end{subfigure}
 \quad 
 \begin{subfigure}[t]{0.29\textwidth}
  \centering
  \includegraphics[width=\textwidth]{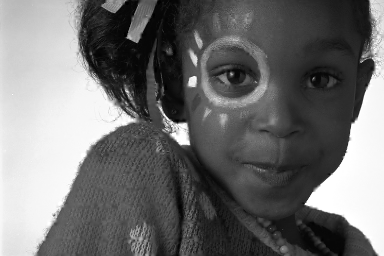}
  \caption{Biharmonic\\
   $\textrm{MSE} = 27.28$}
 \end{subfigure}
   \quad 
 \begin{subfigure}[t]{0.29\textwidth}
  \centering
  \includegraphics[
   width=\textwidth]{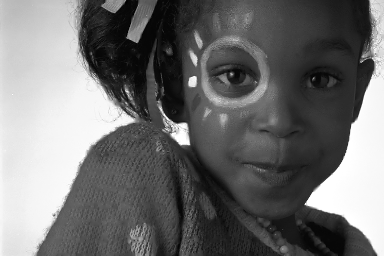}
  \caption{EED\\
   $\textrm{MSE} = 21.10$}
 \end{subfigure}
 \caption[Inpainting of ``girl'' for text removal]{
  Inpainting of 
  image ``girl'' overlaid with text for different inpainting methods. For 
  SPH inpainting, we used an isotropic Gaussian kernel. 
  Parameters for EED are
  $\lambda=1.1$ and $\sigma=1.9$.}
\end{figure}

\begin{figure}[htb]
 \captionsetup[subfigure]{justification=centering}
 \centering
 \begin{subfigure}[t]{0.29\textwidth}
  \centering
  \includegraphics[width=\textwidth]{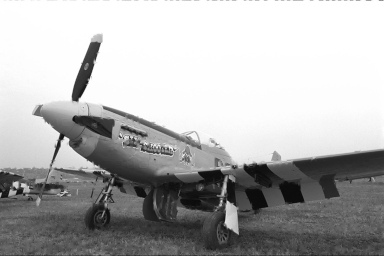}
  \caption{Zero order SPH\\
   $\textrm{MSE} = 36.76$}
 \end{subfigure}
  \quad 
 \begin{subfigure}[t]{0.29\textwidth}
  \centering
  \includegraphics[width=\textwidth]{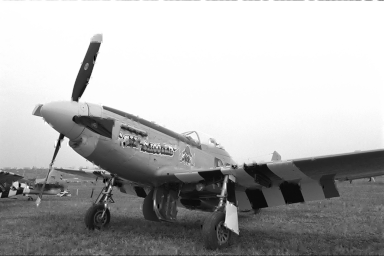}
  \caption{First order SPH\\
   $\textrm{MSE} = 36.71$}
 \end{subfigure}
   \quad 
 \begin{subfigure}[t]{0.29\textwidth}
  \centering
  \includegraphics[width=\textwidth]{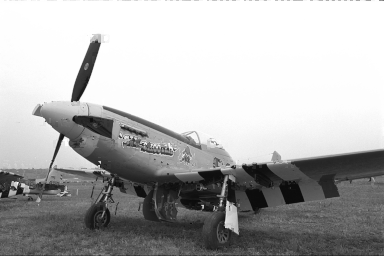}
  \caption{Exemplar-based\\
   $\textrm{MSE} = 54.67$}
 \end{subfigure}
 
 \begin{subfigure}[t]{0.29\textwidth}
  \centering
  \includegraphics[width=\textwidth]{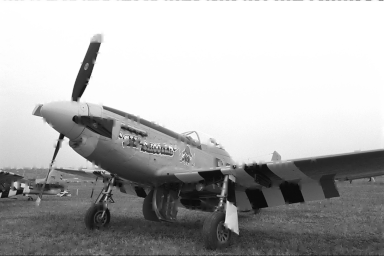}
  \caption{Harmonic\\
   $\textrm{MSE} = 33.18$}
 \end{subfigure}
 \quad 
 \begin{subfigure}[t]{0.29\textwidth}
  \centering
  \includegraphics[width=\textwidth]{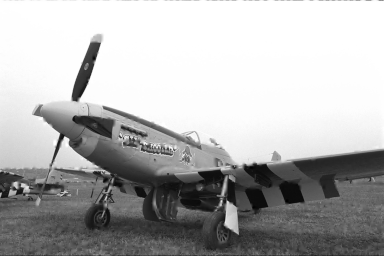}
  \caption{Biharmonic\\
   $\textrm{MSE} = 28.64$}
 \end{subfigure}
   \quad 
 \begin{subfigure}[t]{0.29\textwidth}
  \centering
  \includegraphics[
   width=\textwidth]{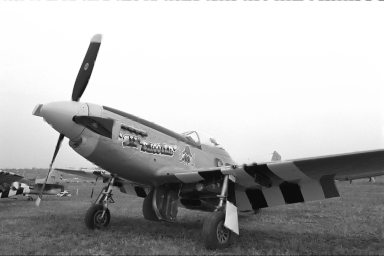}
  \caption{EED\\
   $\textrm{MSE} = 25.94$}
 \end{subfigure}
 \caption[Inpainting of ``plane'' for text removal]{
  Inpainting of 
  image ``plane'' overlaid with text for different inpainting methods. For 
  SPH inpainting, we used an isotropic Gaussian kernel. 
  Parameters for EED are
  $\lambda=2.6$ and $\sigma=0.7$.}
\end{figure}

\begin{figure}[htb]
 \captionsetup[subfigure]{justification=centering}
 \centering
 \begin{subfigure}[t]{0.29\textwidth}
  \centering
  \includegraphics[width=\textwidth]{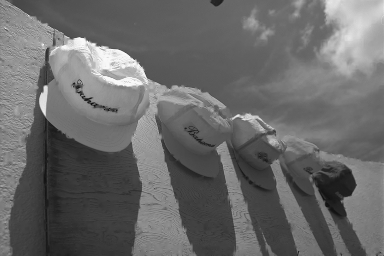}
  \caption{Zero order SPH\\
   $\textrm{MSE} = 31.05$}
 \end{subfigure}
  \quad 
 \begin{subfigure}[t]{0.29\textwidth}
  \centering
  \includegraphics[width=\textwidth]{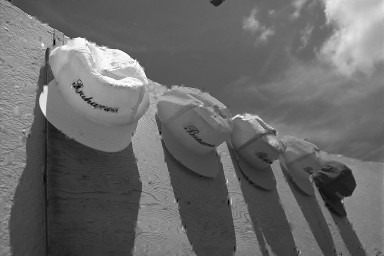}
  \caption{First order SPH\\
   $\textrm{MSE} = 34.67$}
 \end{subfigure}
   \quad 
 \begin{subfigure}[t]{0.29\textwidth}
  \centering
  \includegraphics[width=\textwidth]{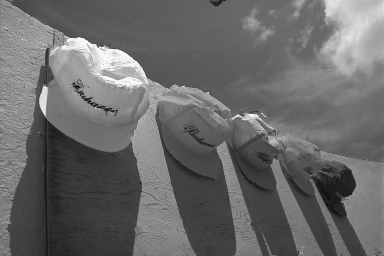}
  \caption{Exemplar-based\\
   $\textrm{MSE} = 42.41$}
 \end{subfigure}
 
 \begin{subfigure}[t]{0.29\textwidth}
  \centering
  \includegraphics[width=\textwidth]{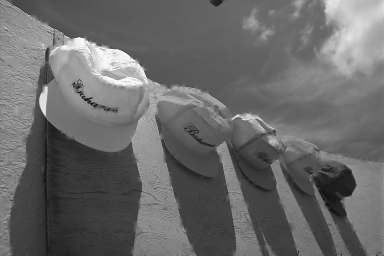}
  \caption{Harmonic\\
   $\textrm{MSE} = 26.53$}
 \end{subfigure}
 \quad 
 \begin{subfigure}[t]{0.29\textwidth}
  \centering
  \includegraphics[width=\textwidth]{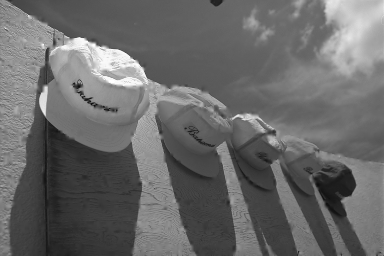}
  \caption{Biharmonic\\
   $\textrm{MSE} = 27.28$}
 \end{subfigure}
   \quad 
 \begin{subfigure}[t]{0.29\textwidth}
  \centering
  \includegraphics[
   width=\textwidth]{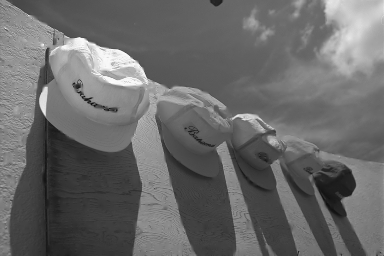}
  \caption{EED\\
   $\textrm{MSE} = 19.41$}
 \end{subfigure}
 \caption[Inpainting of ``hats'' for text removal]{
  Inpainting of 
  image ``hats'' overlaid with text for different inpainting methods. For 
  SPH inpainting, we used an isotropic Gaussian kernel. 
  Parameters for EED are
  $\lambda=0.5$ and $\sigma=2.0$.}
\end{figure}

When it comes to text removal, SPH inpainting again performs 
better than the exemplar-based approach, but not as good as the 
diffusion-based methods. Whether the zero or the first order consistency 
method performs depends on the image under consideration.

\section{Inpainting with Optimized Data}

In this section, we include some more results and comparisons 
for spatially and tonally optimized inpaintings. As kernel for SPH, we mostly
consider the common choice of Gaussian kernels. Only for the images of size
$256 \times 256$ do we also include results for other kernels as differences 
are not very large. For comparison, we also consider the results achieved by
harmonic and biharmonic inpainting, equipped with our Voronoi-based 
densification and tonally optimized gray values.

\begin{figure}[htb]
 \captionsetup[subfigure]{justification=centering}
 \centering
 \begin{subfigure}[t]{0.29\textwidth}
  \centering
  \includegraphics[
   width=\textwidth]{Figures_art/peppers_dens_D005_gauss_mix_to}
  \caption{Mixed order SPH\\
   $\textrm{MSE} = 16.29$}
 \end{subfigure}
 \quad
 \begin{subfigure}[t]{0.29\textwidth}
  \centering
 \includegraphics[
  width=\textwidth]{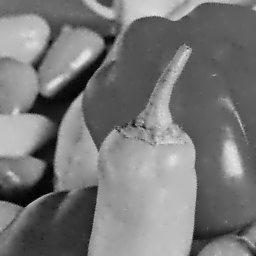}
  \caption{Harmonic\\
   $\textrm{MSE} = 22.66$}
 \end{subfigure}
 \quad
 \begin{subfigure}[t]{0.29\textwidth}
  \centering
 \includegraphics[
  width=\textwidth]{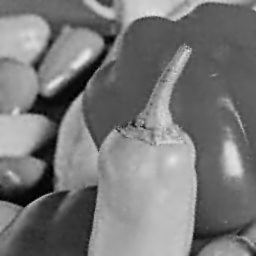}
  \caption{Biharmonic\\
   $\textrm{MSE} = 24.63$}
 \end{subfigure}
 \caption[Inpainting of ``peppers'' with spatially and tonally optimized 
  mask]{Inpainting of ``peppers'' with 5 \% spatially and 
  tonally optimized masks for different inpainting techniques. For SPH 
  inpainting, we used an isotropic Gaussian kernel.} 
\end{figure}

\begin{figure}[htb]
 \captionsetup[subfigure]{justification=centering}
 \centering
 \begin{subfigure}[t]{0.29\textwidth}
  \centering
  \includegraphics[
   width=\textwidth]{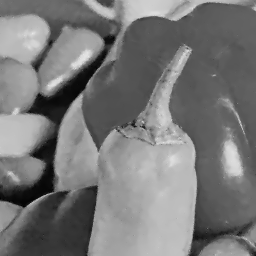}
  \caption{Gaussian\\
   $\textrm{MSE} = 22.46$}
 \end{subfigure}
 \quad
 \begin{subfigure}[t]{0.29\textwidth}
  \centering
  \includegraphics[
   width=\textwidth]{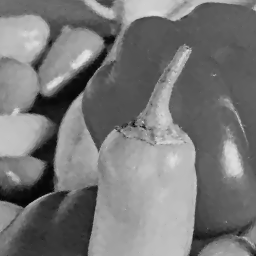}
  \caption{$C^{0}$-Mat\'{e}rn\\
   $\textrm{MSE} = 23.19$}
 \end{subfigure}
 \quad
 \begin{subfigure}[t]{0.29\textwidth}
  \centering
  \includegraphics[
   width=\textwidth]{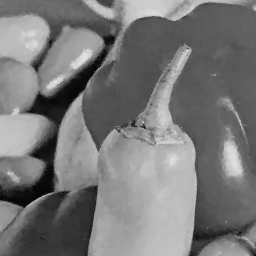}
  \caption{$C^{2}$-Mat\'{e}rn\\
   $\textrm{MSE} = 23.00$}
 \end{subfigure}

 \begin{subfigure}[t]{0.29\textwidth}
  \centering
  \includegraphics[
   width=\textwidth]{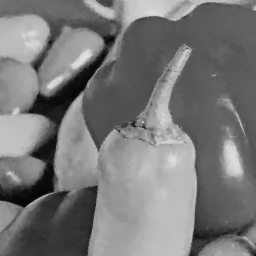}
  \caption{Lucy\\
   $\textrm{MSE} = 23.75$}
 \end{subfigure}
 \quad 
 \begin{subfigure}[t]{0.29\textwidth}
  \centering
  \includegraphics[
   width=\textwidth]{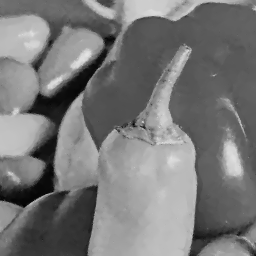}
  \caption{cubic spline\\
   $\textrm{MSE} = 24.17$}
 \end{subfigure}
 \quad
 \begin{subfigure}[t]{0.29\textwidth}
  \centering
  \includegraphics[
   width=\textwidth]{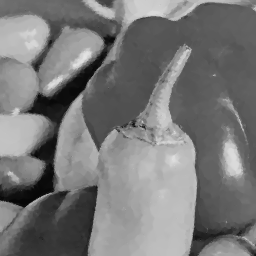}
  \caption{$C^{4}$-Wendland\\
   $\textrm{MSE} = 27.62$}
 \end{subfigure}
 \caption[Inpainting of ``peppers'' with spatially and
 tonally optimized mask]{Inpainting of ``peppers'' with a 5 \% 
 spatially and tonally optimized mask with a zero order consistency method and 
 anisotropic kernels.}  
\end{figure}

\begin{figure}[htb]
 \captionsetup[subfigure]{justification=centering}
 \centering
 \begin{subfigure}[t]{0.29\textwidth}
  \centering
  \includegraphics[
   width=\textwidth]{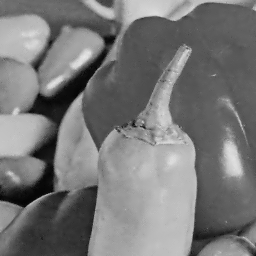}
  \caption{Gaussian\\
   $\textrm{MSE} = 13.69$}
 \end{subfigure}
 \quad
 \begin{subfigure}[t]{0.29\textwidth}
  \centering
  \includegraphics[
   width=\textwidth]{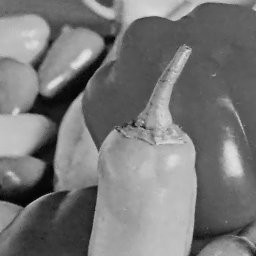}
  \caption{$C^{0}$-Mat\'{e}rn\\
   $\textrm{MSE} = 13.97$}
 \end{subfigure}
 \quad
 \begin{subfigure}[t]{0.29\textwidth}
  \centering
  \includegraphics[
   width=\textwidth]{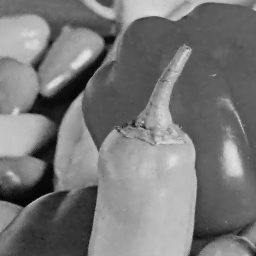}
  \caption{$C^{2}$-Mat\'{e}rn\\
   $\textrm{MSE} = 14.12$}
 \end{subfigure}

 \begin{subfigure}[t]{0.29\textwidth}
  \centering
  \includegraphics[
   width=\textwidth]{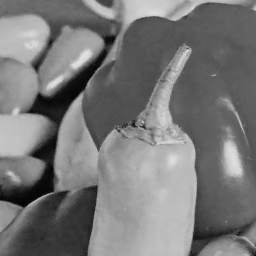}
  \caption{Lucy\\
   $\textrm{MSE} = 14.72$}
 \end{subfigure}
 \quad
 \begin{subfigure}[t]{0.29\textwidth}
  \centering
  \includegraphics[
   width=\textwidth]{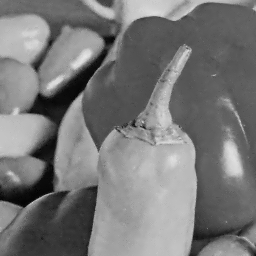}
  \caption{cubic spline\\
   $\textrm{MSE} = 14.75$}
 \end{subfigure}
 \quad
 \begin{subfigure}[t]{0.29\textwidth}
  \centering
  \includegraphics[
   width=\textwidth]{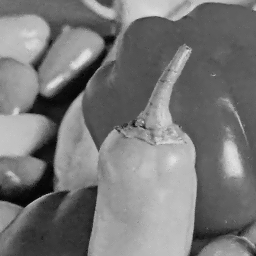}
  \caption{$C^{4}$-Wendland\\
   $\textrm{MSE} = 15.99$}
 \end{subfigure}
 \caption[Inpainting of ``peppers'' with spatially and
 tonally optimized mask]{Inpainting of ``peppers'' with a 5 \% 
 spatially and tonally optimized mask with a mixed order consistency method and 
 anisotropic kernels.}  
\end{figure}

\begin{figure}[htb]
 \captionsetup[subfigure]{justification=centering}
 \centering
 \begin{subfigure}[t]{0.29\textwidth}
  \centering
  \includegraphics[
   width=\textwidth]{Figures_art/walter_dens_D005_gauss_mix_to}
  \caption{Mixed order SPH\\
   $\textrm{MSE} = 5.15$}
 \end{subfigure}
 \quad
 \begin{subfigure}[t]{0.29\textwidth}
  \centering
 \includegraphics[
  width=\textwidth]{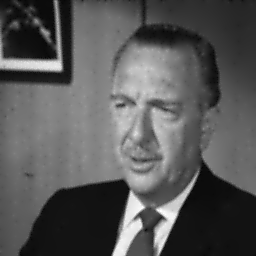}
  \caption{Harmonic\\
   $\textrm{MSE} = 9.20$}
 \end{subfigure}
 \quad
 \begin{subfigure}[t]{0.29\textwidth}
  \centering
 \includegraphics[
  width=\textwidth]{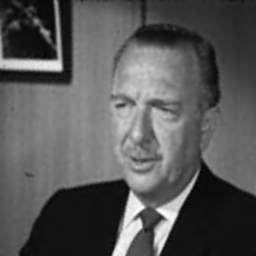}
  \caption{Biharmonic\\
   $\textrm{MSE} = 6.31$}
 \end{subfigure}
 \caption[Inpainting of ``walter'' with spatially and tonally optimized 
  mask]{Inpainting of ``walter'' with 5 \% spatially and 
  tonally optimized masks for different inpainting techniques. For SPH 
  inpainting, we used an isotropic Gaussian kernel.}
\end{figure}

\begin{figure}[htb]
 \captionsetup[subfigure]{justification=centering}
 \centering
 \begin{subfigure}[t]{0.29\textwidth}
  \centering
  \includegraphics[
   width=\textwidth]{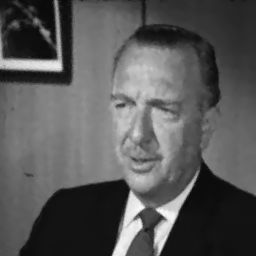}
  \caption{Gaussian\\
   $\textrm{MSE} = 9.10$}
 \end{subfigure}
 \quad
 \begin{subfigure}[t]{0.29\textwidth}
  \centering
  \includegraphics[
   width=\textwidth]{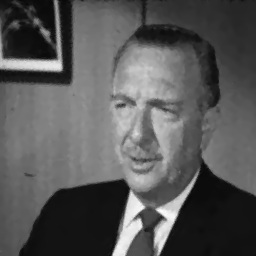}
  \caption{$C^{0}$-Mat\'{e}rn\\
   $\textrm{MSE} = 9.40$}
 \end{subfigure}
 \quad
 \begin{subfigure}[t]{0.29\textwidth}
  \centering
  \includegraphics[
   width=\textwidth]{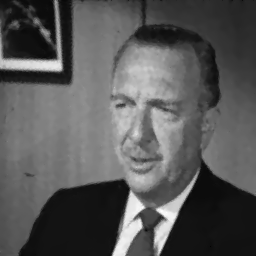}
  \caption{$C^{2}$-Mat\'{e}rn\\
   $\textrm{MSE} = 9.25$}
 \end{subfigure}

 \begin{subfigure}[t]{0.29\textwidth}
  \centering
  \includegraphics[
   width=\textwidth]{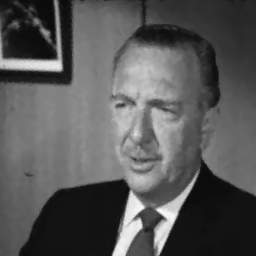}
  \caption{Lucy\\
   $\textrm{MSE} = 10.09$}
 \end{subfigure}
 \quad 
 \begin{subfigure}[t]{0.29\textwidth}
  \centering
  \includegraphics[
   width=\textwidth]{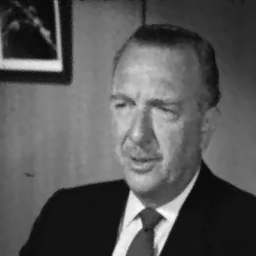}
  \caption{cubic spline\\
   $\textrm{MSE} = 10.60$}
 \end{subfigure}
 \quad
 \begin{subfigure}[t]{0.29\textwidth}
  \centering
  \includegraphics[
   width=\textwidth]{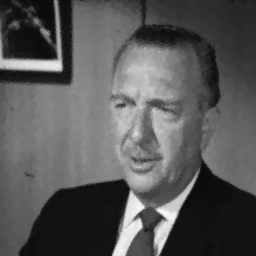}
  \caption{$C^{4}$-Wendland\\
   $\textrm{MSE} = 12.67$}
 \end{subfigure}
 \caption[Inpainting of ``walter'' with spatially and
 tonally optimized mask]{Inpainting of ``walter'' with a 5 \% 
 spatially and tonally optimized mask with a zero order consistency method and
 anisotropic kernels.}
\end{figure}

\begin{figure}[htb]
 \captionsetup[subfigure]{justification=centering}
 \centering
 \begin{subfigure}[t]{0.29\textwidth}
  \centering
  \includegraphics[
   width=\textwidth]{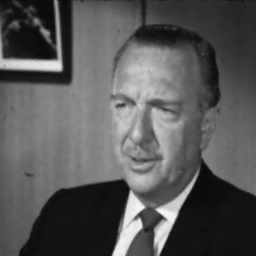}
  \caption{Gaussian\\
   $\textrm{MSE} = 4.38$}
 \end{subfigure}
 \quad
 \begin{subfigure}[t]{0.29\textwidth}
  \centering
  \includegraphics[
   width=\textwidth]{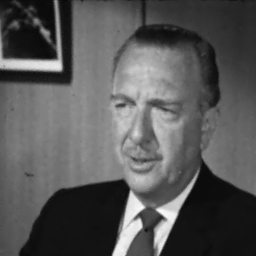}
  \caption{$C^{0}$-Mat\'{e}rn\\
   $\textrm{MSE} = 4.21$}
 \end{subfigure}
 \quad
 \begin{subfigure}[t]{0.29\textwidth}
  \centering
  \includegraphics[
   width=\textwidth]{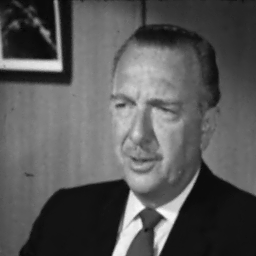}
  \caption{$C^{2}$-Mat\'{e}rn\\
   $\textrm{MSE} = 4.27$}
 \end{subfigure}

 \begin{subfigure}[t]{0.29\textwidth}
  \centering
  \includegraphics[
   width=\textwidth]{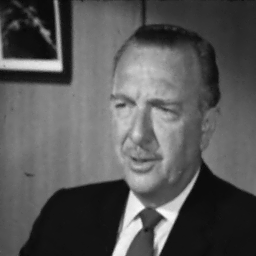}
  \caption{Lucy\\
   $\textrm{MSE} = 4.91$}
 \end{subfigure}
 \quad 
 \begin{subfigure}[t]{0.29\textwidth}
  \centering
  \includegraphics[
   width=\textwidth]{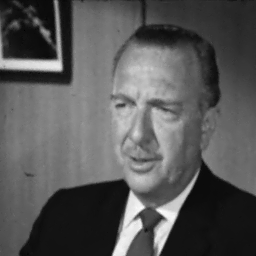}
  \caption{cubic spline\\
   $\textrm{MSE} = 4.94$}
 \end{subfigure}
 \quad
 \begin{subfigure}[t]{0.29\textwidth}
  \centering
  \includegraphics[
   width=\textwidth]{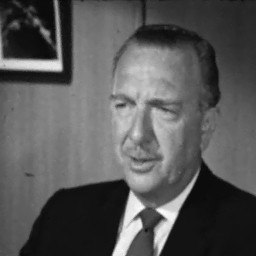}
  \caption{$C^{4}$-Wendland\\
   $\textrm{MSE} = 5.32$}
 \end{subfigure}
 \caption[Inpainting of ``walter'' with spatially and
 tonally optimized mask]{Inpainting of ``walter'' with a 5 \%
 spatially and tonally optimized mask with a mixed order consistency method and
 anisotropic kernels.}
\end{figure}

\begin{figure}[htb]
 \captionsetup[subfigure]{justification=centering}
 \centering
 \begin{subfigure}[t]{0.29\textwidth}
  \centering
  \includegraphics[
   width=\textwidth]{Figures_art/baboon_dens_D010_gauss_mix_to}
  \caption{Mixed order SPH\\
   $\textrm{MSE} = 223.37$}
 \end{subfigure}
 \quad
 \begin{subfigure}[t]{0.29\textwidth}
  \centering
 \includegraphics[
  width=\textwidth]{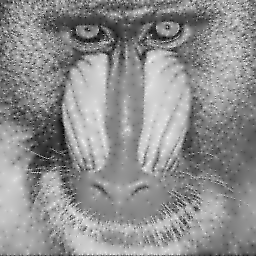}
  \caption{Harmonic\\
   $\textrm{MSE} = 283.96$}
 \end{subfigure}
 \quad
 \begin{subfigure}[t]{0.29\textwidth}
  \centering
 \includegraphics[
  width=\textwidth]{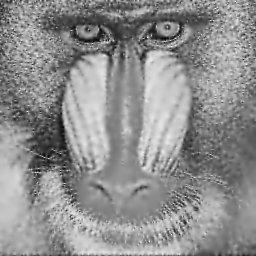}
  \caption{Biharmonic\\
   $\textrm{MSE} = 326.00$}
 \end{subfigure}
 \caption[Inpainting of ``baboon'' with spatially and tonally optimized 
  mask]{Inpainting of ``baboon'' with 10 \% spatially and 
  tonally optimized masks for different inpainting techniques. For SPH 
  inpainting, we used an isotropic Gaussian kernel.}
\end{figure}

\begin{figure}[htb]
 \captionsetup[subfigure]{justification=centering}
 \centering
 \begin{subfigure}[t]{0.29\textwidth}
  \centering
  \includegraphics[
   width=\textwidth]{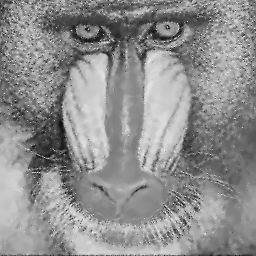}
  \caption{Gaussian\\
   $\textrm{MSE} = 294.17$}
 \end{subfigure}
 \quad
 \begin{subfigure}[t]{0.29\textwidth}
  \centering
  \includegraphics[
   width=\textwidth]{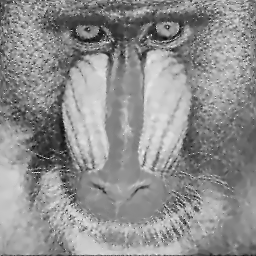}
  \caption{$C^{0}$-Mat\'{e}rn\\
   $\textrm{MSE} = 289.53$}
 \end{subfigure}
 \quad
 \begin{subfigure}[t]{0.29\textwidth}
  \centering
  \includegraphics[
   width=\textwidth]{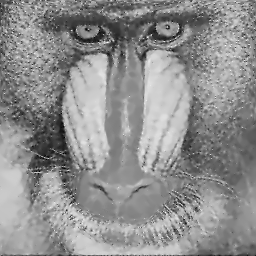}
  \caption{$C^{2}$-Mat\'{e}rn\\
   $\textrm{MSE} = 290.03$}
 \end{subfigure}

 \begin{subfigure}[t]{0.29\textwidth}
  \centering
  \includegraphics[
   width=\textwidth]{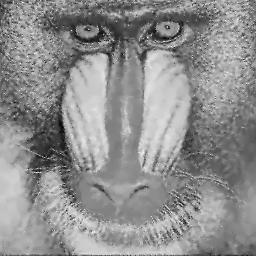}
  \caption{Lucy\\
   $\textrm{MSE} = 305.76$}
 \end{subfigure}
 \quad 
 \begin{subfigure}[t]{0.29\textwidth}
  \centering
  \includegraphics[
   width=\textwidth]{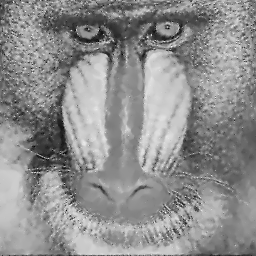}
  \caption{cubic spline\\
   $\textrm{MSE} = 306.82$}
 \end{subfigure}
 \quad
 \begin{subfigure}[t]{0.29\textwidth}
  \centering
  \includegraphics[
   width=\textwidth]{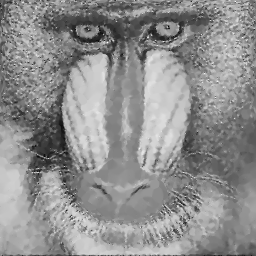}
  \caption{$C^{4}$-Wendland\\
   $\textrm{MSE} = 313.08$}
 \end{subfigure}
 \caption[Inpainting of ``baboon'' with spatially and
 tonally optimized mask]{Inpainting of ``baboon'' with a 10 \% 
 spatially and tonally optimized mask with a zero order consistency method and 
 anisotropic kernels.} 
\end{figure}
 
\begin{figure}[htb]
 \captionsetup[subfigure]{justification=centering}
 \centering
 \begin{subfigure}[t]{0.29\textwidth}
  \centering
  \includegraphics[
   width=\textwidth]{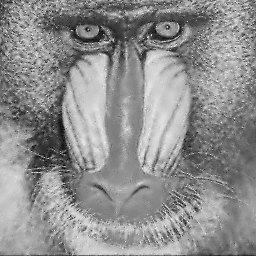}
  \caption{Gaussian\\
   $\textrm{MSE} = 220.82$}
 \end{subfigure}
 \quad
 \begin{subfigure}[t]{0.29\textwidth}
  \centering
  \includegraphics[
   width=\textwidth]{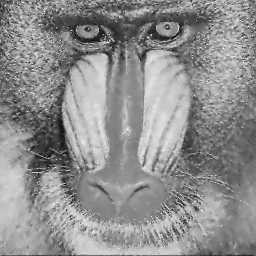}
  \caption{$C^{0}$-Mat\'{e}rn\\
   $\textrm{MSE} = 222.31$}
 \end{subfigure}
 \quad
 \begin{subfigure}[t]{0.29\textwidth}
  \centering
  \includegraphics[
   width=\textwidth]{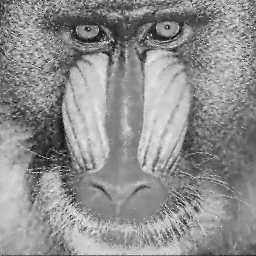}
  \caption{$C^{2}$-Mat\'{e}rn\\
   $\textrm{MSE} = 220.19$}
 \end{subfigure}

 \begin{subfigure}[t]{0.29\textwidth}
  \centering
  \includegraphics[
   width=\textwidth]{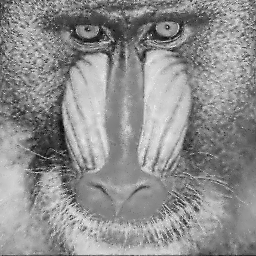}
  \caption{Lucy\\
   $\textrm{MSE} = 226.21$}
 \end{subfigure}
 \quad
 \begin{subfigure}[t]{0.29\textwidth}
  \centering
  \includegraphics[
   width=\textwidth]{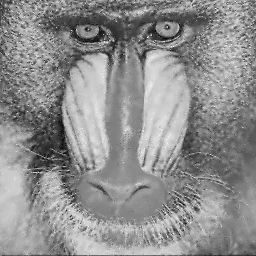}
  \caption{cubic spline\\
   $\textrm{MSE} = 226.60$}
 \end{subfigure}
 \quad
 \begin{subfigure}[t]{0.29\textwidth}
  \centering
  \includegraphics[
   width=\textwidth]{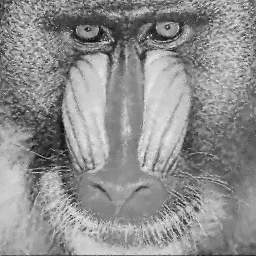}
  \caption{$C^{4}$-Wendland\\
   $\textrm{MSE} = 227.16$}
 \end{subfigure}
 \caption[Inpainting of ``baboon'' with spatially and
 tonally optimized mask]{Inpainting of ``baboon'' with a 10 \% 
 spatially and tonally optimized mask with a mixed order consistency method and
 anisotropic kernels.} 
\end{figure}

\begin{figure}[htb]
 \captionsetup[subfigure]{justification=centering}
 \centering
 \begin{subfigure}[t]{0.29\textwidth}
  \centering
  \includegraphics[
   width=\textwidth]{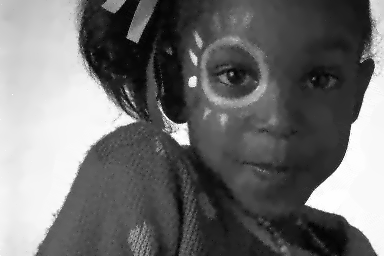}
  \caption{Zero order isotrop.~SPH\\
   $\textrm{MSE} = 36.74$}
 \end{subfigure}
 \quad
 \begin{subfigure}[t]{0.29\textwidth}
  \centering
  \includegraphics[
   width=\textwidth]{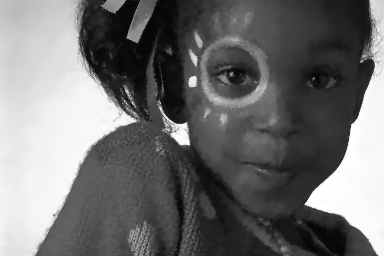}
  \caption{Zero order aniso.~SPH\\
   $\textrm{MSE} = 32.59$}
 \end{subfigure}
 \quad
 \begin{subfigure}[t]{0.29\textwidth}
  \centering
  \includegraphics[
   width=\textwidth]{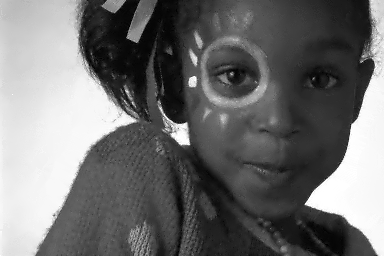}
  \caption{Mixed order isotrop.~SPH\\
   $\textrm{MSE} = 24.49$}
 \end{subfigure}
 
 \begin{subfigure}[t]{0.29\textwidth}
  \centering
 \includegraphics[
  width=\textwidth]{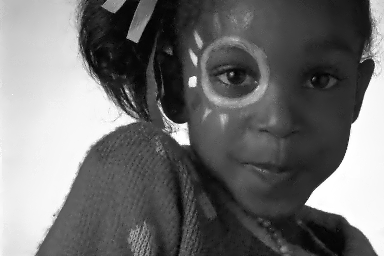}
  \caption{Mixed order aniso.~SPH\\
   $\textrm{MSE} = 21.86$}
 \end{subfigure}
 \quad
 \begin{subfigure}[t]{0.29\textwidth}
  \centering
 \includegraphics[
  width=\textwidth]{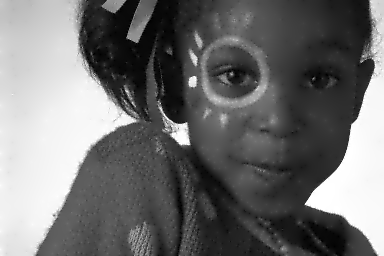}
  \caption{Harmonic\\
   $\textrm{MSE} = 32.77$}
 \end{subfigure}
 \quad
 \begin{subfigure}[t]{0.29\textwidth}
  \centering
  \includegraphics[
   width=\textwidth]{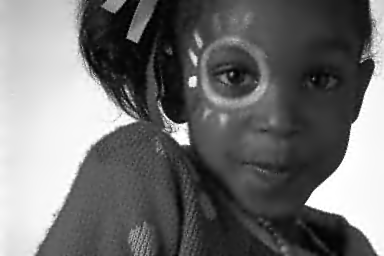}
  \caption{Biharmonic\\
   $\textrm{MSE} = 38.88$}
 \end{subfigure}
 \caption[Inpainting of ``girl'' with spatially and tonally optimized 
  mask]{Inpainting of ``girl'' with 5 \% spatially and tonally
  optimized masks for different inpainting methods. For SPH inpainting, we used 
  a Gaussian kernel.}
\end{figure}

\begin{figure}[htb]
 \captionsetup[subfigure]{justification=centering}
 \centering
 \begin{subfigure}[t]{0.29\textwidth}
  \centering
  \includegraphics[
   width=\textwidth]{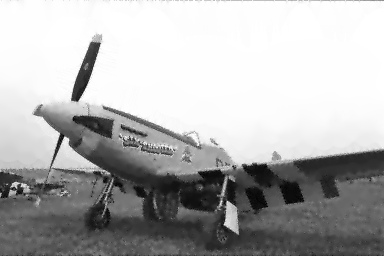}
  \caption{Zero order isotrop.~SPH\\
   $\textrm{MSE} = 33.74$}
 \end{subfigure}
 \quad
 \begin{subfigure}[t]{0.29\textwidth}
  \centering
  \includegraphics[
   width=\textwidth]{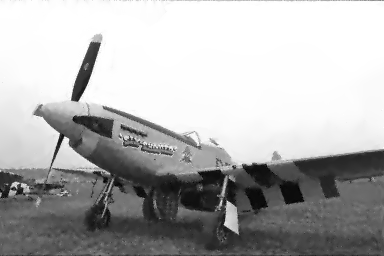}
  \caption{Zero order aniso.~SPH\\
   $\textrm{MSE} = 27.92$}
 \end{subfigure}
 \quad
 \begin{subfigure}[t]{0.29\textwidth}
  \centering
  \includegraphics[
   width=\textwidth]{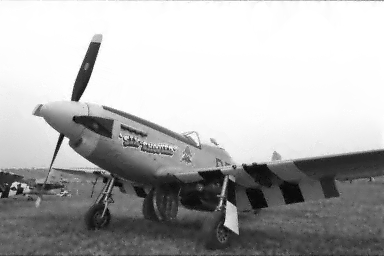}
  \caption{Mixed order isotrop.~SPH\\
   $\textrm{MSE} = 19.60$}
 \end{subfigure}
 
 \begin{subfigure}[t]{0.29\textwidth}
  \centering
 \includegraphics[
  width=\textwidth]{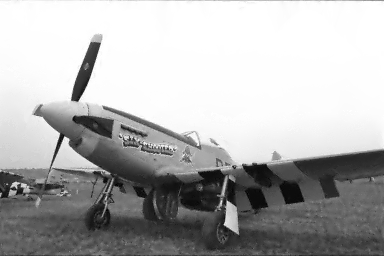}
  \caption{Mixed order aniso.~SPH\\
   $\textrm{MSE} = 16.20$}
 \end{subfigure}
 \quad
 \begin{subfigure}[t]{0.29\textwidth}
  \centering
 \includegraphics[
  width=\textwidth]{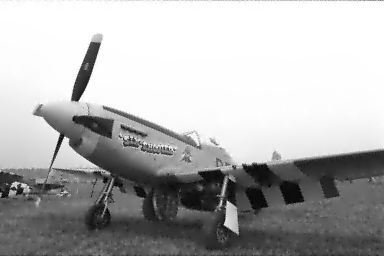}
  \caption{Harmonic\\
   $\textrm{MSE} = 30.28$}
 \end{subfigure}
 \quad
 \begin{subfigure}[t]{0.29\textwidth}
  \centering
  \includegraphics[
   width=\textwidth]{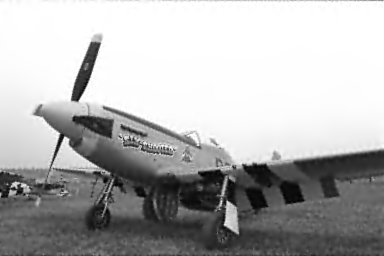}
  \caption{Biharmonic\\
   $\textrm{MSE} = 33.20$}
 \end{subfigure}
 \caption[Inpainting of ``plane'' with spatially and tonally optimized 
  mask]{Inpainting of ``plane'' with 5 \% spatially and tonally
  optimized masks for different inpainting methods. For SPH inpainting, we used
  a Gaussian kernel.}
\end{figure}

\begin{figure}[htb]
 \captionsetup[subfigure]{justification=centering}
 \centering
 \begin{subfigure}[t]{0.29\textwidth}
  \centering
  \includegraphics[
   width=\textwidth]{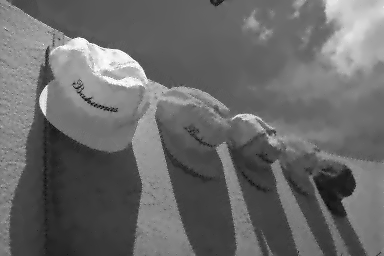}
  \caption{Zero order isotrop.~SPH\\
   $\textrm{MSE} = 30.42$}
 \end{subfigure}
 \quad 
 \begin{subfigure}[t]{0.29\textwidth}
  \centering
  \includegraphics[
   width=\textwidth]{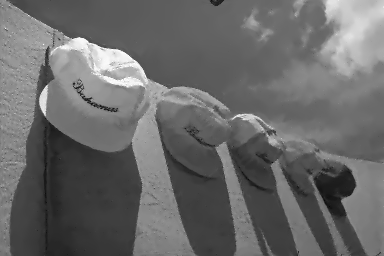}
  \caption{Zero order aniso.~SPH\\
   $\textrm{MSE} = 25.31$}
 \end{subfigure}
 \quad
 \begin{subfigure}[t]{0.29\textwidth}
  \centering
  \includegraphics[
   width=\textwidth]{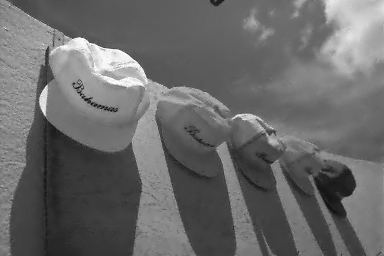}
  \caption{Mixed order isotrop.~SPH\\
   $\textrm{MSE} = 20.42$}
 \end{subfigure}
 
\begin{subfigure}[t]{0.29\textwidth}
  \centering
  \includegraphics[
   width=\textwidth]{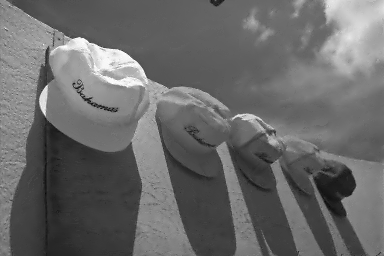}
  \caption{Mixed order aniso.~SPH\\
   $\textrm{MSE} = 17.18$}
 \end{subfigure} 
\quad
 \begin{subfigure}[t]{0.29\textwidth}
  \centering
 \includegraphics[
  width=\textwidth]{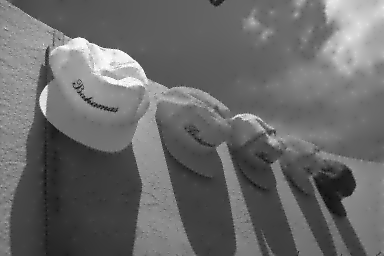}
  \caption{Harmonic\\
   $\textrm{MSE} = 27.19$}
 \end{subfigure}
 \quad
 \begin{subfigure}[t]{0.29\textwidth}
  \centering
 \includegraphics[
  width=\textwidth]{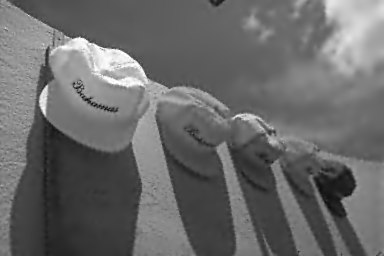}
  \caption{Biharmonic\\
   $\textrm{MSE} = 32.53$}
 \end{subfigure}
 \caption[Inpainting of ``hats'' with spatially and tonally optimized 
  mask]{Inpainting of ``hats'' with 5 \% spatially and tonally
  optimized masks for different inpainting methods. For SPH inpainting, we used 
  a Gaussian kernel.}
\end{figure}

As expected, the mixed order anisotropic SPH inpainting
performs best in all cases with improvements over harmonic or biharmonic 
inpaintings between 22 \% for ``baboon'' and 55 \% for ``walter''.